\documentclass{memo-l}
 \pdfoutput=1 
\usepackage{amssymb,verbatim}
\usepackage{amsmath,amsfonts}
\usepackage[mathscr]{euscript} 
\usepackage{amsthm}
\usepackage{color}
\usepackage{url}
\usepackage{graphicx} 
\usepackage{enumitem} 
\usepackage{lineno,xcolor} 
\usepackage{hyperref} 
\hypersetup{colorlinks} 

\usepackage[lined,algonl,boxed,norelsize]{algorithm2e} 
\usepackage{diagbox}
\usepackage{threeparttable} 
\usepackage{extarrows} 
\usepackage{tikz-cd} 
\usepackage{ltablex} 

\usepackage{lipsum} 

\usepackage{graphicx} 
\usepackage{tikz}
\usetikzlibrary{arrows.meta} 

\usepackage{standalone} 

\setlist[enumerate]{
label=\textnormal{({\roman*})},
ref={\roman*}}

\makeatletter
\def\th@plain{%
  \thm@notefont{}
  \itshape 
}
\def\th@definition{%
  \thm@notefont{}
  \normalfont 
}
\makeatother

\makeatletter
\newtheorem*{rep@theorem}{\rep@title}
\newcommand{\newreptheorem}[2]{%
\newenvironment{rep#1}[1]{%
 \def\rep@title{#2 \ref{##1}}%
 \begin{rep@theorem}}%
 {\end{rep@theorem}}}
\makeatother

\newreptheorem{theorem}{Theorem}
\newreptheorem{corollary}{Corollary}

\newtheorem{theorem}{Theorem}[chapter]
\newtheorem{lemma}[theorem]{Lemma}
\newtheorem{proposition}[theorem]{Proposition}

\newtheorem{corollary}[theorem]{Corollary}
\newtheorem{conjecture}[theorem]{Conjecture}
\newtheorem{question}[theorem]{Question}

\theoremstyle{remark}
\newtheorem*{remark}{Remark}

\theoremstyle{definition}

\newtheorem{examplex}[theorem]{Example}
\newenvironment{example}
  {\pushQED{\qed}\examplex}
  {\popQED\endexamplex}

  \newtheorem{definitionx}[theorem]{Definition}
\newenvironment{definition}
  {\pushQED{\qed}\definitionx}
  {\popQED\enddefinitionx}

\numberwithin{section}{chapter}
\numberwithin{equation}{chapter}
\numberwithin{figure}{chapter}
\numberwithin{table}{chapter}

\makeindex

\makeatletter
\newcommand*{\da@rightarrow}{\mathchar"0\hexnumber@\symAMSa 4B }
\newcommand*{\da@leftarrow}{\mathchar"0\hexnumber@\symAMSa 4C }
\newcommand*{\xdashrightarrow}[2][]{%
  \mathrel{%
    \mathpalette{\da@xarrow{#1}{#2}{}\da@rightarrow{\,}{}}{}%
  }%
}
\newcommand{\xdashleftarrow}[2][]{%
  \mathrel{%
    \mathpalette{\da@xarrow{#1}{#2}\da@leftarrow{}{}{\,}}{}%
  }%
}
\newcommand*{\da@xarrow}[7]{%
  \sbox0{$\ifx#7\scriptstyle\scriptscriptstyle\else\scriptstyle\fi#5#1#6\m@th$}%
  \sbox2{$\ifx#7\scriptstyle\scriptscriptstyle\else\scriptstyle\fi#5#2#6\m@th$}%
  \sbox4{$#7\dabar@\m@th$}%
  \dimen@=\wd0 %
  \ifdim\wd2 >\dimen@
    \dimen@=\wd2 %
  \fi
  \count@=2 %
  \def\da@bars{\dabar@\dabar@}%
  \@whiledim\count@\wd4<\dimen@\do{%
    \advance\count@\@ne
    \expandafter\def\expandafter\da@bars\expandafter{%
      \da@bars
      \dabar@ 
    }%
  }%
  \mathrel{#3}%
  \mathrel{%
    \mathop{\da@bars}\limits
    \ifx\\#1\\%
    \else
      _{\copy0}%
    \fi
    \ifx\\#2\\%
    \else
      ^{\copy2}%
    \fi
  }%
  \mathrel{#4}%
}
\makeatother

\newcommand{\Grt}{\mathscr{K}} 
\newcommand{\Mon}{\mathscr{M}} 
\newcommand{\Msf}{\mathscr{F}} 
\newcommand{\Net}{\mathscr{N}} 
\newcommand{\Subnet}{\mathscr{S}} 
\newcommand{\Agn}{\mathscr{N}} 
\newcommand{\Qcr}{\mathscr{Q}} 
\newcommand{\Gc}{\mathcal{G}} 

\newcommand{\old}[1]{}

\newcommand{\PP}{\mathcal{P}} 
\DeclareMathOperator{\Loc}{{Loc}} 
\DeclareMathOperator{\Rec}{{Rec}} 
\DeclareMathOperator{\Lrec}{\overline{{Rec}}} 
\DeclareMathOperator{\End}{{End}} 
\DeclareMathOperator{\Eig}{\mathcal{E}} 
\DeclareMathOperator{\supp}{{supp}} 
\DeclareMathOperator{\Tor}{{Tor}} 
\DeclareMathOperator{\lvl}{{lvl}} 
\DeclareMathOperator{\cpt}{{cap}} 

\DeclareMathOperator{\Green}{\mathfrak{G}} 

\DeclareMathOperator{\Cb}{\mathbb{C}} 

\DeclareMathOperator{\diff}{\textnormal{diff}}  

\DeclareMathOperator{\Pham}{\textnormal{Pham}} 

\DeclareMathOperator{\st}{\mathsf{ST}} 

\newcommand{\dashrlarrow}{\dashrightarrow\dashleftarrow} 
\newcommand{\rlarrow}{\rightarrow\leftarrow} 

\DeclareMathOperator{\Stop}{\textnormal{Stop}} 

\DeclareMathOperator{\act}{\textnormal{act}} 
\newcommand{\db}{\mathbf{d}} 

\newcommand{\bb}{\mathbf{b}} 
\newcommand{\cc}{\mathbf{c}} 
\newcommand{\ee}{\mathbf{e}} 
\newcommand{\vb}{\mathbf{v}} 
\newcommand{\w}{\mathbf{w}} 
\newcommand{\x}{\mathbf{x}} 
\newcommand{\y}{\mathbf{y}} 
\newcommand{\p}{\mathbf{p}} 
\newcommand{\q}{\mathbf{q}} 
\newcommand{\kk}{\mathbf{k}} 
\newcommand{\satu}{\mathbf{1}} 
\newcommand{\nol}{\mathbf{0}} 
\newcommand{\rr}{\mathbf{r}} 
\newcommand{\s}{\mathbf{s}} 
\newcommand{\uu}{\mathbf{u}} 
\newcommand{\vv}{\mathbf{v}} 
\newcommand{\z}{\mathbf{z}} 
\newcommand{\m}{\mathbf{m}} 
\newcommand{\n}{\mathbf{n}} 
\newcommand{\A}{{A}} 
\newcommand{\D}{{D}} 
\newcommand{\Lap}{{L}} 
\newcommand{\Dartm}{\mathcal{D}} 
\newcommand{\M}{\mathcal{M}} 
\newcommand{\C}{\mathcal{C}} 
\newcommand{\F}{\mathcal{F}} 

\newcommand{\Z}{\mathbb{Z}} 
\newcommand{\N}{\mathbb{N}} 
\newcommand{\Q}{\mathbb{Q}} 
\newcommand{\R}{\mathbb{R}} 

\newcommand{\NN}{\mathbf{M}} 
\DeclareMathOperator{\outdeg}{\textnormal{outdeg}} 
\newcommand{\indeg}{\text{indeg}} 
\DeclareMathOperator{\Out}{{Out}} 

\newcommand{\wt}{\mathsf{wt}}

\begin{document}

\frontmatter

\title[Abelian networks IV.]{Abelian networks IV. Dynamics of nonhalting networks}


\author{Swee Hong Chan}
\address{Swee Hong Chan, Department of Mathematics, Cornell University, Ithaca, NY 14853.
{\tt \url{http://www.math.cornell.edu/~sc2637/}}}
\email{sweehong@math.cornell.edu}
\thanks{}

\author{Lionel Levine}
\address{Lionel Levine, Department of Mathematics, Cornell University, Ithaca, NY 14853.
{\tt \url{http://www.math.cornell.edu/~levine}}}
\email{levine@math.cornell.edu}
\thanks{LL was supported by \href{http://www.nsf.gov/awardsearch/showAward?AWD_ID=1455272}{NSF DMS-1455272} and a Sloan Fellowship.}

\date{\today}



\dedicatory{Dedicated to Sui Lien Peo and Bernard Ackerman}
\maketitle

\setcounter{tocdepth}{2}
\setcounter{secnumdepth}{4}
\tableofcontents

\begin{abstract}
An abelian network is a collection of communicating automata whose state transitions and message passing each satisfy a local commutativity condition. 
This paper is a continuation of the abelian networks series of Bond and Levine (2016), for which 
 we extend the theory of abelian networks that halt on all inputs to networks that can run forever. 
A nonhalting abelian network can be realized as a discrete dynamical system in many different ways, depending on the update order.
We show that certain features of the dynamics, such as minimal period length, have intrinsic definitions that do not require specifying an update order.

We give an intrinsic definition of the \emph{torsion group} of a finite irreducible (halting or nonhalting) abelian network, and show that it coincides with the critical group of Bond and Levine (2016) if the network is halting.
We show that the torsion group acts freely on the set of invertible recurrent components of the trajectory digraph, and identify when this action is transitive.

This perspective leads to new results even in the classical case of sinkless rotor networks (deterministic analogues of random walks). In Holroyd et. al (2008) it was shown that the recurrent configurations of a sinkless rotor network with just one chip are precisely the unicycles (spanning subgraphs with a unique oriented cycle, with the chip on the cycle). We generalize this result to abelian mobile agent networks with any number of chips. We give formulas for generating series such as  
\[ \sum_{n \geq 1} r_n z^n = \det (\frac{1}{1-z}D - A )  \]
where $r_n$ is the number of recurrent chip-and-rotor configurations with $n$ chips; $D$ is the diagonal matrix of outdegrees, and $A$ is the adjacency matrix. A consequence is that the sequence $(r_n)_{n \geq 1}$ completely determines the spectrum of the simple random walk on the network. 
\end{abstract}

%


\mainmatter
\include{Introduction}

\chapter{Introduction}\label{section: intro}

An \emph{abelian network} is a collection of communicating automata 
that live at the vertices of a graph and communicate via the edges, 
 satisfying certain axioms (spelled out in \S\ref{subsection: abelian networks}).
 
 \section{Flashback}
 
 The previous papers in this series developed the theory of \emph{halting} abelian networks. 
  To set the stage we recall a few highlights of this theory. 
 It is proved in \cite{BL16a} that the output and the final state of a halting abelian network depend only on the input and the initial state (and not on the order in which the automata process their inputs).  
 
In \cite{BL16b} the halting abelian networks are characterized as those whose production matrix 
has Perron-Frobenius eigenvalue $\lambda<1$. In \cite{BL16c} the behavior of a halting network on sufficiently large inputs is expressed in terms of a free and transitive action of the finite abelian group 
 	\begin{equation} \label{eq:G} \mathcal{G} := \Z^A / (I-P) K, \end{equation}
where $A$ is the total alphabet, $I$ is the $A \times A$ identity matrix, $P$ is the production matrix, and $K$ is the total kernel of the network (all defined in Chapter \ref{s. abelian networks}). This group generalizes the sandpile group of a finite graph~\cite{Lor89, Dhar90, Biggs99}.

\section{Atemporal dynamics}
\label{s.atemporal}

The protagonists of this paper are the \emph{nonhalting} abelian networks, which come in two flavors: \emph{critical} ($\lambda=1$) and \emph{supercritical} ($\lambda>1$). In either case, there is some input that will cause the network to run forever without halting.
Curiously, the quotient group \eqref{eq:G} is still well-defined for such a network. 
In what sense does this group describe the behavior of the abelian network? 

To make this question more precise, we should say what we mean by ``behavior'' of a nonhalting abelian network.
A usual approach would fix an update rule, such as one of the following.
	\begin{itemize}
	\item Parallel update: All automata update simultaneously at each discrete time step.
	\item Sequential update: The automata update one by one in a fixed periodic order.
	\item Asynchronous update: Each automaton updates at the arrival times of its own independent Poisson process.
	\end{itemize}
Instead, in this paper we take the view that an abelian network is a discrete dynamical system \emph{without a choice of time parametrization}: The trajectory of the system is not a single path but an infinite directed graph encompassing all possible time parameterizations.  An update rule assigns to each starting configuration a directed path in this \emph{trajectory digraph}. The study of the digraph as a whole might be called \emph{atemporal dynamics}: dynamics without time.
An example of a theorem of atemporal dynamics is Theorem~\ref{intro theorem: construction of torsion group for general abelian networks}, which identifies a set of weak connected components of the trajectory digraph on which the torsion subgroup of $\mathcal{G}$ acts freely.

When time is unspecified, what remains of dynamics? Some of the most fundamental dynamical questions are atemporal: Does this computation halt? Is this configuration reachable from that one? Are there periodic trajectories, and of what lengths? 

\section{Relating atemporal dynamics to traditional dynamics}\label{subsection: intro atemporal dynamics}

A concrete example is the discrete time dynamical system known as \emph{parallel update chip-firing} on a finite connected undirected graph $G=(V,E)$. The state of the system is a \emph{chip configuration} $\x : V \to \Z$, and the time evolution is described by 
	\[ \x_{t+1}(v) = \x_t(v) - d_v \satu{\{\x_t(v) \geq d_v \}} + \sum_{u \sim v} \satu\{\x_t(u) \geq d_u \}, \]
where the sum is over the $d_v$ neighbors $u$ of vertex $v$. In words, at each discrete time step, each vertex $v$ with at least as many chips as neighbors simultaneously \emph{fires} by sending one chip to each of its neighbors.

\begin{figure}[tb]
\centering
   \includegraphics[width=1\textwidth]{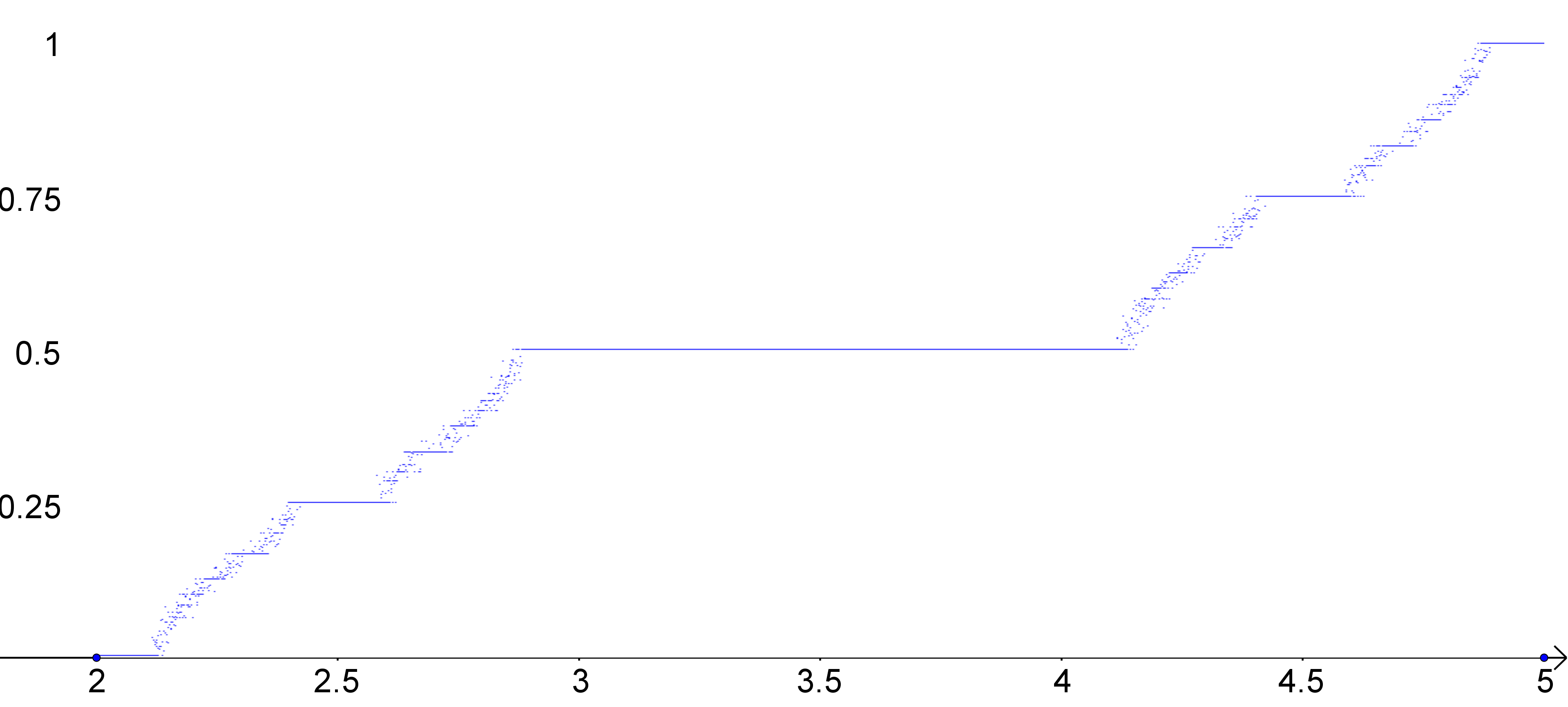}
   \caption{A plot of the firing rate of parallel chip-firing
   on the discrete torus $\Z_{n} \times \Z_{n}$ for $n=32$.
   Each point $(x,y)$ represents a random chip configuration with $xn^2$ chips placed independently with the uniform distribution on the $n^2$ vertices, and eventual firing rate $y$.
   }
   \label{figure: staircase}
\end{figure}

For parallel update chip-firing on discrete torus graphs $\Z_n \times \Z_n$, Bagnolli et al.~\cite{BCF03} plotted the average firing rate as a function of the total number of chips (placed independently at random to form the initial configuration $\x_0$).  
They discovered a mode-locking effect: Instead of increasing gradually, the firing rate remains constant over long intervals between which it increases sharply (Figure~\ref{figure: staircase}). 
The firing rate ``likes'' to be a simple rational number.
This mode-locking has been proved in a special case, when $G$ is a complete graph, by relating it to one of the canonical mode-locking systems, rotation number of a circle map \cite{Lev11}.  

Since $\sum_v \x_t(v)$ (the total number of chips) is conserved, only finitely many chip configurations are reachable from a given $\x_0$, and the sequence $(\x_t)_{t \geq 0}$ is eventually periodic.  In practice one very often observes short periods. Exponentially long periods are possible on some graphs \cite{KNT94}, but not on trees \cite{BG92}, cycles \cite{Dal06}, complete bipartite \cite{Jiang10} 
or complete graphs \cite{Lev11}.

Periodic parallel chip-firing sequences are ``nonclumpy'': if some vertex fires twice in a row, then every vertex fires at least once in any two consecutive time steps \cite{JSZ15}.

Are mode-locking, short periods, and nonclumpiness inherent in the abelian network; or are they artifacts of the parallel update rule? In this paper we find atemporal vestiges of some of these phenomena. For example, despite its definition involving parallel update, the firing rate is constant on components of the trajectory digraph (Proposition~\ref{proposition: activity vector}).

Abelian networks have the \emph{confluence} property: any two legal executions are joinable. The Exchange Lemma~\ref{l. weak exchange lemma} says that any two legal executions are joinable in the minimum possible number of steps. In the case of a critical network, we show that the number of additional steps needed is upper bounded by a constant that does not depend on the executions (Theorem~\ref{theorem: confluence bound}).
 
 \section{Computational questions}
 
 Goles and Margenstern \cite{GM97}
showed that parallel update chip-firing on a suitably constructed infinite graph is capable of universal computation. 
The choice of parallel update is essential for the circuits in \cite{GM97}, which rely on the relative timing of signals along pairs of wires.
Using the circuit designs of Moore and Nilsson \cite{MN01}, Cairns \cite{Cairns15} proved that 
\emph{regardless of the time parameterization},
chip-firing on the cubic lattice $\Z^3$ can emulate a Turing machine. Hence, even atemporal questions about chip-firing can be algorithmically undecidable. An example of such a question is: Given a triply periodic configuration of chips on $\Z^3$ plus finitely many additional chips, will the origin fire infinitely often?
 
What kinds of computation can be performed in a finite abelian network?  In the atemporal viewpoint, a halting abelian network with $k$ input wires and one output wire computes a function $f: \N^k \to \N$: If $x_i$ chips are sent along the $i$th input wire for each $i=1,\ldots,k$, then regardless of the order in which the input chips arrive, exactly $f(x_1,\ldots,x_k)$ chips arrive at the end of the output wire. Holroyd, Levine and Winkler \cite{HLW15} classify the functions $f$ computable by a finite network of finite abelian processors: these are precisely the increasing functions of the form 
	\[ f = L+P, \]
where $L$ is a linear function with rational coefficients, and $P$ is an eventually periodic function. 
Any such function can be computed by a finite halting abelian network of certain simple gates.
An example that shows all gate types is 
  \[ f(x,y,z) = \max(0,x-1) + \min(1,y)+ \left\lfloor \frac{x+\lfloor 2z / 3 \rfloor}{4}\right\rfloor. \]
 
 
The next subsections survey a few highlights of the paper. We have sacrificed some generality in order to state them with a minimum of notation.
The abelian network $\Net$ in our main results is assumed to be finite and locally irreducible.
We also assume that $\Net$ is strongly connected for the latter half of the paper~(Chapter \ref{s. critical networks}-\ref{s. abelian mobile agents}).

\section{The torsion group of a nonhalting abelian network} \label{subsection: intro torsion group of abelian networks}

\old{
In the halting case, the critical group \eqref{eq:G} acts freely and transitively on recurrent states of the network, and 
the critical group arises naturally from the action of the {global monoid}~(defined in \S\ref{subsection: torsion group of subcritical networks}) on recurrent states of the network~\cite{BL16c}.

In this paper we take a new approach in defining the critical group of a (halting or nonhalting) abelian network. 
The group  acts on weak components of the trajectory digraph  (in keeping with our ``atemporal'' viewpoint, we choose not to distinguish between vertices in the same component).  
}

We are going to associate a finite abelian group $\Tor(\Net)$ to any finite, irreducible abelian network $\Net$. In the case $\Net$ is halting, $\Tor(\Net)$ coincides with the critical group of \cite{BL16c}, which acts freely and transitively on the recurrent states of $\Net$. 

What does $\Tor(\Net)$ act on in the nonhalting case? 
Here it is more natural to work with weak connected components of the trajectory digraph. Sending input to $\Net$ can shift it between components, and these shifts are quantified by the \emph{shift monoid} $\Mon(\Net)$.
The torsion group arises from the action of $\Mon(\Net)$ on the \emph{invertible recurrent components} of the trajectory digraph. 
These are components that contain either a cycle or an infinite path, and such that the inverse action of $\Mon(\Net)$ on these components is well defined (see Definitions~\ref{definition: recurrent class} and~\ref{definition: invertible recurrent class} for details).

Now we can answer our motivating question about the dynamical significance of the group $\mathcal{G}$ defined in \eqref{eq:G}.

\begin{theorem}\label{intro theorem: construction of torsion group for general abelian networks}
$\mathcal{G}$ is isomorphic to the Grothendieck group of the shift monoid $\Mon(\Net)$, and
the torsion part of $\mathcal{G}$ acts freely on the invertible recurrent components of the trajectory digraph.
\end{theorem}

Theorem~\ref{intro theorem: construction of torsion group for general abelian networks} is proved in \S\ref{subsection: construction of torsion group for all abelian networks} as a corollary of  Theorem~\ref{theorem: construction of torsion group for general abelian networks}.  In the case that $\Net$ is halting,
the invertible recurrent components are in bijection with recurrent states, and this bijection preserves the group action (Theorem~\ref{theorem: torsion group for BL and CL constructions are equal for subcritical networks}).

\section{Critical networks}\label{subsection: intro critical networks}

 The critical networks (those with Perron-Frobenius eigenvalue $\lambda=1$) are particularly interesting. They include sinkless chip-firing, rotor-routing, and their respective generalizations, arithmetical networks and agent networks (Figure~\ref{figure: Venn diagram}).
 
 A critical network has a conserved quantity which we call \emph{level}; for example, the level of a chip-firing configuration is the total number of chips. We define the \emph{capacity} of a critical network as the maximum level of a configuration that halts.  A problem mentioned in \cite{BL16c} is to find algebraic invariants that can distinguish between ``homotopic'' networks (those with the same production matrix $P$ and total kernel $K$). Capacity is such an invariant: Rotor and chip-firing networks on the same graph have the same $P$ and $K$, but different capacities.

A halting network has recurrent \emph{states}, and so far we have generalized this notion to recurrent \emph{components} of the trajectory digraph. Can we choose a representative configuration in each component? In the halting case, yes: each recurrent component contains a unique configuration of the form $0.\q$ where $\q$ is a recurrent state. 
In a general nonhalting network it is not clear how to define recurrent \emph{configurations} $\x.\q$. But we are able to define them in the critical case, and show that the recurrent components are precisely the components that contain a recurrent configuration.  We then prove a recurrence test, Theorem~\ref{t. recurrence test}, for configurations in a critical network, analogous to Dhar's burning test for states \cite{Dhar90} (and Speer's extension of it to directed graphs, \cite{Speer93}, further extended to halting networks in \cite{BL16c}). This answers another problem posed at the end of~\cite{BL16c}.
\old{
We call this test the \emph{burning test}~(Theorem~\ref{t. recurrence test}), as it has similarities to the burning test for other types of abelian networks~\cite{Dhar90, Speer93, AB11, BL16c}.
The running time of the test ranges from being very fast (linear time for sandpile model on Eulerian digraphs) to being very slow (exponential time for sandpile model on general digraphs).
A more efficient test for a special family of critical networks is given in Theorem~\ref{theorem: cycle test}.
}

Our second main result for critical networks is a combinatorial description 
for the orbits of the action of the torsion group. 

\begin{theorem}\label{intro theorem: torsion group for critical networks}
Let $\Net$ be a critical network.
Then for all but finitely many positive $m$
the action of the torsion group on the recurrent components of level $m$
\[\Tor(\Net) \times \Lrec(\Net,m) \to \Lrec(\Net,m),  \] 
is free and transitive.
\end{theorem}

Theorem~\ref{intro theorem: torsion group for critical networks} is proved in \S\ref{subsection: torsion group for critical networks} as a corollary of Theorem~\ref{theorem: torsion group for critical networks}. 
The exceptional values of $m$ are those for which there exists a halting configuration of level $m$.

\section{Example: Rotor networks and abelian mobile agents}
\label{subsection: intro rotor-routing}

The critical networks of zero capacity (i.e.,  those that run forever on any positive input) are precisely the ``abelian mobile agents'' defined in \cite{BL16a} (see Lemma~\ref{l. agent network is critical networks with capacity 0}).

In particular these include the \emph{sinkless rotor networks}, whose defining property is that each vertex serves its neighbors in a fixed periodic order. 
The walk performed by a single chip input to a sinkless rotor network has variously been called ant walk \cite{WLB96}, Eulerian walk \cite{PDDK96}, rotor walk \cite{HP10}, quasirandom rumor spreading \cite{DF11}, and ``deterministic random walk'' \cite{CDST07}.

Let $G=(V,E)$ be a finite, strongly connected directed graph with multiple edges permitted. For each vertex $v$, fix a cyclic permutation $t_v$ of the outgoing edges from $v$. The role of $t_v$ is to specify the order in which $v$ serves its neighbors.

A \emph{chip-and-rotor configuration} is a pair $\x.\rho$, where $\x:V \to \Z$ indicates the number of chips at each vertex, and $\rho : V \to E$ assigns an outgoing edge to each vertex.
The legal moves in a sinkless rotor network are as follows:
For a vertex  $v$ such that $\x(v) \geq 1$, replace $\rho(v)$ by $\rho'(v) := t_v(\rho(v))$, and then transfer one chip from $v$ to the other endpoint of $\rho'(v)$.
 
 A \emph{cycle} of $\rho$ is a minimal nonempty set of vertices $C \subset V$ such that $\rho(v) \in C$ for all $v \in C$.
 T\'othm\'er\'esz \cite[Theorem~2.4]{Toth18} proved the following test for recurrence; the special case when $\x$ has just one chip goes back to \cite[Theorem 3.8]{HLM08}.

\begin{theorem}[Cycle test for recurrence in a sinkless rotor network, \cite{Toth18}]\label{intro theorem: cycle test}
A chip-and-rotor configuration $\x.\rho$ is recurrent if and only if $\x \in \N^V$ and $\sum_{v \in C} \x(v) \geq 1$ for every cycle $C$ of $\rho$.
\end{theorem}

For the general statement when $G$ is not strongly connected, see \cite[Theorem~2.4]{Toth18}.
In \S\ref{ss. cycle test}  we present a new proof of Theorem~\ref{intro theorem: cycle test}  
that extends to all abelian mobile agent networks (see Theorem~\ref{theorem: cycle test} for details).

Using the cycle test, it becomes a problem of pure combinatorics to enumerate the recurrent chip-and-rotor configurations. Their generating function has a determinantal form resembling the matrix-tree theorem.

\begin{theorem}\label{intro theorem: determinantal formula for sinkless rotor networks}
For $n \geq 1$, let $r_n$ be the number of recurrent chip-and-rotor configurations with exactly $n$ chips on a finite, strongly connected digraph $G$.  Then we have the following identity (in $\mathbb{C}$ for $|z|<1$, and also in the ring of formal power series $\Z[[z]]$):
\[\sum_{n \geq 1 } r_n z^n=   \det \left( \frac{\D}{1-z}  -\A \right), \]
where $\D$ is the diagonal matrix of outdegrees, and $\A$ is the adjacency matrix of $G$.  
\end{theorem}

In particular, it follows from Theorem~\ref{intro theorem: determinantal formula for sinkless rotor networks} that  the sequence $(r_n)_{n \geq 1}$ determines the characteristic polynomial of the Markov transition matrix $(AD^{-1})^\top$ for random walk on $G$.
A multivariate version (in $\#V + \#E$ variables) of Theorem~\ref{intro theorem: determinantal formula for sinkless rotor networks}  is given in Theorem~\ref{theorem: determinantal formula for sinkless rotor networks}.

\section{Proof ideas}

A basic tool underlying many of our results is the Removal Lemma~\ref{lemma: removal lemma}, which extends both the exchange lemma of Bj\"orner, Lov\'{a}sz, and Shor~\cite{BLS91} and the least action principle \cite{FLP10, BL16a}.  It implies that if $m$ is the minimal length of a periodic path in the trajectory digraph of a (finite, irreducible) critical abelian network, then any periodic path can be shortened to a periodic path of length $m$, and any two periodic paths of length $m$ have the same multiset of edge labels.  One could view this fact as an atemporal version of the short period phenomenon described in \S\ref{s.atemporal}. 

 The proof of Theorem~\ref{intro theorem: cycle test} uses an idea of~\cite{Lev15,Chan18} relating the chip-firing  with sinks to its sinkless counterpart. One motivation for the present paper is to see how far this technique can be generalized.
To that end, we introduce \emph{thief networks}, which are halting networks constructed from a given critical network.
 We  show that the recurrent configurations of an agent network can be determined 
 from the recurrent states of its thief networks, and vice versa~(Lemma~\ref{lemma: recurrent configurations of agent network is essentially a recurrent state of subcritical networks}).

The rest of the paper is organized as follows:
In Chapter \ref{section: background on commutative monoid theory} we discuss the relevant commutative monoid theory that used to construct the torsion group.
In Chapter \ref{s. abelian networks} we review the theory of halting abelian networks from~\cite{BL16a,BL16b,BL16c}.
In Chapter \ref{section: torsion group for abelian networks}, Chapter \ref{s. critical networks}, Chapter~\ref{section: critical networks dynamics}, and Chapter \ref{s. abelian mobile agents}  we prove the theorems in \S\ref{subsection: intro torsion group of abelian networks}, \S\ref{subsection: intro critical networks}, \S\ref{subsection: intro atemporal dynamics}  and \S\ref{subsection: intro rotor-routing}, respectively.


\section{Summary of notation}\label{subsection:summary of notation}~

\begin{tabularx}{\textwidth}{l X}
$\Mon$ & a commutative monoid  \\
$\Grt$ & the Grothendieck group of $\Mon$ \\
 $\tau(\Grt)$ &  the torsion subgroup of $\Grt$\\
 $X^\times$ & the set of $\tau(\Grt)$-invertible elements of $X$ (Def.~\ref{definition: invertible element}) \\
 $\Msf$ & a finite commutative monoid\\
 $e$ & the minimal idempotent of $\Msf$ (Def. \ref{definition: minimal idempotent})\\
 $G=(V,E)$  & a directed graph\\
 $\A_G$ & the adjacency matrix of $G$\\
$\D_G$ & the outdegree matrix of $G$\\
$\PP_v$ & the processor at vertex $v$ (\S\ref{subsection: abelian networks}) \\
$A_v$  & the input alphabet of $\PP_v$ (\S\ref{subsection: abelian networks})\\
$Q_v$ & the  state space of $\PP_v$ (\S\ref{subsection: abelian networks})\\
$\Net$ & an abelian network (\S\ref{subsection: abelian networks})\\
$A$ & the total alphabet of $\Net$ (\S\ref{subsection: abelian networks})\\
$Q$ & the total state space of $\Net$ (\S\ref{subsection: abelian networks})\\
$A^*$ & the free monoid on  $A$\\
$\N$ & the set $\{0,1,2,\ldots\}$ of nonnegative integers\\
$\nol$ & the vector in $\Z^A$ with all entries equal to 0  \\
$\satu$ & the vector in $\Z^A$ with all entries equal to 1\\
$\m,\n$ & vectors in $\N^A$ \\
$\x,\y,\z$ & vectors in $\Z^A$ \\
$\x^+, \x^-$ &   the positive and negative part of $\x\in \Z^A$\\
$w$ & a word  in the alphabet $A$\\
$|w|$ &  the vector in $\N^A$ counting the number of  each letter  in~$w$\\
$T_v$ & the transition function of vertex $v$ (\S\ref{subsection: abelian networks})\\
$T_{(v,u)}$ & the message passing function of edge $(v,u)$ (\S\ref{subsection: abelian networks})\\
$t_w(\q)$ &  the state after $\Net$  in state $\q$ processes  $w$ (\S\ref{subsection: abelian networks})\\
$\NN_w(\q)$ &  the message passing vector of  $w$ and   $\q$ (\S\ref{subsection: abelian networks})\\
$\p,\q$ & states in $Q$\\
$\x.\q$ & a configuration of  $\Net$ (\S\ref{subsection: abelian networks})\\
$\pi_w(\x.\q)$ &  the configuration $(\x+\NN_w(\q)-|w|). t_w(\q)$\\
$\x.\q \xdashrightarrow{w} \x'.\q'$ & $w$ is an  execution from $\x.\q$ to $\x'.\q'$ (\S\ref{subsection: legal and complete executions})\\
$\x.\q \xlongrightarrow{w} \x'.\q'$ & $w$ is a legal  execution from $\x.\q$ to $\x'.\q'$ (\S\ref{subsection: legal and complete executions})\\
$\x.\q \dashrightarrow \x'.\q'$ & there exists an execution from $\x.\q$ to $\x'.\q'$\\
$\x.\q \longrightarrow \x'.\q'$ & there exists a legal execution from $\x.\q$ to $\x'.\q'$\\
$\Loc(\Net)$ & locally recurrent states of $\Net$~(\S\ref{ss. local recurrence})\\
$K$ & the total kernel of $\Net$~(Def.~\ref{definition: total kernel})\\
$P$ & the production matrix of $\Net$~(Def.~\ref{definition: production matrix})\\
$\lambda(P)$ & the spectral radius of $P$ \\
$\supp(\x)$ &  the set $\{a \in A \mid \x(a)\neq 0 \}$\\
$w \setminus \n$  & the removal of $\n$ from $w$ (\S\ref{subsection: removal lemma})\\
$\x.\q \dashrlarrow \x'.\q'$ &  $\x.\q$ and $\x'.\q'$ are quasi-legally related (Def.~\ref{definition: weak and strong relation})\\
$\x.\q \rlarrow \x'.\q'$ &  $\x.\q$ and $\x'.\q'$ are legally related (Def.~\ref{definition: weak and strong relation})\\
$\overline{\x.\q}$ & the equivalence class for $\rlarrow$  that contains $\x.\q$\\
$\Lrec(\Net)$ & the set of recurrent components of $\Net$~(Def.~\ref{definition: recurrent class})\\ 
$\Mon(\Net)$ & the shift monoid of $\Net$~(Def. \ref{definition: monoid action on recurrent classes})\\
$\Grt(\Net)$ & the Grothendieck group of $\Net$ (\S\ref{subsection: construction of torsion group for all abelian networks})\\
$\Tor(\Net)$ & the torsion group of $\Net$~(Def.~\ref{definition: torsion group})\\
$\Lrec(\Net)^\times$  & the set of invertible recurrent components of $\Net$~(Def.~\ref{definition: invertible recurrent class})\\
$\Subnet$ & a subcritical abelian network\\
$\Msf(\Subnet)$ & the global monoid of  $\Subnet$ (\S\ref{subsection: torsion group of subcritical networks})\\
$\rr$ & the period vector of $\Net$ (Def.~\ref{definition: period vector})\\
$\Net_R$ &  the thief network on $\Net$ restricted to $R\subseteq A$~(\S\ref{ss. abelian network with thief})\\
$\satu_R$ & the indicator vector for $R \subseteq A$ in $\Z^A$\\
$\x_R$ & the vector in $\Z^A$ given by   $\x_R(\cdot):=\satu_R(\cdot)\x(\cdot)$ \\
$\s$ & the exchange rate vector of $\Net$~(Def.~\ref{definition: exchange rate vector})\\
 $\cpt$ & the capacity of an object (Def.~\ref{definition: capacity})\\
$\lvl$ & the level of an object~(Def.~\ref{definition: level})\\
$\Lrec(\Net,m)$ & the set of recurrent components with level $m$\\
$\Stop(\Net)$ & the set of stoppable levels of $\Net$~(Def.~\ref{definition: stoppable level})\\
 $\Z^A_0$ &  the set $\{\z \in \Z^A \mid  \s^\top \z=0\}$\\
$\varrho_\q$ &  the  rotor digraph of $\q$~(Def.~\ref{definition: rotor digraph})\\
$M_R$ &  the  $A \times A$ matrix   $(\satu_{R}(a) M({a,a'}))_{a, a' \in A}$ \\
$\Rec(\Net,\n)$ & the set of recurrent configurations  with input $\n$\\
$\Rec(\Net,m)$ & the set of recurrent configurations with level $m$
\end{tabularx}

\chapter[Commutative Monoid]{Commutative Monoid Actions}\label{section: background on commutative monoid theory}

In this chapter we  review some commutative monoid theory that will be used in Chapter \ref{section: torsion group for abelian networks} to construct  the torsion group of an abelian network.
Parts of this material are covered in greater generality in~\cite{Gri01,Lang02, Gri07, Ste10}.

\section{Injective actions and Grothendieck group}\label{subsection: commutative monoid that acts injectively}
Let $\Mon$ be a \emph{commutative monoid}, i.e.,  a set equipped with an associative and commutative operation $(m,n) \mapsto mn$ with an identity element $\epsilon \in \Mon$ satisfying $\epsilon m=m$ for all $m \in \Mon$.

The \emph{Grothendieck group} $\Grt$ of $\Mon$ is  $(\Mon \times \Mon) / \sim$,
where $(m_1,m_1') \sim (m_2,m_2')$ if there is $m\in \Mon$ such that  $mm_1m_2'=mm_1'm_2$.
The multiplication of $\Grt$ is defined coordinate-wise.
 The set $\Grt$ is an abelian group under this operation.
 
The  Grothendieck group satisfies the \emph{universal enveloping property}:
 If $f:\Mon \to H$ is a monoid homomorphism into an abelian group $H$, then there exists a unique group homomorphism   $f_*: \Grt \to H$ such that the following diagram commutes:
 \begin{equation*}\label{equation: universal enveloping property}
  \begin{tikzcd}
 \Mon  \arrow[r,"f"] \arrow[d,"\iota"]& H\\
 \Grt \arrow[ru,"f_*"]
 \end{tikzcd},
 \end{equation*}
 where $\iota: \Mon \to \Grt$ is the map $m\mapsto \overline{(m,\epsilon)}$.

An \emph{action} of a monoid $\Mon$ on a  set $X$ is an operation $(m,x) \mapsto mx$ such that $\epsilon x=x$ and $m(m'x)=(mm')x$ for all $x \in X$ and $m,m'\in \Mon$.

\begin{definition}[Injective action]\label{definition: injective action}
Let $\Mon$ be a commutative monoid. 
An action of $\Mon$ on $X$ is \emph{injective} if, for 
all $x,x' \in X$ and all $m \in \Mon$, we have that $mx=mx'$ implies $x=x'$.
\end{definition}

\begin{definition}[Invertible element]\label{definition: invertible element}
Let $\Mon$ be a commutative monoid that acts on  $X$.
Let $H$ be a subgroup of the Grothendieck group $\Grt$ of $\Mon$.
 An element $x \in X$ is $H$-\emph{invertible} if,   for any $g \in H$,  there exists $x_g\in X$ such that
\[ mx=m'x_g, \]
for any representative $(m,m')$ of  $g$.
We denote by $X_H$  the set of  $H$-invertible elements of $X$.
\end{definition}


For any subgroup $H$ of $\Grt$, 
we define the group action of  $H$  on  $X_H$ by 
\begin{align*}
H \times X_H &\to X_H\\
(g,x) &\mapsto x_g,
\end{align*}
where $x_g$ is as in Definition~\ref{definition: invertible element}.
In the next lemma we show that this is a well-defined group action if $\Mon$ acts injectively on $X$.

\begin{lemma}\label{lemma: H-invertible elements let you define a group action}
Let $\Mon$ be a commutative monoid that acts injectively  on  $X$, and let $H$ be a subgroup of the Grothendieck group $\Grt$ of $\Mon$. 
For any $g\in H$ and any $H$-invertible element $x$,
\begin{enumerate}
\item The corresponding element $x_g$ is unique.
\item The element $x_g$ is $H$-invertible.
\item For any $h\in H$, we have $h(gx)=(hg)x$. 
\end{enumerate}
\end{lemma}
\begin{proof}
\begin{enumerate}[wide, labelwidth=!, labelindent=10pt]
\item

 Let $(m,m')$ be a representative of $g$ and let 
  $x_1,x_2\in X_H$ be such that $mx=m'x_1$ and $mx=m'x_2$.
This implies that $ m'x_1= mx=m'x_2$.
Since $\Mon$ acts injectively on $X$,
this implies that $x_1=x_2$.
This completes the proof.

\item

Let $h$ be an arbitrary element of $H$ and $(n,n')$ an arbitrary  representative of $h$.
Let $x_{hg}$ be an element of $X$ such that
$nmx=n'm'x_{hg}$.  
Note that $x_{hg}$ exists because  $x$ is $H$-invertible and $hg=\overline{(nm,n'm')}\in H$.
Then
\[m'n'x_{hg}=n'm'x_{hg}=nmx=nm'x_g=m'nx_g.  \] 
 Since $\Mon$ acts injectively on $X$, the equation above implies that $n'x_{hg}=nx_g$.
Since the choice of $h$ and $(n,n')$ are arbitrary, the claim now follows.
\item Let $(n,n') \in h$ and 
$x_{hg}\in X$ be such that  $nmx=n'm'x_{hg}$.
 It suffices to show that  $x_{hg}$ satisfies $nx_g=n'x_{hg}$, and note that
this has been done in the proof of part (ii).
 \qedhere
\end{enumerate}
\end{proof}

The action of $\Mon$ on $X$ is \emph{free} if, for any $x \in X$ and $m,m' \in \Mon$, we have
$mx=m'x$ implies that $m=m'$.

\begin{lemma}\label{lemma: free monoid action gives rise to free group action, and finite subgroup gives rise to finite set}
Let $\Mon$ be a commutative monoid that acts injectively  on  $X$, and let $H$ be a subgroup of the Grothendieck group $\Grt$ of $\Mon$.
\begin{enumerate}
\item If $\Mon$ acts freely on $X$, then $H$ acts freely on $X_H$.
\item\label{item: finite set X gives rise finite invertible set} If $H$ is finite and $X$ is nonempty, then $X_H$ is nonempty.
\end{enumerate}
\end{lemma}
\begin{proof}
\begin{enumerate}[wide, labelwidth=!, labelindent=10pt]
\item
Suppose that $\overline{(m_1,m_1')},\overline{(m_2,m_2')} \in H$ and $x \in X_H$ are such that $\overline{(m_1,m_1')}x=\overline{(m_2,m_2')}x$.
Then   
\begin{align*}
  & \quad  m_1m_2'x=m_1'm_2x \qquad \text{(by Definition~\ref{definition: invertible element})}\\
 \Longrightarrow &\quad m_1m_2'=m_1'm_2 \qquad \text{(because $\Mon$ acts freely on $X$})\\
 \Longrightarrow &\quad \overline{(m_1,m_1')} =\overline{(m_2,m_2')} \qquad \text{(by the definition of Grothendieck group)}.
\end{align*}
This proves the claim.

\item Let $g_1,\ldots, g_k$ be an enumeration of the elements of $H$.
For each $i \in \{1,\ldots, k\}$, choose a representative  $(m_i,m_i')$ of  $g_i$, and write $m_H:=m_1'\cdots m'_k$.
Since $X$ is nonempty, the set $m_HX$ is also nonempty.
Hence it suffices to show that $m_H X \subseteq X_H$.

For any  $i \in \{1,\ldots,k\}$ and any   $x \in X$, write $x_i:=m_i m_1'\cdots\widehat{m_i'}\cdots m_k'x$.
Then
\begin{equation}\label{equation: lemma free monoid}
 m_i m_Hx=m_i m_1' \cdots m_k' x= m_i'm_i m_1'\cdots\widehat{m_i'}\cdots m_k'x=m_i'x_i, 
 \end{equation}
by the commutativity of the monoid.

Let $i$ be an arbitrary element of $\{1,\ldots,k\}$, and let
 $(n_i,n_i')$ be an arbitrary representative of $g_i$.
Since $(m_i,m_i')$ and $(n_i,n_i')$ are contained in $g_i$, 
there exists  $m \in \Mon$  such that $mm_in_i'=mm_i'n_i$.
Then, continuing from equation~\eqref{equation: lemma free monoid},
\begin{align*}
& m_i m_Hx=m_i'x_i \quad \Longrightarrow \quad mn_i'm_i m_Hx=mn_i'm_i'x_i\\
&  \Longrightarrow \quad m m_i n_i' m_Hx=mm_i'n_i'x_i  \quad  \Longrightarrow \quad m m_i' n_i m_Hx=mm_i'n_i' x_i  \\
&\Longrightarrow \quad n_i m_Hx=n_i'x_i \qquad \text{(because } \Mon \text{ acts injectively}).
\end{align*}
Since the choice of $i$ and $(n_i,n_i')$ are arbitrary, 
it then follows from Definition~\ref{definition: invertible element} that  $m_Hx$ is $H$-invertible. \qedhere
\end{enumerate}
\end{proof}

Let $\tau(\Grt)$  be the \emph{torsion subgroup} of $\Grt$, 
\[\tau(\Grt):=\{g \in \Grt \mid g \text{ has finite order} \}.  \]
The monoid $\Mon$ is \emph{finitely generated} if there is a finite  subset $A$ of $\Mon$  such that every $m \in \Mon$ can be written as a product of finitely many  elements in $A$.
Note that $\tau(\Grt)$ is a finite group if $\Mon$ is finitely generated by the fundamental theorem of finitely generated abelian groups.
We denote by $X^\times$ the set of $\tau(\Grt)$-invertible elements of $X$.

The following proposition is  a corollary of Lemma \ref{lemma: free monoid action gives rise to free group action, and finite subgroup gives rise to finite set}. 

\begin{proposition}\label{proposition: abstract construction of torsion groups}
Let $\Mon$ be a finitely generated commutative monoid that acts freely and injectively on a nonempty set $X$.
Then 
 $X^\times$ is a nonempty set; and 
 $\tau(\Grt)$ is a finite abelian group that acts freely on  $X^\times$. \qed
\end{proposition}

\section[Finite commutative monoid]{The case of finite commutative monoids}\label{subsection: finite commutative monoid}
Here we refine the results of the previous section  to the case when the monoid is finite.

Let $\Msf$ be a finite commutative monoid that acts on a set $Y$.

\begin{definition}[Minimal idempotent]\label{definition: minimal idempotent}
The \emph{minimal idempotent} of  a finite commutative monoid $\Msf$ is
\[ e:=\prod_{f \in \Msf, ff=f}f. \qedhere \]
\end{definition}

The action of $\Msf$ on $Y$ is \emph{irreducible} if for any $y,y' \in Y$ there exist $m,m' \in \Msf$ such that $my=m'y'$.

\begin{lemma}[{\cite[Lemma~2.2, Lemma~2.3, Lemma~2.4]{BL16b}}]
\label{lemma: monoid lemma from BL16}
Let $\Msf$ be a finite commutative monoid that acts on $Y$, and let $e$ be the minimal idempotent of $\Msf$.
\begin{enumerate}
\item \label{item: eM is a group} The set $e\Msf$ is a finite abelian group with identity element $e$.

\item \label{item: locally recurrent elements are reachable from any other elements} If the action of $\Msf$ on $Y$ is irreducible and $y \in eY$, then for any $y' \in Y$ there exists $m' \in \Msf$ such that
$m'y'=y$.
\item \label{item: the monoid acts invertibly on locally recurrent elements}
For every $m \in \Msf$, the map   defined by $y \mapsto my$ is a bijection from $eY$ to $eY$. \qed
\end{enumerate}
\end{lemma}

Let $X:=eY$, and let  $\eta:\Msf \to \End(X)$ be the (monoid) homomorphism induced by the action of $\Msf$ on $X$.
We denote by $\Mon$  the image of $\Msf$ under the map $\eta$.
Just like in \S\ref{subsection: commutative monoid that acts injectively}, we denote by $\Grt$ the Grothendieck group of $\Mon$, and by $X^\times$ the set of $\tau(\Grt)$-invertible elements of $X$.

The action of $\Msf$ on $Y$ is \emph{faithful} if there do not exist distinct $m,m' \in \Msf$ such that $my=m'y$ for all $y \in Y$.
A set $Y' \subseteq Y$ is \emph{closed} under the action of $\Msf$ if $mY' \subseteq Y'$ for all $m \in \Msf$.

\begin{proposition}\label{proposition: finite commutative monoid has a unique injective set}
Let $\Msf$ be a finite commutative monoid that acts faithfully and irreducibly on a nonempty set $Y$, and let $X:=eY$.
Then
\begin{enumerate}
\item  $X$ is the unique nonempty closed subset of $Y$ on which $\Msf$ acts injectively.
\item The group  $e\Msf$ is isomorphic to $\tau(\Grt)$   by the map $\varphi:e\Msf \to \tau(\Grt)$ defined by $em \mapsto \overline{(\eta(em),\epsilon)}$.
 \item  $X^\times$ is equal to $X$. 
\item The isomorphism $\varphi:e\Msf \to \tau(\Grt)$ preserves the action of
$e\Msf$ and $\tau(\Grt)$ on $X=X^\times$. 
\end{enumerate}
\end{proposition}

\begin{proof}
\begin{enumerate}[wide, labelwidth=!, labelindent=10pt]
\item The set $X$ is closed since $mX=m(eY)=e(mY)=eY=X$ by commutativity.
The set $X$
   is nonempty since $Y$ is nonempty.   
By  Lemma~\ref{lemma: monoid lemma from BL16}\eqref{item: the monoid acts invertibly on locally recurrent elements},
the action of $\Msf$ on $X=eY$ is injective.

Suppose that $X'$ is another nonempty closed subset of $Y$ such that $\Msf$ acts injectively on $X'$.
Let $x'$ be an arbitrary element of $X'$.
Note that $ex'=eex'$ since $e$ is an idempotent.
The injectivity assumption then implies that $x' =ex'$.
This shows that $X' \subseteq eY=X$.

Let $y$ be  any element of $Y$, and let $x'$ be an element of $X'$ (note that $x'$ exists because $X'$ is nonempty).
By the irreducibility assumption, there exist $m,m'\in \Msf$ such that $my=m'x'$.
Applying Lemma~\ref{lemma: monoid lemma from BL16}\eqref{item: locally recurrent elements are reachable from any other elements} to $ey \in eY$ and $my \in Y$, there exists $m'' \in \Msf$ such that $m''my=ey$.
Hence we have
\[ ey=m''my =m''m'x'.  \]
Now note that $m''m'x'$ is in $X'$ since $X'$ is closed.
Since the choice of $y$ is arbitrary, we conclude that
 $X=eY \subseteq X'$.
This proves the claim.

\item 
We first  show that the map $\eta$ sends $e\Msf$ to $\Mon$  bijectively.
Note that the action of $e$ on $eY=X$ is trivial as $e$ is idempotent, and 
hence $\eta(e)$ is the identity element of $\Mon$.
Then  
\[\eta(e\Msf)=\eta(e) \eta(\Msf)=\Mon,\]
 which shows  surjectivity.
For injectivity, let $m,m' \in \Msf$ be  such that $\eta(em)=\eta(em')$.
Then
\begin{align*}
em(ey)=em'(ey) \quad  \forall y \in Y \quad \Longrightarrow \quad emy=em'y \quad  \forall y \in Y.
\end{align*}
Since  the action of $\Msf$ on $Y$ is faithful, the equation above implies that $em=em'$.
This shows  injectivity.

Since $e\Msf$ is a finite group   by Lemma~\ref{lemma: monoid lemma from BL16}\eqref{item: eM is a group} and 
$\eta: e\Msf \to \Mon$ is a bijective monoid homomorphism,
we conclude that   $\Mon$ is a finite group and  $\eta$ is a group isomorphism. 
Since $\Mon$ is a group, 
 the map $\iota: \Mon \to \Grt$ is a group isomorphism by the universal enveloping property of Grothendieck group.
Since $\Mon$ is finite, we have the group $\Grt$ is finite, and hence $\Grt=\tau(\Grt)$.
Now note that 
\begin{equation*}
\begin{tikzcd}
e\Msf \arrow[r, "\eta"] \arrow[rr,"\varphi", bend right=90] & \Mon \arrow[r,"\iota"]& \Grt = \tau(\Grt)\end{tikzcd} \  {.}         
\end{equation*}
Since $\eta$ and $\iota$ are group isomorphisms, it follows that $\varphi$ is a group isomorphism, as desired.

\item Since $\Mon$ is a group,  all elements of $X$ are $\tau(\Grt)$-invertible, as desired.

\item This follows  from the definition of $\eta$.
\qedhere
\end{enumerate}
\end{proof}

\chapter[Abelian Networks]{Review of Abelian Networks}\label{s. abelian networks}

The expert reader can skim  this section.
Here we recall the basic setup of abelian networks,
referring the reader to 
 \cite{BL16a,BL16b} for details. 
 Sinkless rotor and sinkless sandpile networks~(Examples~\ref{e. rotor network} and~\ref{e. sandpile network}) are the basic examples to keep in mind when reading this chapter.

\section[Abelian Networks]{Definition of abelian networks}
\label{subsection: abelian networks}

Let $G=(V(G),E(G))$ be a directed graph (or a \emph{digraph} for short), which may have self-loops and multiple edges.
 We will write $V$ and $E$ instead of $V(G)$ and $E(G)$ if the digraph $G$ is evident from the context.
An \emph{outgoing edge} of $v$ is an edge with source vertex $v$,
 and the \emph{outdegree} $\outdeg(v)$ of  $v$ is the number of outgoing edges of $v$. 
We denote by $\Out(v)$ the set of outgoing edges of $v$.
An \emph{out-neighbor} of  $v$ is the target vertex of an outgoing edge of $v$.
  The \emph{indegree} and the \emph{in-neighbors} of $v$ are defined analogously.

In an \emph{abelian network} $\Net$ with underlying digraph $G$, each vertex $v \in V$ has a \emph{processor} $\PP_v$, which is an automaton with  (nonempty) input alphabet $A_v$ and (nonempty) state space $Q_v$.
The data specifying the automaton are:
\begin{enumerate}
\item  A \emph{transition function} $T_a:Q_v\to Q_v$ for each $a \in A_v$; and 
\item A \emph{message-passing function} $T_{e}: Q_v \times A_v \to A_u^*$ for each edge $e$ directed from $v$ to $u$,
\end{enumerate}
where $A_u^*$ denotes the free monoid of all finite words in the alphabet $A_u$.
In the event that the processor $\PP_v$ in state $q \in Q_v$ processes a letter $a \in A_v$, 
the automaton transitions to the state $T_a(q)$ and 
sends the message $T_{e}(q,a)$ to $\PP_u$ for each edge $e$ directed from $v$ to $u$.

We require these functions to satisfy commutativity conditions,
i.e.,  for any $a, b \in A_v$ and any $q \in Q_v$,
\begin{enumerate}
\item $T_a \circ T_b= T_b \circ T_a$; and 
\item  For any outgoing edge $e$ of $v$, the word $T_{e}(q,a) T_{e}(T_{a}(q),b)$ is equal to
$T_{e}(q,b) T_{e}(T_{b}(q),a)$ up to permuting  the letters.

\end{enumerate}
Described in words,  permuting the letters processed by $\PP_v$ does not change the resulting state of
the processor $\PP_v$, and may change the output sent to $\PP_u$ only by permuting
its letters.

The \emph{(total)  state space}  is $Q:=\prod_{v \in V} Q_v$, and the \emph{(total) alphabet}  is $A:=\sqcup_{v \in V} A_v$.
An \emph{input} of $\Net$ is  
 a vector $\x \in \Z^A$, where  $\x(a)$ indicates the number of   $a$'s that are waiting to be processed.
 A \emph{state} $\q$ of $\Net$ is an element  of the total state space $Q$, where $\q(v)$ indicates the state of the processor $\PP_v$.
A \emph{configuration} of $\Net$ is a pair $\x.\q$, where $\x$ is an input and $\q$ is a state of $\Net$.

Let $a \in A$, and let $v \in V$ be such that $a \in A_v$.
The \emph{(total) transition function} $t_a:Q \to Q$ is given by
\begin{align*}
t_a\q(u):=\begin{cases}
T_{a}(\q(u)) & \text{ if } u=v;\\
\q(u) & \text{ otherwise.}
\end{cases}
\end{align*}
(Note that we write $t_a\q$ instead of $t_a(\q)$ to simplify the notation.)
The \emph{message-passing vector} $\NN_{a}:Q \to \N^A$ is given by
\begin{align*}
\NN_{a}(\q):=\sum_{e \in \Out(v)} |T_{e}(\q(v),a)|,
\end{align*}
where  $|w|$ is the  vector in $\N^A$ such that  $|w|(a)$ is the number of  $a$'s  in the word $w$ ($a \in A$).
(We adopt the convention that $\N$ denotes the set $\{0,1,\ldots\}$ of nonnegative integers.)
Described in words,   $\NN_{a}(\q)(b)$ is the number of  $b$'s produced when  network $\Net$ in state $\q$ processes the letter $a$.

In the event that $\Net$  processes a copy of the letter $a$ on the configuration $\x.\q$,  the following three things happen:
\begin{enumerate}
\item The state of $\Net$ changes to $t_a \q \in Q$;
\item  $\NN_a(\q)(b)$ many $b$'s are created  for each $b \in A$;  and
\item The processed letter $a$  is removed from $\Net$.
\end{enumerate}
This process can be described formally 
 by 
  the  \emph{configuration transition function}
$\pi_{a}: \Z^A \times Q \to \Z^A \times Q$, given by
\[\pi_a(\x.\q):=(\x +\NN_a(\q)-|a|).t_a\q.\]

We  extend the  transition functions defined above to any finite word  $w=a_1\ldots a_\ell$ over $A$ by:
\begin{align*}
t_w \q &:=t_{a_\ell}\cdots t_{a_1}\q, \\
  \NN_w(\q) &:=\sum_{i=1}^\ell  \NN_{a_i}( t_{a_{i-1}} \cdots t_{a_1}\q),\\
\pi_w(\x.\q) &:=\pi_{a_\ell}   \cdots \pi_{a_1} (\x.\q) =(\x+\NN_w(\q)-|w|).t_{w} \q,
\end{align*}
which encode the state, the generated letters, 
and the configuration obtained after processing the word $w$, respectively.

For any $\x, \y \in \Z^A$, 
we write  $\x\leq \y$ if $\y-\x$ is a  vector with nonnegative entries.
\begin{lemma}
 [{\cite[Lemma 4.1, Lemma 4.2]{BL16a}}]\label{l. abelian enumerate}
Let $\Net$ be an abelian network,  and let $w,w'\in A^*$. 
\begin{enumerate}
\item \label{l. monotonicity} \textnormal{(Monotonicity)} If $|w|\leq |w'|$, then $\NN_w(\q)\leq \NN_{w'}(\q)$ for all $\q \in Q$.
\item \label{l. abelian property} \textnormal{(Abelian property)} If $|w|=|w'|$, then $t_w=t_{w'}$, $\pi_w=\pi_{w'}$, and $\NN_w= \NN_{w'}$.  \qed 
\end{enumerate}
\end{lemma}

Lemma \ref{l. abelian enumerate}(\ref{l. abelian property}) implies that the functions $t_w, \pi_{w}$, and $\NN_{w}$  depend only on the vector $|w|$.
Therefore, we can extend these transition functions to any  vector $\w \in \N^A$ by
\[t_{\w}:= t_{w}, \qquad \pi_{\w}:=\pi_{w}, \qquad  \NN_{\w}:=\NN_{w},  \]
where $w$ is any word such that $\w=|w|$.

\section{Legal and complete executions}\label{subsection: legal and complete executions}
An \emph{execution} is a word $w \in A^*$,
which prescribes an order in which the letters in $w$ are to be processed.
We assume that an execution is finite, unless stated otherwise.

Let $w=a_1\cdots a_\ell$, and let $\x.\q$ be a configuration of $\Net$.
We write  $\x_i.\q_i:=\pi_{a_i} \cdots \pi_{a_1} (\x.\q)$ for $i \in \{0,1,\ldots,\ell\}$. 
 We say that  $w$ is  a \emph{legal execution} for $\x.\q$ if  $\x_{i-1}(a_i)\geq 1$ for all $i \in \{1,\ldots, \ell\}$.
   We say that $w$ is  a \emph{complete execution} for $\x.\q$ if  $\x_{\ell}(a)\geq 0$ for all $a \in A$.

 \begin{definition}[$\dashrightarrow$ and  $\longrightarrow$]
 Let $\Net$ be an abelian network.
 We write $\x.\q \xdashrightarrow{w} \x'.\q'$
 if $\pi_{w}(\x.\q)=\x'.\q'$.  
We write $\x.\q \xlongrightarrow{w} \x'.\q'$
 if $\pi_{w}(\x.\q)=\x'.\q'$ and $w$ is a legal execution for $\x.\q$.
 \end{definition}
In order to simplify the notation,
we will  write $\dashrightarrow$ and $\longrightarrow$
when the word $w$ is not a major component of the discussion.
We remark that $\x.\q \xlongrightarrow{} \x.\q$ since the empty word is a legal execution that sends $\x.\q$ to $\x.\q$. 

In the next lemma, we list several properties of $\dashrightarrow$ and  $\longrightarrow$.
The \emph{support} of a vector $\uu \in \Z^A$ is 
$\supp(\uu):=\{a \in A \mid \uu(a)\neq 0 \}$.

\begin{lemma}\label{o. to legal properties}
Let $\Net$ be an abelian network. 
\begin{enumerate}
\item \label{item: weak arrow if and only if} If $\x.\q \xdashrightarrow{w} \x'.\q'$,  then    $(\x+\z).\q \xdashrightarrow{w} (\x'+\z).\q'$ for all $\z \in \Z^A$.
 \item \label{o. to contagious}
If $\x.\q \xlongrightarrow{w} \x'.\q'$ and $\z \in \Z^A$ satisfies 
$\z(a) \geq 0$ for all $a \in \supp(|w|)$,
then $(\x+\z).\q \xlongrightarrow{w} (\x'+\z).\q'$.
\item \label{o. nonnegativity} For any $a \in A$, if $\x.\q \xlongrightarrow{w} \x'.\q'$ and $|w|(a)>0$, then  $\x'(a)\geq 0$.
\item \label{o. to transitive}  If $\x.\q \xlongrightarrow{w} \x'.\q'$ and  $\x'.\q' \xlongrightarrow{w'} \x''.\q''$, then $\x.\q \xlongrightarrow{ww'} \x''.\q''$.
\end{enumerate}
\end{lemma}
\begin{proof}
This  follows directly from the definition of  $\dashrightarrow$ and $\longrightarrow$.
\end{proof}

\section{Locally recurrent states}\label{ss. local recurrence}
An abelian network $\Net$ is \emph{finite} if both the (total) state space $Q$ and the (total) alphabet $A$ are finite sets.
All abelian networks in this paper are assumed to be finite, unless stated otherwise.

We denote by   $M \subseteq \End(Q)$  the \emph{transition monoid}  
$\langle t_a \rangle_{a \in A}$.
Note  that $M$ is a finite commutative monoid as $\Net$ is  finite.
Since $M$ is finite, it has a (unique) minimal idempotent $e$
(Definition~\ref{definition: minimal idempotent}).

A state $\q \in Q$ is \emph{locally recurrent} if  $\q \in eQ$. 
We denote by  $\Loc(\Net)$ the set of locally recurrent states of $\Net$.
For maximum generality we don't assume local recurrence, but the reader will not lose much by restricting the state space of the network to $\Loc(\Net)$.
Note that $\Loc(\Net)$ is a nonempty set (since $Q$ is nonempty by definition of $\Net$).

Here we list properties of locally recurrent states that will be used in this paper.
We denote by $\satu$ the vector $(1,\ldots,1)^\top$ in $\Z^A$.
\begin{lemma}\label{l. local recurrence}
Let $\Net$ be a finite abelian network. 
Then
\begin{enumerate}
\item \label{item: existence of idempotent elements that make every states locally recurrent} There exists  $\ee \in \N^A$ such that $t_{\ee}\q$ is locally recurrent for all $\q \in Q$.
\item \label{i. local recurrence sufficient and necessary condition} A state $\q$ is locally recurrent if  there exists  $\n \in \N^A$ such that $\n\geq \satu$ and $t_\n\q=\q$.  
\end{enumerate}
\end{lemma}
\begin{proof}
\begin{enumerate}[wide, labelwidth=!, labelindent=10pt]
\item 
The claim follows by taking $\ee$ to be a   vector in $\N^A$ such that $t_{\ee}$ is the minimal idempotent of $M$.

\item 
Since $\n \geq \satu$, we can  without loss of generality assume that $t_{\n} \in eM$~(by taking a finite multiple of $\n$ if necessary).
Then $\q=t_{\n}\q\in t_{\n}Q \subseteq eQ$,
and hence $\q$ is locally recurrent. \qedhere
\end{enumerate}
\end{proof}


\begin{lemma}\label{lemma: the monoid $N^A$ acts invertibly on locally recurrent configurations}
Let $\Net$ be a finite abelian network.
For any $\n \in \N^A$, 
\begin{enumerate}
\item\label{item: monoid acts invertible on locally recurrent states 1} The function $t_{\n}$ restricted to  $\Loc(\Net)$ is a bijection from $\Loc(\Net)$ to $\Loc(\Net)$.
\item \label{item: monoid acts invertible on locally recurrent states 2} The function $\pi_{\n}$ restricted  to $\Z^A \times \Loc(\Net)$ is a bijection from  $\Z^A \times \Loc(\Net)$ to $\Z^A \times \Loc(\Net)$.
\end{enumerate}
\end{lemma}
\begin{proof}
The first part of the lemma follows directly from Lemma~\ref{lemma: monoid lemma from BL16}\eqref{item: the monoid acts invertibly on locally recurrent elements}.
The second part of the lemma is a consequence  of the first part.
\end{proof}

\section{The production matrix}

For any vector $\z \in \Z^A$,
the  \emph{positive part} $\z^+$ and  \emph{negative part} $\z^-$ of $\z$ are the unique  vectors in $\N^A$ such that $\z=\z^+-\z^-$ and $\supp(\z^+) \cap \supp(\z^-)=\varnothing$.

\begin{definition}[Total kernel]\label{definition: total kernel}
Let $\Net$ be a finite abelian network.
The \emph{total kernel} $K \subseteq \Z^A$ 
is 
\[K:=\{\z \in \Z^A \mid t_{\z^+}\q=t_{\z^-}\q \text{ for all } \q \in  \Loc(\Net)   \}. \qedhere\]
\end{definition}

We say that $\Net$ is \emph{locally irreducible} if for any $\q,\q'\in Q$
there exist $w,w' \in A^*$ such that $t_w\q=t_{w'}\q'$.

\begin{lemma}[{\cite[Lemma~4.5, Lemma~4.6]{BL16b}}]\label{lemma: the total kernel is a subgroup of finite index}
Let $\Net$ be a finite abelian network.
\begin{enumerate}
\item \label{item: the total kernel is a subgroup of finite index}
The  total kernel $K$ is a subgroup of $\Z^A$ of finite index. 
\item \label{item: x is in K if and only if it sends locally recurrent q to q} If $\Net$ is locally irreducible, then for any $\q \in \Loc(\Net)$,
\[ 
\pushQED{\qed} 
K \cap \N^A=\{ \x \in \N^A \mid t_\x\q=\q \}. \qedhere
 \popQED \]
\end{enumerate}
\end{lemma}

For   $\q \in \Loc(\Net)$,  we define   $P_\q: K\cap \N^A\to \Z^A$ to be 
\[P_\q(\kk):= \NN_{\kk}(\q). \]
  The map $P_\q$ extends uniquely to a group homomorphism  $K\to \Z^A$~\cite[Lemma 4.6]{BL16b}.
  Since $K$ is a subgroup of $\Z^A$ of finite index~(by Lemma~\ref{lemma: the total kernel is a subgroup of finite index}\eqref{item: the total kernel is a subgroup of finite index}), 
  we get a linear map $P_\q:\Q^A\to \Q^A$ by   tensoring the  group homomorphism $P_\q$ with $\Q$.

If $\Net$ is locally irreducible, then the matrix $P_\q:\Q^A \to \Q^A$ does not depend on the choice of $\q$ \cite[Lemma 4.9]{BL16b}.

\begin{definition}[Production matrix]\label{definition: production matrix}
Let $\Net$ be a finite and locally irreducible abelian network.
The \emph{production matrix} of $\Net$ is the matrix $P:=P_\q$, where $\q$ is any locally recurrent state of $\Net$.
\end{definition}

\begin{lemma}
\label{l. N locally irreducible}
Let $\Net$ be a finite and  locally irreducible abelian network.
If  $\q\in Q$ and  $\n,\n'\in \N^A$ satisfy  
  $t_{\n}\q=t_{\n'}\q$, then  
  \[\n-\n' \in K \qquad \text{and} \qquad \NN_{\n}(\q)-\NN_{\n'}(\q)=P(\n-\n').\] 
\end{lemma}
\begin{proof}
By Lemma \ref{l. local recurrence}(\ref{item: existence of idempotent elements that make every states locally recurrent}), there exists $\ee \in \N^A$ such that $\p:=t_{\ee} \q$ is locally recurrent.
Write $\p':=t_{\n}\p=t_{\n'}\p$.
Since $\Net$ is locally irreducible and $\p \in \Loc(\Net)$,
by Lemma~\ref{lemma: monoid lemma from BL16}\eqref{item: locally recurrent elements are reachable from any other elements}
there exists $\m \in \N^A$ such that $t_{\m}\p'=\p$.

By the abelian property~(Lemma \ref{l. abelian enumerate}\eqref{l. abelian property}), we have:
\begin{equation}
\begin{tikzcd} \label{diagram: diagram for lemma production matrix}
\q \arrow[d,"t_{\ee}"]\arrow[r, bend left,"t_{\n}"] \arrow[r, bend right,"t_{\n'}"]& \q' \arrow[d,"t_{\ee}"]\\
\p \arrow[r, bend left,"t_{\n}"] \arrow[r, bend right,"t_{\n'}"]& \p' \arrow[r,"t_{\m}"] & \p.
\end{tikzcd}.
\end{equation}
In particular, the bottom row of Diagram~\eqref{diagram: diagram for lemma production matrix} above gives us $t_{\n+\m}\p=t_{\n'+\m}\p=\p$.
By Lemma~\ref{lemma: the total kernel is a subgroup of finite index}\eqref{item: x is in K if and only if it sends locally recurrent q to q} these equations
  imply 
 that both $\n+\m$ and $\n'+\m$ are in  $K$, 
 and hence
   $\n-\n' \in K$.

By the abelian property and the commutativity of Diagram~\eqref{diagram: diagram for lemma production matrix}, 
\begin{align*}
\NN_{\n}(\q)+ \NN_{\ee+\m}(\q')=&\NN_{\ee}(\q)+\NN_{\n+\m}(\p);\\
\NN_{\n'}(\q)+ \NN_{\ee+\m}(\q')=&\NN_{\ee}(\q)+\NN_{\n'+\m}(\p).
\end{align*}
By subtracting one equation from the other,
\begin{align*}
\NN_{\n}(\q)- \NN_{\n'}(\q)=\NN_{\n+\m}(\p)-\NN_{\n'+\m}(\p).
\end{align*}
Since  $\n+\m$ and  $\n'+\m$ are in  $K$ and $\p \in \Loc(\Net)$,
 \begin{align*}
 \NN_{\n+\m}(\p)-\NN_{\n'+\m}(\p)=P(\n+\m)-P(\n'+\m)=P(\n-\n').
 \end{align*}
This completes the proof.
\end{proof}

\section[Subcritical, critical, and supercritical]{Subcritical, critical, and supercritical abelian networks}\label{subsection: a classification of abelian networks} 
Let $\Net$ be a finite and locally irreducible abelian network.
The \emph{production digraph} $\Gamma$ is the directed graph with vertex set $A$ and edge set $\{(a,b): P_{ba}>0 \}$.

We define an equivalence relation on $A$ by  considering $a$ and $b$ to be equivalent  if there exists a directed path  from $a$ to $b$ and a directed path  from $b$ to $a$ in $\Gamma$.
The \emph{strong components} of $\Gamma$ are the equivalence classes  of this relation.
A network $\Net$ is  \emph{strongly connected} if $\Gamma$ has only one strong component.

The  \emph{spectral radius} of the production matrix $P$ is
\[  \lambda(P):=\max \{|\lambda| \,:\, \lambda \text{ is an eigenvalue of } P \}.\]
{
We distinguish  (finite, locally irreducible) abelian networks by the value of $\lambda(P)$:
\begin{itemize}
\item The network $\Net$ is \emph{subcritical} if $\lambda(P)<1$.
Subcritical networks are studied in~\cite{BL16b, BL16c}.

\item The network $\Net$ is \emph{critical} if $\lambda(P)=1$.
We will study critical networks in more detail in the latter half of this paper.  
 
 \item The network $\Net$ is \emph{supercritical} if $\lambda(P)>1$.
\end{itemize}
 See Example~\ref{example: toppling networks} for a concrete example of each network.
}

Let $A_1, \ldots, A_s$ be the strong components of $\Gamma$.
Denote by $P_i$ the matrix obtained by restricting the production matrix $P$ to rows and columns from $A_i$.
We say that $A_i$ is a \emph{subcritical} component if $\lambda(P_i)<1$,
and a letter $a \in A$ is \emph{subcritical} if it is contained in a subcritical component.
\emph{Critical} and \emph{supercritical} components/letters are defined analogously.

We denote by $A_{<}$ the set of subcritical letters, and by $A_{\leq}$ the set of subcritical and critical letters.
The sets $A_{=}$, $A_>$, and $A_{\geq}$ are defined analogously.
Recall that  the \emph{support} of $\uu \in \R^A$ is 
$\supp(\uu):=\{a \in A \mid \uu(a)\neq 0 \}$.

A real matrix $P$ is \emph{nonnegative} if all its entries are nonnegative,
and is \emph{positive} if all of its entries are positive.  
For all matrices $P$ and $Q$ of the same dimension, we write   $Q \leq P$  if $P-Q$ is a nonnegative matrix.
\emph{Nonnegative vectors} and \emph{positive vectors} are defined analogously.

We now present variants  of the Perron-Frobenius theorem that will be used in this paper, referring  to \cite{BP79} for most of the proof.
\begin{lemma}[Perron-Frobenius]
\label{lemma: Perron-Frobenius theorem}
Let $A$ be a finite set, and let $P$ be an $A \times A$ matrix whose entries are nonnegative rational numbers. 
\begin{enumerate}
\item \label{item: Perron-Frobenius 1}
$P$ has a nonnegative real eigenvector with eigenvalue $\lambda(P)$.

\item \label{item: Perron-Frobenius 2}  If $\alpha$ is a real number such that    $P\uu=\alpha \uu$ for some positive vector $\uu \in \R^A$,
then $\alpha=\lambda(P)$.

\item \label{item: Perron-Frobenius spectral from inequality}
Let $P$ be strongly connected, and let $\alpha$  be a real number such that $ P\uu \geq \alpha \uu$  for some  nonzero nonnegative vector $\uu \in \R^A$.
Then $\lambda(P) \geq \alpha $, 
and equality holds if and only if 
 $P\uu=\alpha \uu $.
Furthermore, the claim is still true if
 ``$\geq$''  is replaced with ``$\leq$''.

\item \label{item: Perron-Frobenius lambda is strictly increasing}
If $P$ is strongly connected and $Q$ is a nonnegative matrix such that $Q \leq P$ and $Q \neq P$, then $\lambda(Q)<\lambda(P)$.
\item \label{item: Perron-Frobenius 4} If $P$ is strongly connected, then  the eigenspace of $\lambda(P)$ is spanned by a positive real vector.

\item  \label{item: Perron-Frobenius 5} If $P$ is strongly connected and $\lambda(P) \in \Q$, then the eigenspace of $\lambda(P)$ is spanned by a positive integer vector.

\item  \label{item: Perron-Frobenius 3} 
There exists  $\n,\n',\n'' \in \N^A$ such that 
\begin{itemize}
\item $\supp(\n)=A_{<}$ 
and 
$P\n(a)<\n(a)$ for all $a \in A_{<}$;

\item $\supp(\n')=A_{=}$ 
and 
$P\n'(a)\geq\n'(a)$ for all $a \in A_{=}$; and

\item $\supp(\n'')=A_{>}$ 
and 
$P\n''(a)>\n''(a)$ for all $a \in A_{>}$.
\end{itemize}

\item \label{item: Perron-Frobenius 3.5}
There exists $\m \in \N^A$ such that 
$\supp(\m)=A_{\geq}$ 
and 
$P\m(a)\geq \m(a)$ for all $a \in A_{\geq}$. 
\end{enumerate}
\end{lemma}
\begin{proof}
\begin{enumerate}[wide, labelwidth=!, labelindent=10pt]
\item This follows from \cite[Theorem~2.1.1]{BP79}.

\item This follows from \cite[Theorem~2.1.11]{BP79}.

\item This follows from \cite[Theorem~2.1.11]{BP79}.

\item This follows from~\cite[Theorem~2.1.5(b)]{BP79}.

\item This follows from~\cite[Theorem~2.1.4(b)]{BP79}.
\item 
Since both $P$ and $\lambda(P)$ are  rational, the eigenspace $\Eig$ of $\lambda(P)$ has a basis that consists of integer vectors.
It then follows from part \eqref{item: Perron-Frobenius 4} that $\Eig$ is spanned by a positive integer vector.

\item We prove only the subcritical case, as the other two cases are analogous.
Let  $A_1,\ldots ,A_k$ be the  subcritical  components of $\Gamma$.
Write $\lambda_i:=\lambda(P_i)$ ($i\in \{1,\ldots,k\}$).
Note that $\lambda_i<1$ by assumption.

It follows from part \eqref{item: Perron-Frobenius 4}
 that for each $i\in \{1,\ldots,k\}$ there exists a nonnegative vector $\uu_i \in \R^A$  such that $\supp(\uu_i)=A_i$ and $P\uu_i(a)=\lambda_i\uu_i(a)$ for all $a \in A_i$.
 By scaling and rounding $\uu_i$ if necessary, there exist $\n_i \in \N^A$ and sufficiently small $\epsilon_i >0$ such that $\supp(\n_i)=A_i$ and $P\n_i(a)<(1-\epsilon_i)\n_i(a)$ for all $a \in A_i$.
 By scaling $\n_1,\ldots,\n_k$ if necessary, we can assume that $\n:=\n_1+\ldots+\n_k$  satisfies $P\n(a)<\n(a)$ for all $a \in A_{<}=A_1 \sqcup \ldots \sqcup A_k$.
This proves the lemma.

\item  Let  $\m:=\n'+\n''$, where $\n'$ and $\n''$ are as in part \eqref{item: Perron-Frobenius 3}.
Then for any critical letter $a$,
\[ P\m(a)=P\n'(a)+ P\n''(a)\geq P\n'(a) \geq \n'(a)=\m(a),  \]
and  for any supercritical letter $a$,
\[ P\m(a)=P\n'(a)+ P\n''(a)\geq P\n''(a) >\n''(a)=\m(a).  \]
This proves the lemma.\qedhere
\end{enumerate}
\end{proof}

\begin{remark}
We would like to warn the reader that the subcritical variant of  part \eqref{item: Perron-Frobenius 3.5} (i.e., there  exists $\m \in \N^A$ such that 
$\supp(\m)=A_{\leq}$ 
and 
$P\m(a)\leq \m(a)$ for all $a \in A_{\leq}$) is false.
Indeed, let $P$ be the matrix
\[P=\begin{bmatrix}
1& 1\\ 0 & 1
\end{bmatrix}.\]
A direct computation then shows that 
the inequality $P\m\leq \m$ is always false for any positive vector $\m$.
\end{remark}

\section[Examples]{Examples: sandpiles, rotor-routing, toppling, etc}\label{subsection: examples of abelian networks}
\begin{figure}[tb]
\includegraphics[width=\textwidth]{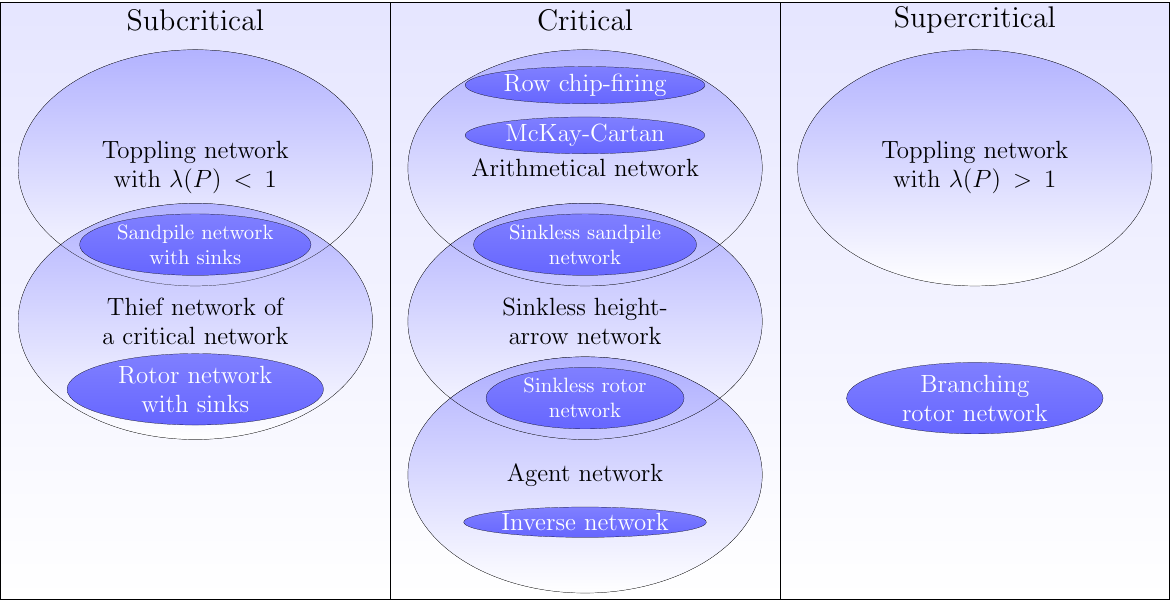}
\caption{A Venn diagram illustrating several classes of (finite and locally irreducible) abelian networks. The Perron-Frobenius eigenvalue $\lambda$ increases from left to right. In the middle bubble in the critical column, the capacity~(see Definition~\ref{definition: capacity}) increases from bottom to top.
Note that an arithmetical network is equivalent to a strongly connected toppling network with $\lambda=1$;  hence the latter is not listed in the diagram.
}
\label{figure: Venn diagram}
\end{figure}

In this section we present several  examples of abelian networks.
The relationship between these  networks is illustrated in Figure~\ref{figure: Venn diagram}.

We use the following graph theory terminology throughout this paper.
Recall that $G$ is a directed graph with vertex set $V$ and edge set $E$.
A digraph is \emph{Eulerian} if for all $v \in V$ the outdegree of $v$ is equal to the indegree of $v$. 
Any undirected graph can be changed into an Eulerian directed graph by replacing each  undirected edge $\{v,u\}$  
with a pair  of directed edges $(v,u)$ and $(u,v)$. 
We call such a digraph \emph{bidirected}.

The \emph{adjacency matrix} $\A_G$ of   $G$ is the matrix $(a_{v,v'})_{v,v' \in V}$, where $a_{v,v'}$ is  the number of edges directed from $v'$ to $v$.
The \emph{outdegree matrix} $\D_G$ of $G$  is the $V \times V$ diagonal matrix with $D_{G}(v,v):=\outdeg(v)$ $(v \in V)$.
The \emph{Laplacian matrix} $\Lap_G$ of  $G$ is the matrix $\D_G-\A_G$.

The digraph $G$ is \emph{strongly connected} if for any $v,v' \in V$  there exists a directed path in $G$ from $v$ to $v'$.

The following digraph will be our main running example  for  the underlying digraph of an abelian network.
For $n\geq 3$, the \emph{bidirected cycle} $C_n$  is
\[V(C_n):=\{v_k \mid  k \in \Z_n\}, \qquad  E(C_n):= \bigcup_{k \in \Z_n} \{(v_k,v_{k-1}),(v_k,v_{k+1})  \}. \]

All networks presented in this section are \emph{irreducible}, i.e. they satisfy these two properties:
\begin{itemize}
\item The network is locally irreducible; and 
\item The minimal idempotent of the transition monoid $M$  is the identity element of $M$.
\end{itemize}
In particular, any state of an irreducible network is locally recurrent.

\begin{example}
[Sinkless  rotor network~\cite{PDDK96,WLB96,Propp03}]
\label{e. rotor network}
For each vertex $v \in V$, fix a cyclic total order  on the set of  the outgoing edges $\Out(v)$ of  $v$, i.e.
an enumeration $e_0^v,e_1^v,\ldots, e_{\outdeg(v)-1}^v$  indexed by $\Z_{\outdeg(v)}$.
The alphabet, state space, and state transition of the processor $\PP_v$ are given by
\begin{align*}
& A_v:=\{v\}, \qquad  Q_v:=\Out(v), \qquad  T_v(e_i^v):=e_{i+1}^v \quad (i \in \Z_{\outdeg(v)}).
\end{align*}

For each edge $e_j^v$ directed from $v$ to $u_j^v$, the message-passing function is given by
\[ T_{e_j^v}(e_i^{v},v):=\begin{cases} u_j^v & \text{ if } i= j-1; \\
\epsilon & \text{ otherwise.} \end{cases} \]

A state of the full network is described by a \emph{rotor configuration} of  $G$, that is, a function $V \to E$ assigning to each vertex $v$ an outgoing edge from $v$.
When a chip/letter at vertex $v$ is processed, the  edge/state $e_i^v$ assigned to $v$  changes to $e_{i+1}^v$ (the next edge in the cyclic total order), and the processed chip  is moved from $v$ to the target vertex of $e_{i+1}^v$.
See Figure~\ref{figure:rotor network example} for an illustration of the process.

\begin{figure}[tb]
\centering
   \includegraphics[width=1\textwidth]{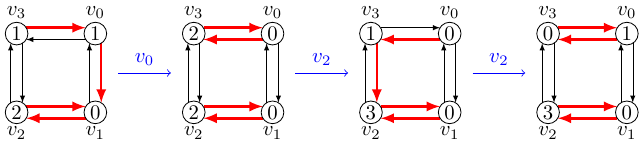}
   \caption{A three-step legal execution in the  sinkless rotor network on the bidirected cycle $C_4$.
   The  number on each vertex records the number of letters waiting to be processed, and the (red) thick outgoing edge  records the state of the processor.}  
   \label{figure:rotor network example}
\end{figure}

 Any sinkless rotor network is strongly connected if the underlying digraph $G$ is  strongly connected.
  The total kernel and the production matrix of this network are given by 
 \[ K=\{  \z \in \Z^V \mid  \z(v) \text{ is divisible by } \outdeg(v) \text{ for all } v \in V  \}; \qquad P=\A_G  \D_G^{-1}, \]
 where $\A_G$  is the adjacency matrix of $G$ and $\D_G$ is the outdegree matrix  
 of $G$.
 Because $\satu  \A_G  \D_G^{-1}= \satu$,
  the Perron-Frobenius theorem (Lemma~\ref{lemma: Perron-Frobenius theorem}\eqref{item: Perron-Frobenius 2}) implies that $\lambda(P)=1$.
  Hence this network is a critical network.
\end{example}

\begin{example}
[Sinkless  sandpile network/chip-fi\-ring~\cite{Dhar90, BLS91}]
\label{e. sandpile network}
For each vertex $v \in V$ of the underlying digraph,
 the processor $\PP_v$ is given by
 \begin{align*}
& A_v:=\{v\}, \qquad Q_v:=\{0,1, \ldots, \outdeg(v)-1\},  \qquad  T_{v}(i):=i+1 \ \text{mod } \outdeg(v).
 \end{align*} 
For each edge $e$ directed from $v$ to $u$, the message-passing function 
is given by
\[ T_{e}(i,v):=\begin{cases} u & \text{ if } i=\outdeg(v) -1; \\
\epsilon & \text{ otherwise.} \end{cases} \]

We can think of each processor $\PP_v$ as a ``locker'' that can store up to $\outdeg(v)-1$ chips, and its state $q_v$ represents the number of chips it is currently storing.
When $\PP_v$ receives a new chip,
the chip is stored in the locker if it has  unallocated space (i.e.,  if $\q(v) < \outdeg(v)-1$).
If the locker is already full (i.e.,  $\q(v) = \outdeg(v)-1$), 
then $\PP_v$ sends all $\outdeg(v)-1$ stored chips  plus the extra chip
to its neighbors by sending one chip along each outgoing edge from $v$.
See Figure~\ref{figure:sandpile network example} for an illustration of this process.

\begin{figure}[tb]
\centering
   \includegraphics[width=1\textwidth]{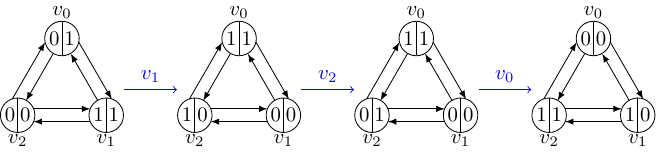}
   \caption{A three-step legal execution in the sinkless sandpile network on the bidirected cycle $C_3$.
  In the figure,  the left part of a vertex records the number of letters waiting to be processed, and the right part records the state of the processor.}  
   \label{figure:sandpile network example}
\end{figure}

The total kernel and the production matrix of this network are equal to the corresponding objects in the sinkless rotor network (with the same underlying digraph).
Hence by the same reasoning as Example~\ref{e. rotor network}, a sinkless sandpile network on a strongly connected digraph is a critical network.

\begin{remark}
We would like to warn the reader that   (network) configurations in this paper have a subtle difference when compared to  (chip) configurations in the literature.
A (chip) configuration in the usual sense is a vector $\cc \in \Z^V$
that records the number of chips at each vertex.
By contrast, a (network) configuration in this paper 
is a pair $\x.\q$,
where the vector $\x \in \Z^V$ records  the number of chips that are not  stored in the lockers,
and the  state $\q \in \prod_{v \in V} \Z_{\outdeg(v)}$
records the number of chips currently stored in the lockers.

Identifying $\Z_{\outdeg(v)}$ with $\{0,1,\ldots,\outdeg(v)-1\}$, the chip configuration corresponding to $\x.\q$ is the vector sum $\x+\q$.
 Note that there is more than one  way to represent a  chip configuration as a network configuration.
  \qedhere
\end{remark} 
\end{example}

\begin{example}
[{Sinkless height-arrow network~\cite{DR04}}]
\label{e. height-arrow}
In this network,
 each vertex $v \in V$ of the underlying digraph $G$ is  assigned  \emph{threshold value} $\tau_v \in \{1,\ldots, \outdeg(v)\}$.
The  processor $\PP_v$ is given by 
\begin{align*}
A_v:=&\{v\}, \\
Q_v:=&\{ (d,c) \in \{0,\ldots, \outdeg(v)-1\} \times  \{0,\ldots, \tau_v-1\} \mid    \\
& \phantom{\{} d\equiv k \tau_v   \text{ (mod} \outdeg(v)) \text{ for some } k \in \Z \},\\
 T_v(d,c):=&\begin{cases} (d,c+1)  & \text{if } c<\tau_v-1;\\
(d+\tau_v \text{ mod }\outdeg(v),0)  & \text{if } c=\tau_v-1.
\end{cases} 
\end{align*}
For each $v \in V$,   fix a  cyclic total order  
$\{e_j^v \mid j \in \Z_{\outdeg(v)} \}$   on the set of  outgoing edges  of  $v$.
 The message-passing function for the edge $e_j^v$ directed from $v$ to $u^v_j$  is given by
\[T_{e_j^v}(d,c,v):=\begin{cases} u_j^v  & \text{if } c=\tau_v-1 \text{ and } j -  d \in  \{1,\ldots, \tau_v\} \ \text{ (mod } \outdeg(v));\\
\epsilon  & \text{otherwise. }
\end{cases} \]

For each $v \in V$,
the state $(d,c)$ of $\PP_v$ represents  an arrow  pointing from $v$ to $u_d^v$,
 and with $c$ chips sitting on  $v$.
When the vertex  $v$ collects $\tau_v$ chips, the arrow is incremented $\tau_v$ times,  and one chip is sent to each vertex in  $\{u_{d+j}^v \mid 1\leq j \leq \tau_v\}$.
See Figure~\ref{figure: height arrow network example} for an illustration of this process.

\begin{figure}[tb]
\centering
   \includegraphics[width=1\textwidth]{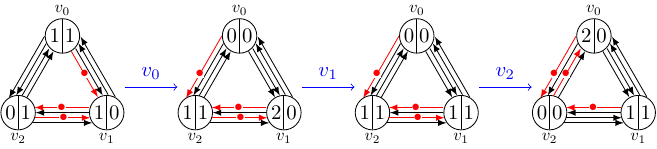}
   \caption{A three-step legal execution in a sinkless height-arrow network.
   For every $v \in V$, the 
 threshold $\tau_v$ is equal to $2$, and the cyclic total order on $\Out(v)$ is the  counterclockwise ordering.
  In the figure,  the left part of a vertex records the number of letters waiting to be processed,  the right part records the height $c_v$ of the processor, and the marked (red) outgoing edge  records the arrow $d_v$ of the processor.}  
   \label{figure: height arrow network example}
\end{figure}

Note that
sinkless rotor networks  are height-arrow networks with $\tau_v=1$ for all $v \in V$,
and  sinkless sandpile networks are height-arrow networks 
with $\tau_v=\outdeg(v)$  for all $v \in V$.

Height-arrow networks have the same  total kernel and  production matrix as sinkless rotor and sandpile networks. 
In particular, height-arrow networks on a strongly connected digraph are critical networks.

\begin{remark}
Note that   height-arrow networks as originally defined in \cite{DR04} have  state space $Q_v=\Z_{\outdeg(v)} \times \Z_{\tau_v}$ instead.
Note that this  choice of state space is in general not  locally irreducible, and  our choice of $Q_v$    restricts the state space to an irreducible component of the network. \qedhere 
\end{remark}
\end{example}

\begin{example}
[Height-arrow network with sinks]
\label{e. sandpile network with sinks}
Fix a nonempty set $S \subseteq V$ that we designate as \emph{sinks}.
For each $v \in V$,  assign a threshold value $\tau_v\in \{1,\ldots, \outdeg(v)\}$ and  a  cyclic total order  $\{e_j^v \mid j \in \Z_{\outdeg(v)}\}$ to  the  out-going edges of  $v$.

The alphabet $A_v$, the state space $Q_v$,
and the transition function $T_v$ are the same as in sinkless height-arrow networks.
 The message-passing function for the edge $e_j^v$ directed from $v$ to $u^v_j$  is given by
\[T_{e_j^v}(d,c,v):=\begin{cases} u_j^v  & \text{if }   c=\tau_v-1, \ j -  d \in \{ 1,\ldots, \tau_v \} \text{ (mod } \outdeg(v)), \text{ and } v_j \notin S;\\
\epsilon  & \text{otherwise. }
\end{cases} \]
This network is identical to the sinkless height-arrow network, except that  letters passing through any  edge pointing to the sink are removed from the network.
See Figure~\ref{figure:sandpile network with sink example} for an illustration of this process.

\begin{figure}[tb]
\centering
   \includegraphics[width=0.9\textwidth]{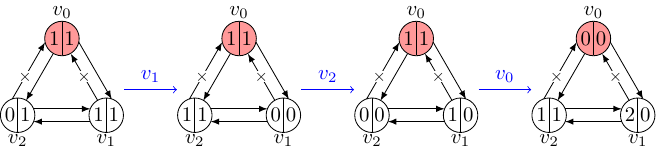}
   \caption{A three-step legal execution in the sandpile network with a sink at $S=\{v_0\}$.
  The incoming edges of a sink are marked with ``$\times$''.
   (Note that the left part of  $v\in V$ records $\x(v)$, while the right part records $\q(v)$.)
    }  
   \label{figure:sandpile network with sink example}
\end{figure}

The total kernel of a height-arrow network with sinks is equal to the total kernel of the corresponding sinkless height-arrow network.
The production matrix $P$ of this network is equal 
to the matrix $\A_G \D_G^{-1}$ with rows corresponding to $S$ replaced with zero vectors.
Since $P \leq \A_G \D_G^{-1}$ and $\lambda(\A_G \D_G^{-1})=1$,
we have by the Perron-Frobenius theorem (Lemma~\ref{lemma: Perron-Frobenius theorem}\eqref{item: Perron-Frobenius lambda is strictly increasing})  that $\lambda(P)<1$ (if $G$ is strongly connected).
 Hence a height-arrow network with sinks on a strongly connected digraph is subcritical, unlike its sinkless counterpart.

\begin{remark}
In \cite{BL16a} a sink is defined as a processor with one state that sends no messages. 
However, in this paper we follow the convention from \cite{Chan18} that places sinks  on the \emph{incoming edges} to each $s \in S$ instead. The user can still opt to send input to $s$, and the processor $\PP_s$ can still send messages to its out-neighbors.  This extra flexibility comes in handy when we relate critical and subcritical networks in \S\ref{ss. abelian network with thief}. \qedhere
\end{remark}
\end{example}

\begin{example}[Arithmetical network~\cite{Lor89}]
\label{e. arithmetical graphs}
This network is determined  by the pair $(\Dartm,\bb)$,
where  $\Dartm$ is a diagonal matrix with positive integer diagonal entries,  and 
$\bb$ is a positive vector in the kernel of $\Dartm-A_G$ that satisfies $\gcd_{v \in V} (\bb(v))=1$.

For each  vertex $v \in V$, the processor 
$\PP_v$ is given by:
 \begin{align*}
& A_v:=\{v\}, \qquad Q_v:=\{0,1,\ldots, d_v-1\},  \qquad  T_{v}(i):=i+1 \ \text{mod } d_v,
 \end{align*} 
where $d_v$ is the diagonal entry of $\Dartm$ that corresponds to $v$.
For each edge $e$ directed from $v$ to $u$,  the message-passing function is given 
by
\[T_{e}(c,v):=\begin{cases} u  & \text{if } c=d_v -1;\\
\epsilon & \text{otherwise}.
\end{cases} 
\]

Similar to  sandpile networks, we can think of each processor $\PP_v$ of this network as a locker that   can store up to $d_v-1$ chips.
Once it has $d_v$ chips, all these $d_v$ chips in $\PP_v$ are  
removed, and then $\PP_v$ sends one chip along each of its outgoing edges to its out-neighbors.
Note that the total number of chips in this network may decrease or increase, 
depending on the quantity $\outdeg(v)-d_v$.
See Figure~\ref{figure:row chip firing network example} for an example of this process.

\begin{figure}[tb]
\centering
   \includegraphics[width=1\textwidth]{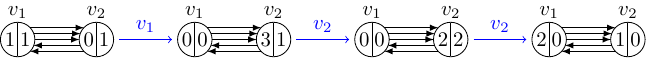}
   \caption{A three-step legal execution in a row chip-firing network (i.e.,  $d_{v_1}=2$ and $d_{v_2}=3$).
  In the figure,  the left part of a vertex records the number of letters waiting to be processed, and the right part records the state of the processor.}  
   \label{figure:row chip firing network example}
\end{figure}

  If $\Dartm$ is the outdegree matrix of $G$, then $\Net$ is  the sinkless sandpile network on $G$.
 If $\Dartm$ is the indegree matrix of $G$,
 then $\Net$ is called the \emph{row chip-firing network}~\cite{PS04,AB11}
 (Note that due to   a different convention  for matrix indexing,
$\Net$ is called the    column chip-firing network in \cite{AB11}).

 Any arithmetical network is strongly connected if the underlying digraph $G$ is  strongly connected.
  The total kernel and the production matrix of this network are given by
 \[ K=\{  \z \in \Z^V \mid  \z(v) \text{ is divisible by } d_v \text{ for all } v \in V  \}; \qquad P=\A_G\Dartm^{-1}. \]
   Because $P(\Dartm\bb)=\Dartm\bb$ by definition, 
 the spectral radius $\lambda(P)$ is   1 by
 the Perron-Frobenius theorem (Lemma~\ref{lemma: Perron-Frobenius theorem}\eqref{item: Perron-Frobenius 2}).
  Hence  an arithmetical network on a strongly connected digraph is a critical network.

  There exist only finitely many arithmetical networks on a fixed  strongly connected digraph~\cite{CV16}.
 For  example, the bidirected cycle $C_3$ has ten arithmetical structures~\cite{CV16}, namely all the   permutations of these three structures:
 \begin{align*}
&\Dartm_1:=\begin{bmatrix}
2 & 0 & 0\\
0 & 2 & 0\\
0 & 0 & 2
\end{bmatrix},
\
\bb_1:=\begin{bmatrix}
1 \\ 1\\ 1
\end{bmatrix};
\qquad
\Dartm_2:=\begin{bmatrix}
1 & 0 & 0\\
0 & 3 & 0\\
0 & 0 & 3
\end{bmatrix},
\
\bb_2:=\begin{bmatrix}
2 \\ 1\\ 1
\end{bmatrix};\\
&\Dartm_3:=\begin{bmatrix}
1 & 0 & 0\\
0 & 2 & 0\\
0 & 0 & 5
\end{bmatrix},
\
\bb_3:=\begin{bmatrix}
3 \\ 2\\ 1
\end{bmatrix}. 
\end{align*}

For a study  of  arithmetical structures on bidirected paths and cycles, we refer the reader to \cite{CV16} and \cite{BCC17}.
\end{example}

All examples presented so far are either subcritical or  critical networks.
In the following example we present a family of abelian networks that includes  supercritical networks.

\begin{example}
[Branching  rotor network]
\label{example: branching rotor network}
Just like for sinkless rotor networks, we  assign to each  $v \in V$  a cyclic total order   $\{e_i^v \mid i \in \Z_{\outdeg(v)}\}$ to the outgoing edges of $v$.
The processor $\PP_v$ is given by
\begin{align*}
& A_v:=\{v\}, \qquad  Q_v:=\{ e_{2i}^v \mid i \in \Z_{\outdeg(v)} \}, \\
&  T_v(e_{2i}^v):=e_{2i+2}^v \quad (i \in \Z_{\outdeg(v)}).
\end{align*}
(Note that $|Q_v|$ is equal to $\frac{\outdeg(v)}{2}$  if $\outdeg(v)$ is even, and is equal to $\outdeg(v)$ otherwise.) 

For each edge $e_j^v$ directed from $v$ to $u_j^v$, the message-passing function is given by
\[ T_{e}(e_{2i}^{v},v):=\begin{cases} u_j^v & \text{ if } 2i-j\in \{1,2\} \  \text{ (mod } {\outdeg(v)}); \\
\epsilon & \text{ otherwise.} \end{cases} \]

Similar to sinkless rotor networks, 
a state  of this network  can be thought as a function $V \to E$ assigning a vertex $v$ to an outgoing edge of $v$.
When  a chip/letter at vertex $v$ is processed, the  edge/state $e_{2i}^v$ assigned to $v$  first moves to $e_{2i+1}^v$ and then to $e_{2i+2}^v$, and drops one chip at the target vertex of every visited edge.
Note that  branching rotor networks create two new chips
 for each processed chip.
See Figure~\ref{figure:branching rotor network example} for an illustration of this process.

\begin{figure}[tb]
\centering
   \includegraphics[width=1\textwidth]{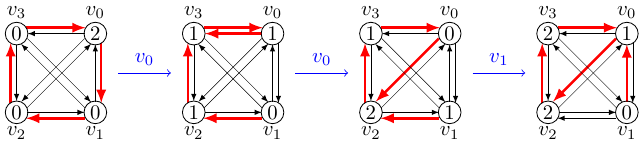}
   \caption{A three-step legal execution in the  branching rotor network on the  complete digraph with four vertices.
Each vertex is assigned the counterclockwise ordering for the    cyclic total order on its outgoing edges.
    Note that the circled number  records the number of letters waiting to be processed, and the (red) thick outgoing edge  records the state of the processor.}  
   \label{figure:branching rotor network example}
\end{figure}

 Any branching rotor network is strongly connected if the underlying digraph $G$ is  strongly connected.
The total kernel and the production matrix  of this network are given by 
 \[ K=\{  \z \in \Z^V \mid  \z(v) \text{ is divisible by } |Q_v| \text{ for all } v \in V  \}; \qquad P=2\A_G  \D_G^{-1}. \] 
 Because $\satu  \A_G  \D_G^{-1}= \satu$,
 the Perron-Frobenius theorem  (Lemma~\ref{lemma: Perron-Frobenius theorem}\eqref{item: Perron-Frobenius 2}) implies that $\lambda(P)=2$, 
 and hence  this network is  supercritical.
\end{example}

\begin{example}[Toppling network~\cite{Gab93,BL16a}]
\label{example: toppling networks}
In a toppling network, each vertex $v \in V$ of the underlying digraph $G$ is assigned a \emph{threshold} $t_v \in \N$.
For each  $v \in V$,
 the processor $\PP_v$ is given by
 \begin{align*}
& A_v:=\{v\}, \qquad Q_v:=\{0,1,\ldots, t_v-1\},  \qquad  T_{v}(i):=i+1 \ \text{mod } t_v.
 \end{align*} 
For each edge $e$ directed from $v$ to $u$, the message-passing function 
is given by
\[ T_{e}(i,v):=\begin{cases} u & \text{ if } i=t_v-1; \\
\epsilon & \text{ otherwise.} \end{cases} \]

Consider now the toppling network on the bidirected cycle $C_3$
with $t_{v_0}=t_{v_1}=t_{v_2}=:t$.
 The production matrix of this network is given by:
\[P=\frac{1}{t}\begin{bmatrix}
0&1&1\\
1& 0 & 1\\
1& 1 &0 \end{bmatrix}.  \]
It follows that $\lambda(P)=\frac{2}{t}$,
so this network is subcritical if $t>2$, is critical if $t=2$, and is supercritical if $t=1$.

We remark that subcritical toppling networks are also known as 
\emph{avalanche-finite} networks, and we refer to \cite{GK15} for more discussions on this network.
We also remark that,
 on a strongly connected digraph,
critical toppling networks are equal to arithmetical networks from Example~\ref{e. arithmetical graphs}.
\end{example}

 The following example is an instance of  toppling networks that 
arises naturally from the representation theory.

\begin{example}[McKay-Cartan network~\cite{BKR18}]
\label{example: McKay-Cartan network}
Let $\Gc$ be finite group, and let $\gamma: \Gc \hookrightarrow \texttt{GL}_n(\Cb)$
be a faithful representation.
The underlying digraph of the McKay-Cartan network is the \emph{McKay quiver} with vertices
the complex irreducible characters  $\chi_0,\ldots, \chi_k$   of $\Gc$,
and  with $m_{ij}$ edges from $\chi_i$ to $\chi_j$ if 
\[ \chi_{\gamma} \chi_i = \sum_{j=0}^k m_{ij} \chi_j,    \]
where $\chi_{\gamma}$ is the character of $\gamma$.
The \emph{McKay-Cartan} network of $(\Gc,\gamma)$ is the toppling network on the McKay quiver with threshold $n$ for every vertex.

The production matrix of this network is equal to $\frac{1}{n} M$,
where $M:=(m_{i,j})_{0\leq i,j \leq k}$ is the \emph{extended McKay-Cartan} matrix of $(\Gc,\gamma)$.
This network is strongly connected since $\gamma$ is faithful
\cite[Proposition~5.3(c)]{BKR18}.
Moreover, $P\mathbf{d} =\mathbf{d}$, where $\mathbf{d}(\chi_i)$ is the dimension of $\chi_i$~\cite[Proposition~5.3(b)]{BKR18}.
Hence this network is a critical network.

When $\gamma$ is a faithful representation of $G$ into the special linear group $\texttt{SL}_{n}(\Cb)$, the torsion group (to be defined in \S\ref{subsection: construction of torsion group for all abelian networks}) of this network is isomorphic to the  abelianization of $\Gc$ ~\cite[Theorem~1.3]{BKR18}.
\end{example}

All the  examples presented so far are  \emph{unary networks}, i.e.,  the alphabet  of each processor contains exactly one letter.
In the following example we present a non-unary  network.

\begin{example}[Inverse network]
\label{e. inverse}
For each vertex $v \in V$, fix a positive integer $m_v$.
 The processor $\PP_v$ is given by:
\begin{alignat*}{2}
& A_v:=\{a_v,b_v\}, \qquad && Q_v:= \Z_{m_v}, \\
& T_{a_v}(i):=i+1 \mod m_v , \quad  && T_{b_v}(i):=i-1 \mod m_v   \quad (i \in \Z_{m_v}).
\end{alignat*}

Let $c_v$ and $d_v$ be two distinct letters 
in $\bigsqcup_{w \in \Out(v)} A_w$.
For each $i \in \Z_{m_v}$, fix an   element $x_i$  from $\{c_v,d_v\}$.
We define $x_i^*$ to be
\[x_i^*:=\begin{cases} c_v &\text{ if } x_i=d_v;\\ d_v &\text{ if } x_i=c_v. \end{cases} \]
  
The processor $\PP_v$ operates as follows:
\begin{itemize}
\item 
 Processing the letter $a_v$ on state $i$ produces the letter $x_i$; and 
 \item Processing the letter $b_v$ on state $i$ produces the letter $x_{i-1}^*$.
\end{itemize}

{    \renewcommand{\arraystretch}{1.2}
\begin{table}[tb]
\caption{Example of a message-passing function of an inverse network on the digraph with one vertex and one loop.
The alphabet is $\{a,b\}$ and the state space is $\Z_{7}$.
The $(i,\alpha)$-th entry of the table represents the letter produced when a precessor in state $i$ processes the letter $\alpha$.
Note that the $(i,a)$-th entry is always different from the $(i+1,b)$-th entry.
  }
{\small
\begin{tabular}{|c|c |c |c|c|c|c|c|}
\hline
 \backslashbox{$A$ }{$Q$}  & 0 & 1 &2 &3 &4 &5&6\\\hline
a  &  a & b  & a & a& b &b &b   \\
\hline 
b  & a & b & a & b & b &a &a  \\
\hline 
\end{tabular}
}

  \label{table: inverse network example}
\end{table}
}

For each $v\in V$, 
note that  $t_{a_v}\circ t_{b_v}=t_{b_v}\circ t_{a_v}=\text{id}$.
Also note that, for all $i \in \Z_{m_v}$, 
\begin{align*}
\NN_{a_v b_v}( i)= &\NN_{a_v}(i)+ \NN_{b_v}(t_{a_v}(i))=|x_{i}| +|x_{i}^*|=|c_v|+|d_v|,\\
\NN_{b_v a_v} (i)= &\NN_{b_v}(i)+ \NN_{a_v}(t_{b_v}(i))=|x_{i-1}^*| +|x_{i-1}|=|c_v|+|d_v|.
\end{align*}
This shows that  inverse network  is an abelian network.

The total kernel of this network is  
 \[ K=\{  \z \in \Z^A \mid  \z(a_v)=\z(b_v) \mod m_v \text{ for all } v \in V  \}. \] 
The production matrix $P$ of  any inverse  network satisfies
   $\satu  P=\satu$ since executing any letter in $A$ produces exactly one new (not necessarily the same) letter.
   By the Perron-Frobenius theorem (Lemma~\ref{lemma: Perron-Frobenius theorem}\eqref{item: Perron-Frobenius 2}) the spectral radius
   $\lambda(P)$ is equal to 1, and hence this network is    critical.
\end{example}

\chapter[Torsion Group]{The Torsion Group of an Abelian Network}\label{section: torsion group for abelian networks}
We start  this chapter  with a fundamental lemma that we call the removal lemma.
We then use the removal lemma  and the monoid theory from Chapter \ref{section: background on commutative monoid theory} to construct the torsion group for any  abelian network.
Finally,  we show that the torsion group is equal to the critical group from \cite{BL16c} if the network is subcritical.

\section{The removal lemma}\label{subsection: removal lemma}

\begin{definition}[Removal of a vector from a word]\label{definition: removal}
For  $w\in A^*$  and $\n \in \N^A$, the \emph{removal of} $\n$ \emph{from} $w$, denoted  
$w\setminus \n$,
is the word obtained from $w$ by deleting the first $\n(a)$ occurrences of $a$ for all $a \in A$. 
(If $a$ appears for less than $\n(a)$ times in $w$, then delete all occurrences of $a$.)
\end{definition}

Recall the definition of $\dashrightarrow$, $\longrightarrow$, and legal executions from \S\ref{subsection: legal and complete executions}.
Also recall that, for any $w \in A^*$,
we denote by  $|w|$ the vector in $\N^A$ that counts the number of occurrences of each letter in $w$.  

The following lemma is called the \emph{removal lemma},  as it  removes some letters from a legal execution to get a shorter legal execution.
A special case of this lemma when $\Net$ is a sinkless sandpile network and  $\n$ is  the period vector~ (to be defined in \S\ref{ss. recurrent configurations and burning test})  is proved in   \cite{BL92}.

\begin{lemma}[Removal lemma]\label{lemma: removal lemma}
Let $\Net$ be an  abelian network, and let $\x.\q$ be a configuration of $\Net$.
Then for any $\n \in \N^A$ and any legal execution $w$ for $\x.\q$,
the word $w \setminus \n$ is a legal execution for $\pi_{\n}(\x.\q)$.
\end{lemma}

  \begin{proof}
By induction on the length of the vector $\n$,
it suffices to show that, for any $a \in A$, the word  
$w \setminus |a|$ is a legal execution for 
$\pi_a(\x.\q)$.
  
  Fix $a \in A$ throughout this proof.
  Let  $w=a_1\cdots a_\ell$ be the given legal execution for $\x.\q$.
 Let $k$ be equal to the smallest number such that $a_k=a$ if $w$ contains $a$, and   equal to $\ell+1$ if $w$ doesn't contain $a$.  
For  $i \in \{0,\ldots,\ell \}$, we write $\x_i.\q_i:=\pi_{a_1\cdots a_i}(\x.\q)$ and $\y_i.\p_i:=\pi_{a_1 \cdots a_i \setminus |a| } (\pi_{a}(\x.\q))$. We need to show that 
     $\y_{i-1}(a_i)\geq 1$ for $i \in \{1,\ldots,\ell\} \setminus\{k\}$.

If $i \in \{1,\ldots, k-1\}$, then 
  \begin{align*}
    \y_{i-1}=& \x+ \NN_{aa_1\cdots a_{i-1}}(\q)- |a|
 - \sum_{j=1}^{i-1} |a_j| \\
 \geq&  \x+ \NN_{a_1\cdots a_{i-1}}(\q)- |a|
 -\sum_{j=1}^{i-1}   |a_j|   \qquad \text{(by the monotonicity property (Lemma \ref{l. abelian enumerate}\eqref{l. monotonicity}))}\\
 =& \x_{i-1}-|a|.
 \end{align*}  
Note that $|a|(a_i)=0$ by the minimality of $k$, and also note that $\x_{i-1}(a_i)\geq 1$ since $w$ is legal for $\x.\q$.
Hence  $\y_{i-1}(a_i)\geq  \x_{i-1}(a_i)-|a|(a_i)\geq 1$.
  
  If $i \in \{k+1,\ldots,\ell\}$,  then
\begin{align*}
\y_{i-1}=& \x+ \NN_{aa_1\cdots \widehat{a_k} \cdots a_{i-1} }(\q) - |a|
 -\sum_{j \in \{1,\ldots, i\} \setminus \{k\} }   |a_j|\\
 =& \x+ \NN_{a_1\cdots  a_i }(\q)
 -\sum_{j=1}^{i-1}  |a_j| \qquad \text{(by the abelian  property (Lemma~\ref{l. abelian enumerate}\eqref{l. abelian property}))}\\
 =&\x_{i-1}.
\end{align*}
Then $\y_{i-1}(a_i)=\x_{i-1}(a_i)\geq 1$ since $w$ is legal for $\x.\q$.
This completes the proof. 
  \end{proof}

Described using a diagram, the removal lemma says that
\begin{center}
\begin{tikzcd}
\x.\q \arrow[r,"w"] \arrow[d,"\n",dashed]& \pi_w(\x.\q) \\
\pi_{\n}(\x.\q)
\end{tikzcd}
 \  \ implies \ \ 
\begin{tikzcd}
\x.\q \arrow[r,"w"] \arrow[d,"\n",dashed]& \pi_w(\x.\q) \arrow[d,"\n \setminus |w|",dashed]\\
\pi_{\n}(\x.\q) \arrow[r,"w\setminus \n"] &\pi_{\max(|w|,\n)}(\x.\q)
\end{tikzcd},
\end{center}
 where   $\max(\x,\y)$ of two vectors $\x,\y \in \Z^A$ denotes the coordinatewise maximum of $\x$ and $\y$.

Despite the apparent simplicity of  the removal lemma, its consequences are very useful.
One such consequence is  the \emph{least action principle}.

Recall the definition of complete execution from \S\ref{subsection: legal and complete executions}.
\begin{corollary} [Least action principle~{\cite[Lemma 4.3]{BL16a}}]
 \label{c. least action principle}
Let $\Net$ be an  abelian network.
 If $w$ is a legal execution for $\x.\q$ and $w'$ is a complete execution for $\x.\q$, then $|w| \leq |w'|$. 
\end{corollary}
\begin{proof}
Since $w$ is legal for $\x.\q$, the removal lemma implies that $w \setminus |w'|$ is a legal execution for $\pi_{w'}(\x.\q)$.
On the other hand, the only legal execution for $\pi_{w'}(\x.\q)$ is the empty word since $w'$ is complete for $\x.\q$.
 Hence $w \setminus |w'|$ is the empty word, which  implies that $|w| \leq |w'|$.
\end{proof}

The second consequence of the removal lemma is  the {exchange lemma}, presented below.

\begin{lemma}[Exchange lemma, c.f. {\cite[Lemma 1.2]{BLS91}}]
\label{l. weak exchange lemma}
Let $\Net$ be an abelian network.
If $w_1$ and $w_2$ are two legal executions for $\x.\q$, then there exists $w \in A^*$ such that $w_1w$ is a legal execution for $\x.\q$ and $|w_1|+|w|=\max(|w_1|,|w_2|)$. 
\end{lemma}
\begin{proof}
This follows  from the removal lemma by taking $w$ to be  $w_2 \setminus |w_1|$.
\end{proof}

Described using a diagram, the exchange lemma says that
\begin{equation}\label{diagram: exchange lemma}
\begin{tikzcd}
 & \x_2.\q_2\\
\x_1.\q_1 \arrow[ru,"w_1"] \arrow[rd,"w_2"]& \\
 & \x_3.\q_3
\end{tikzcd}
\ \  \textnormal{implies} \ \   
\begin{tikzcd}
 & \x_2.\q_2 \arrow[rd, "w_2 \setminus |w_1|"]\\
\x_1.\q_1 \arrow[ru, "w_1"] \arrow[rd, "w_2"]&  & \x_4.\q_4\\
 & \x_3.\q_3 \arrow[ru, "w_1 \setminus |w_2|"]
\end{tikzcd}.
\end{equation}

The exchange lemma is named after a similar property of antimatroids with repetition~\cite{BZ92}.
It was proved by Bj\"orner, Lov\'asz and Shor \cite[Lemma~1.2]{BLS91} for sandpile networks on undirected graphs,
and extended to directed graphs by Bj\"orner and Lov\'asz \cite[Proposition~1.2]{BL92}.

One consequence of the exchange lemma is that all abelian networks are \emph{confluent} in the sense of Huet \cite{Huet80}: that is,
any two legal executions $w_1$ and $w_2$  for the same configuration $\x.\q$ can be extended to longer legal executions that are equal up to a permutation of their letters (see Diagram~\eqref{diagram: exchange lemma} for an illustration).
Furthermore, 
if the abelian network $\Net$ is critical, 
then we will show that the extended execution can be taken to be of length $\max(|w_1|,|w_2|)+C$ for a constant $C$ that depends only on the network (see Theorem~\ref{theorem: confluence bound}).

\section{Recurrent components}
\label{subsection: recurrent components}
\begin{figure}
\begin{tabular}{l}
 (i) $t_{v_0}=t_{v_1}=t_{v_2}=3$ ($\Net$ is subcritical):\\
\begin{tikzcd}
{  \color{blue} {\mathbf{\bullet\bullet\bullet}}} \arrow[rr,"v_2^3v_1^3v_0^3", blue,very thick] & &
{\color{blue}\begin{bmatrix}
1 \\ 2 \\ 3
\end{bmatrix}.
\begin{bmatrix}
0 \\ 0 \\ 0
\end{bmatrix} }   
 \arrow[rr,"v_2^3v_1^3v_0^3", blue,very thick] & &
{\color{blue} \begin{bmatrix}
0 \\ 1 \\ 2
\end{bmatrix}.
\begin{bmatrix}
0 \\ 0 \\ 0
\end{bmatrix}    }
\arrow[rr,"v_1v_2^2", blue,very thick] & &
{\color{blue} \begin{bmatrix}
0 \\ 0 \\ 0
\end{bmatrix}.
\begin{bmatrix}
0 \\ 1 \\ 2
\end{bmatrix}}\\
\cdots \arrow[rr] & &
\begin{bmatrix}
1 \\ 1 \\0
\end{bmatrix}.
\begin{bmatrix}
1 \\ 2 \\ 0
\end{bmatrix}
\arrow[rr,"v_0"]
&  &
\begin{bmatrix}
0 \\ 1 \\0
\end{bmatrix}.
\begin{bmatrix}
2 \\ 2 \\ 0
\end{bmatrix}
\arrow[u,"v_1v_0"]
\\
\cdots \arrow[rr] & &\begin{bmatrix}
-1 \\ 0 \\ 1
\end{bmatrix}.
\begin{bmatrix}
2 \\ 2 \\ 2
\end{bmatrix} \arrow[rru, "v_2"] 
\end{tikzcd}
\\

\\
 (ii) $t_{v_0}=t_{v_1}=t_{v_2}=2$ ($\Net$ is critical):\\

\begin{tikzcd}
\cdots \arrow[r] & 
\begin{bmatrix}
3 \\ 1 \\ 0
\end{bmatrix}.
\begin{bmatrix}
1 \\ 0 \\ 0
\end{bmatrix}    
\arrow[r, "v_0" ]   
&    
{\color{blue}
\begin{bmatrix}
2 \\ 1 \\ 0
\end{bmatrix}.
\begin{bmatrix}
0 \\ 0 \\ 0
\end{bmatrix}}
 \arrow[r, "v_0",blue,very thick ] 
 & 
{\color{blue} 
 \begin{bmatrix}
1 \\ 1 \\ 0
\end{bmatrix}.
\begin{bmatrix}
1 \\ 0 \\ 0
\end{bmatrix}}
 \arrow[rd, "v_0",blue,very thick]\\
& 
{\color{blue}
\begin{bmatrix}
1 \\ 0 \\ 1
\end{bmatrix}.
\begin{bmatrix}
0 \\ 0 \\ 1
\end{bmatrix} 
}   \arrow[ru, "v_2",blue,very thick] & & &
{\color{blue}
\begin{bmatrix}
0 \\ 2 \\ 1
\end{bmatrix}.
\begin{bmatrix}
0 \\ 0 \\ 0
\end{bmatrix} }
 \arrow[ld, "v_1",blue,very thick,swap]  \\
& & 
{\color{blue}
\begin{bmatrix}
1 \\ 0 \\ 2
\end{bmatrix}.
\begin{bmatrix}
0 \\ 0 \\ 0
\end{bmatrix} }
  \arrow[lu,"v_2",blue,very thick,swap] &
{\color{blue}
\begin{bmatrix}
0 \\ 1 \\ 1
\end{bmatrix}.
\begin{bmatrix}
0 \\ 1 \\ 0
\end{bmatrix}  }
 \arrow[l,"v_1",blue,very thick,swap] 
\end{tikzcd}
\\

\\
 (iii) $t_{v_0}=t_{v_1}=t_{v_2}=1$ ($\Net$ is supercritical):\\

 \begin{tikzcd}
\begin{bmatrix}
2 \\0\\0
\end{bmatrix}.
\begin{bmatrix}
0 \\0\\0
\end{bmatrix}
\arrow[rd,"v_0"]
& 
&
\begin{bmatrix}
2 \\0\\2
\end{bmatrix}.
\begin{bmatrix}
0 \\0\\0
\end{bmatrix}
\arrow[r]
&\cdots\\
{\color{blue}
\begin{bmatrix}
0 \\2\\0
\end{bmatrix}.
\begin{bmatrix}
0 \\0\\0
\end{bmatrix}}
\arrow[r,blue,very thick,"v_1"]
 &
 {\color{blue}
 \begin{bmatrix}
1 \\1\\1
\end{bmatrix}.
\begin{bmatrix}
0 \\0\\0
\end{bmatrix}}
\arrow[ru,"v_1"]
\arrow[r,blue,very thick, "v_2"]
\arrow[rd,"v_0"]
 &
 {\color{blue}
 \begin{bmatrix}
2 \\2\\0
\end{bmatrix}.
\begin{bmatrix}
0 \\0\\0
\end{bmatrix}}
\arrow[r,blue,very thick,"v_0"]
 &
 {\color{blue}
 \begin{bmatrix}
1 \\3\\1
\end{bmatrix}.
\begin{bmatrix}
0 \\0\\0
\end{bmatrix}}
\arrow[r,blue,very thick,"v_1"]
 &
{\color{blue} \bullet\bullet \bullet}\\
 \begin{bmatrix}
0 \\0\\2
\end{bmatrix}.
\begin{bmatrix}
0 \\0\\0
\end{bmatrix}
\arrow[ru,"v_2"]
& &
 \begin{bmatrix}
0 \\2\\2
\end{bmatrix}.
\begin{bmatrix}
0 \\0\\0
\end{bmatrix}
\arrow[r]
& \cdots
\end{tikzcd}
\\
\end{tabular}
\caption{Three different toppling networks on the bidirected cycle $C_3$. In each case, a portion of one component of the trajectory graph is shown.  
 The presence of a backward infinite path / cycle / forward infinite path shows that the component is recurrent.}  
\label{figure: diverse infinite walks}
\end{figure}
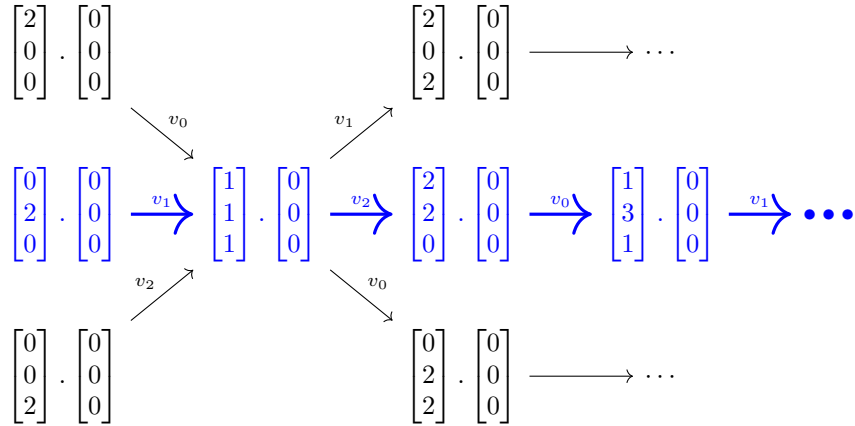

In this section we discuss recurrent components, which will be an integral ingredient in the construction of  the torsion group.
The reader can use the illustrations in Figure~\ref{figure: diverse infinite walks}  to develop intuition  when reading this section.

We start  with the definition of  recurrent components, which requires the notion of the trajectory digraph given below.
 
 \begin{definition}[Trajectory digraph]\label{definition: trajectory digraph}
 Let $\Net$ be an abelian network.
 The \emph{trajectory digraph} of $\Net$ is the digraph with edges labeled by $A$ given by
\begin{align*}
V:=&\{ \x.\q \mid x \in \Z^A, \q \in Q     \};\\
E:=& \bigsqcup_{a \in A} E_a;\\
E_a:=& \{  (\x.\q, \x'.\q') \mid \x.\q \xlongrightarrow{a} \x'.\q'  \} \quad (a \in A).\qedhere
\end{align*} 
 \end{definition}

\begin{definition}[Quasi-legal  and legal relation]\label{definition: weak and strong relation}
Let $\Net$ be an abelian network.
Two configurations $\x_1.\q_1$ and $\x_2.\q_2$ of $\Net$ are \emph{quasi-legally related}, denoted  $\x_1.\q_1 \dashrlarrow \x_2.\q_2$, 
if there exists $\x_3.\q_3$ such that 
$\x_1.\q_1 \dashrightarrow \x_3.\q_3$ and $\x_2.\q_2 \dashrightarrow \x_3.\q_3$.
Two configurations  $\x_1.\q_1$ and $\x_2.\q_2$ are \emph{legally related}, denoted  $\x_1.\q_1 \rlarrow \x_2.\q_2$, 
if there exists $\x_3.\q_3$ such that 
$\x_1.\q_1 \longrightarrow \x_3.\q_3$ and $\x_2.\q_2 \longrightarrow \x_3.\q_3$.
\end{definition}

The symmetry and reflexivity of these two relations follow from the definition.
The transitivity of  $\rlarrow$ follows from the  exchange lemma~(Lemma~\ref{l. weak exchange lemma}), because
\begin{equation}\label{diagram: diagram transitivity}
\begin{tikzcd}
\x_1.\q_1 \arrow[r] & \x_4.\q_4\\
\x_2.\q_2 \arrow[ru,"w_1"] \arrow[rd,"w_2"]& \\
\x_3.\q_3 \arrow[r] & \x_5.\q_5
\end{tikzcd}
\ \  \textnormal{implies} \ \   
\begin{tikzcd}
\x_1.\q_1 \arrow[r] & \x_4.\q_4 \arrow[rd, "w_2 \setminus |w_1|"]\\
\x_2.\q_2 \arrow[ru, "w_1"] \arrow[rd, "w_2"]&  & \x_6.\q_6\\
\x_3.\q_3 \arrow[r] & \x_5.\q_5 \arrow[ru, "w_1 \setminus |w_2|"]
\end{tikzcd}.
\end{equation}
The transitivity of the quasi-legal relation is proved by an analogous diagram.
Hence both  relations are equivalence relations on the configurations of $\Net$.

\begin{definition}[Component of the trajectory digraph]\label{definition: component}
Let $\Net$ be an abelian network.
A \emph{component} of the trajectory digraph of $\Net$ is an induced subgraph of the trajectory digraph formed by an  equivalence class for the legal relation. 
\end{definition}
See Figure~\ref{figure: diverse infinite walks} for an illustration.

A \emph{forward infinite walk} in $\Net$ is an infinite  legal execution of the form  $\x_0.\q_0 \xlongrightarrow{a_1} \x_1.\q_1 \xlongrightarrow{a_2} \cdots$ ($a_i \in A$).
A \emph{backward infinite walk} is an infinite legal execution $\cdots \xlongrightarrow{a_{-1}} \x_{-1}.\q_{-1} \xlongrightarrow{a_{0}} \x_0.\q_0$.
A \emph{bidirectional infinite walk} is an infinite legal execution 
$\cdots \xlongrightarrow{a_{0}} \x_{0}.\q_{0} \xlongrightarrow{a_{1}}  \cdots $.
A bidirectional infinite walk  is a \emph{cycle} if there is a positive $k$ such that $\x_{i+k}.\q_{i+k}=\x_i.\q_i$ and $a_{i+k}=a_i$ for all $i \in \Z$.
An \emph{infinite walk} in $\Net$ means either one of those three walks, i.e.,  a sequence $\cdots \xlongrightarrow{a_{i}} \x_{i}.\q_{i} \xlongrightarrow{a_{i+1}} \cdots $ indexed by $I$, where $I$ is either $\Z_{\leq 0}$, $\Z_{\geq 0}$, or $\Z$.
An \emph{infinite path} is an infinite walk in which all $\x_i.\q_i$'s  are  distinct.

\begin{definition}[Recurrent component]\label{definition: recurrent class}
Let $\Net$ be an abelian network.
An infinite walk  indexed by a set $I$ is \emph{diverse} if  for all $a \in A$ the set 
$\{ i \in I \mid a_i=a  \}$
is infinite.
A component of the trajectory digraph is a \emph{recurrent component} if it contains  a diverse infinite walk.
\end{definition}

We denote by $\Lrec(\Net)$ the set of recurrent components of $\Net$, and
 by $\overline{\x.\q}$ the component of the trajectory digraph that contains the configuration $\x.\q$.

Assume throughout the rest of this section that $\Net$ is finite and locally irreducible.
The first main result of this section is that, assuming recurrence,
we have the quasi-legal relation implies the legal relation.

\begin{proposition}\label{proposition: weak relation implies strong relation for recurrent class}
Let $\Net$ be a finite and locally irreducible  abelian network.
If $\x_1.\q_1$ and $\x_2.\q_2$ are configurations such that 
$\overline{\x_1.\q_1}$ and $\overline{\x_2.\q_2}$ are recurrent components,
then  $\x_1.\q_1 \dashrlarrow \x_2.\q_2$ implies $\x_1.\q_1 \rlarrow \x_2.\q_2$.
\end{proposition}

We remark that Proposition~\ref{proposition: weak relation implies strong relation for recurrent class} for the special case of sinkless rotor networks was proved in \cite[Proposition~3.7]{Toth18}.

The second main result of this section is a trichotomy   of the recurrent components of $\Net$ that depends on the value of $\lambda(P)$.

\begin{proposition}
\label{proposition: recurrent component combinatorial description}
Let $\Net$ be a finite, locally irreducible, and strongly connected abelian network.
Then  the following are equivalent:
\begin{enumerate}
\item \label{item: recurrent combinatorial 1} $\Net$ is a subcritical  network;

\item   \label{item: recurrent combinatorial 2}  All recurrent components of $\Net$ contain a diverse backward infinite path; and 
\item  \label{item: recurrent combinatorial 3}  There exists a recurrent component of $\Net$ that contains a diverse backward infinite path.
\end{enumerate}
Furthermore, the same statement holds if subcritical is replaced with critical (resp. supercritical) and diverse backward infinite path is replaced with diverse cycle (resp. diverse forward infinite path).
\end{proposition}

An illustration of  recurrent components for each case (subcritical, critical, supercritical) is shown in Figure~\ref{figure: diverse infinite walks}.

We now build towards the proof of Proposition~\ref{proposition: weak relation implies strong relation for recurrent class}.

Recall from \S\ref{subsection: a classification of abelian networks} that $A_{<}$ denotes the set of subcritical letters of $\Net$,
and $A_{\geq}$ denotes the set of critical and supercritical letters of $\Net$.
We say that $\vb,\w \in \N^A$ are \emph{extendable vectors} of $\Net$ if 
\begin{enumerate}
[{label=\textnormal{({E\arabic*})},
ref=\textnormal{E\arabic*}}]
\item \label{item: extendable 1} $\supp(\vb)=A_{<}$ and $P\vb(a)\leq \vb(a)$ for all $a \in A_{<}$;  
\item \label{item: extendable 2}  $\supp(\w)=A_{\geq}$ and $P\w(a)\geq \w(a)$ for all $a \in A_{\geq}$; and
\item \label{item: extendable 3}  $\vb$ and $\w$ are contained in $K$.
\end{enumerate}
Note that extendable vectors always exist.
Indeed, there exist $\vb,\w \in \N^A$ that satisfy \eqref{item: extendable 1} and \eqref{item: extendable 2} by   the  Perron-Frobenius theorem (Lemma~\ref{lemma: Perron-Frobenius theorem}\eqref{item: Perron-Frobenius 3}-\eqref{item: Perron-Frobenius 3.5}). 
Since the total kernel $K$ is a subgroup of $\Z^A$ of finite index (by Lemma~\ref{lemma: the total kernel is a subgroup of finite index}\eqref{item: the total kernel is a subgroup of finite index}),
 we can assume that  $\vb,\w$ satisfy \eqref{item: extendable 3}  (by taking their finite multiple if necessary).

Let $\ee \in \N^A$ be a  vector satisfying the conclusion of  Lemma~\ref{l. local recurrence}\eqref{item: existence of idempotent elements that make every states locally recurrent}, 
 i.e. for any $\q \in Q$ the state $t_{\ee}(\q)$ is locally recurrent 
 (Note that $\ee$ exists by Lemma~\ref{l. local recurrence}\eqref{item: existence of idempotent elements that make every states locally recurrent}).
The following lemma provides  a method to construct diverse infinite walks.

\begin{lemma}\label{lemma: construction of infinite walk}
Let $\Net$ be a finite and locally irreducible abelian network.
Let $\vb,\w $ be extendable vectors of $\Net$,
let $\x.\q$ and $\x'.\q'$ be configurations of $\Net$, and let $u \in A^*$ be a word such that  $\x'.\q' \xlongrightarrow{u} \x.\q$.
\begin{enumerate}
\item \label{item: construction of infinite walk 1} 
If $|u|\geq \vb+\ee$,
then 
there exist $v \in A^*$  and  $\x_{-1},\x_{-2}, \ldots  \in \Z^A$    such that  $|v|=\vb$ and  the   infinite execution
\[\cdots  \xlongrightarrow{v}  \x_{-2}.\q \xlongrightarrow{v}   \x_{-1}.\q \xlongrightarrow{v} \x.\q\]
is legal.
\item \label{item: construction of infinite walk 2} 
If $|u|\geq \w+\ee$,
then 
there exist $w \in A^*$  and  $\x_1,\x_2, \ldots  \in \Z^A$  such that    $|w|=\w$ and the  infinite execution
\[\x.\q  \xlongrightarrow{w}  \x_{1}.\q \xlongrightarrow{w}   \x_{2}.\q \xlongrightarrow{w} \cdots\]
is legal.
\end{enumerate}
\end{lemma}
\begin{proof}
We present only the proof of \eqref{item: construction of infinite walk 1}, 
as the proof of \eqref{item: construction of infinite walk 2} is analogous.

Write
\[v:=u \setminus (|u|-\vb); \qquad \y.\p:=\pi_{|u|-|v|}(\x'.\q'); \qquad \x_{-i}:=\x+i(\y-\x) \quad (i\geq 0).  \]
Note that $|v|=\vb$ since $|u|\geq \vb$.
It suffices to show that $\x_{-(i+1)}.\q \xlongrightarrow{v}\x_{-i}.\q$ for all $i \geq 0$.

   Since $|u|-|v| \geq \ee$ and $\p=t_{|u|-|v|}(\q')$, 
   it follows from  Lemma~\ref{l. local recurrence}\eqref{item: existence of idempotent elements that make every states locally recurrent}
that  $\p$  is locally recurrent.
Since $\pi_{v}(\y.\p)=\pi_{u}(\x'.\q')=\x.\q$ and $\vb \in K$,
we then have $\q=t_{\vb}(\p)=\p \in \Loc(\Net)$ and $\y-\x=(I-P)\vb$.
Then  for all $i\geq 0$,
\[\pi_{v}(\x_{-(i+1)}.\q)= (\x_{-(i+1)}-(I-P)\vb).\q=\x_{-i}.\q. \]

Since $\x'.\q' \xlongrightarrow{u} \x.\q$ and $\pi_{|u|-|v|}(\x'.\q')=\y.\q$,  
 the removal lemma  (Lemma \ref{lemma: removal lemma}) implies that $\y.\q \xlongrightarrow{v} \x.\q$.
   Also note that
 $(\y-\x)(a) =((I-P)\vb)(a)\geq 0$ for all $a \in \supp(\vb)$ by \eqref{item: extendable 1}.
It then follows from Lemma~\ref{o. to legal properties}\eqref{o. to contagious}
that 
  \[\x_{-(i+1)}.\q=(\y+i(\y-\x)).\q \xlongrightarrow{v} (\x+i(\y-\x)).\q  =\x_{-i}.\q,\] 
  for all $i\geq 0$.
 This completes the proof.
\end{proof}

As a consequence of Lemma~\ref{lemma: construction of infinite walk}, we show that recurrent components always exist.

\begin{corollary}\label{corollary: the set of recurrent classes is nonempty}
Let $\Net$ be a finite and locally irreducible abelian network.
Then the set $\Lrec(\Net)$ is nonempty.
\end{corollary}
\begin{proof}
Let $\q' \in Q$ and let  $\x' := \max (\vb,\w)+\ee$, where $\vb,\w$ are extendable vectors of $\Net$.
Let $u$ be a word such that $|u|=\x'$.
Write $\x.\q:=\pi_{u}(\x'.\q')$,
and note that
 $\x'.\q' \xlongrightarrow{u} \x.\q$ since $|u|=\x'$.

 Since $\vb,\w$ are extendable vectors, it follows from 
 Lemma~\ref{lemma: construction of infinite walk} that 
there exist $v,w\in A^*$ and vectors $\x_i'$ ($i \in \Z \setminus \{0\}$) such that $|v|=\vv$, $|w|=\w$, and the following infinite execution
\[ \cdots \xlongrightarrow{v} \x'_{-1}.\q' \xlongrightarrow{v} \x'.\q'  \xlongrightarrow{w} \x'_1.\q' \xlongrightarrow{w} \cdots  \]
is legal.
It follows from the construction that the infinite execution above is a diverse infinite walk in $\overline{\x.\q}$.
Hence $\overline{\x.\q}$ is a recurrent component, which shows that $\Lrec(\Net)$ is nonempty.
\end{proof}

A \emph{strongly diverse} infinite walk in $\Net$ is 
a sequence of legal executions
\[\cdots \xlongrightarrow{v} \x_{-2}.\q\xlongrightarrow{v} \x_{-1}.\q   \xlongrightarrow{v} \x_0.\q_0 \xlongrightarrow{w} \x_{1}.\q \xlongrightarrow{w} \x_{2}.\q \xlongrightarrow{w} \cdots \]
such that
\begin{enumerate}
\item The state $\q$ is locally recurrent;
\item $\supp(|v|)=A_{<}$ and $P|v|(a)\leq |v|(a)$ for all $a \in A_<$; and
\item $\supp(|w|)=A_{\geq}$ and $P|w|(a)\geq |w|(a)$ for all $a \in A_{\geq}$.
\end{enumerate}

\begin{lemma}\label{lemma: recurrent classes has a strongly diverse infinite walk}
Let $\Net$ be a finite and locally irreducible abelian network.
A component of the trajectory digraph is a recurrent component if and only if it contains a strongly diverse infinite walk.
\end{lemma}
\begin{proof} 
It suffices to prove the only if direction, as the if direction 
follows from the fact that a strongly diverse infinite walk is also diverse.

Let 
$\cdots \xlongrightarrow{a_{i}} \x_{i}.\q_{i} \xlongrightarrow{a_{i+1}} \cdots $ $(i \in I)$ 
be a diverse infinite walk in the recurrent component.
Since the walk is diverse, there exist $j \in I$ and $k\geq 1$
such that $u:=a_{j+1}\cdots a_{j+k}$ satisfies $|u|\geq \max(\vv,\w)+\ee$, where $\vb,\w$ are extendable vectors of $\Net$.

Write $\x'.\q':=\x_j.\q_j$ and $\x.\q:=\x_{j+k}.\q_{j+k}$, and note that 
$\x'.\q' \xlongrightarrow{u}\x.\q$. 
Also note that 
we have $\q=t_{u}\q_j$ is locally recurrent by Lemma~\ref{l. local recurrence}\eqref{item: existence of idempotent elements that make every states locally recurrent} since $|u| \geq  \ee$.

By Lemma~\ref{lemma: construction of infinite walk},
there exist $v,w\in A^*$ and  $\x_i$ ($i \in \Z \setminus \{0\}$) such that $|v|=\vv$,  $|w|=\w$, and  the following infinite execution
\[\cdots \xlongrightarrow{v} \x_{-2}.\q\xlongrightarrow{v} \x_{-1}.\q   \xlongrightarrow{v} \x.\q \xlongrightarrow{w} \x_{1}.\q \xlongrightarrow{w} \x_{2}.\q \xlongrightarrow{w} \cdots \]
is legal.
This infinite execution  is a strongly diverse infinite walk in the given recurrent component, which proves the claim.
\end{proof}

We now present the proof of Proposition~\ref{proposition: weak relation implies strong relation for recurrent class}. 

\begin{proof}[Proof of Proposition~\ref{proposition: weak relation implies strong relation for recurrent class}]
By Lemma~\ref{lemma: recurrent classes has a strongly diverse infinite walk} and the transitivity of $\dashrlarrow$ and $\rlarrow$,
 we can without loss of generality assume that $\x_i.\q_i$ is contained in a strongly diverse infinite walk for $i \in \{1,2\}$ (by taking another configuration in the recurrent component if necessary).
 In particular, each $\q_i$ is a locally recurrent state.

For $i \in \{1,2\}$ let $\vb_i,\w_i \in \N^A$ and  $\x_3.\q_3$ be configurations such that
$\supp(\vb_i)=A_{<}$, $\supp(\w_i)=A_{\geq}$, and 
$\x_i.\q_i \xdashrightarrow{\vb_i+\w_i} \x_3.\q_3$.
(Note that $\vv_i,\w_i$, and $\x_3.\q_3$ exist because $\x_1.\q_1 \dashrlarrow \x_2.\q_2$.)
By the abelian property~(Lemma~\ref{l. abelian enumerate}\eqref{l. abelian property}) and Lemma~\ref{lemma: the monoid $N^A$ acts invertibly on locally recurrent configurations}\eqref{item: monoid acts invertible on locally recurrent states 2},
there exist (unique) $\x_i'.\q_i'$ with $\q_i' \in \Loc(\Net)$   ($i\in \{1,2,3\}$)  such that this diagram commutes. 
\begin{center}
\begin{tikzcd}
 & \x_1.\q_1 \arrow[rd, "\w_1",dashed] & &\\
  \x_1'.\q_1' \arrow[ru, "\vb_2",dashed] \arrow[dashed, rd , "\w_1"] & &\phantom{\x_1.\q_1} \arrow[rd, "\vb_1", dashed]& \\
  &  {\x_3'.\q_3'} \arrow[ru, "\vb_2", dashed]  \arrow[rd, "\vb_1", dashed] & &  \x_3.\q_3\\
  \x_2'.\q_2' \arrow[ru, "\w_2", dashed]  \arrow[rd, "\vb_1", dashed] & &  \phantom{\x_1.\q_1}  \arrow[ru, "\vb_2", dashed]& \\
  & \x_2.\q_2  \arrow[ru, "\w_2", dashed]& & 
\end{tikzcd}.
\end{center}

For $i\in \{1,2\}$, there exist $v_i,v_i',w_i' \in A^*$, $\y_i,\y_i' \in \Z^A$, and $\p_i,\p_i' \in \Loc(\Net)$ such that (details are given after Diagram~\eqref{diagram: diagram 2 for lemma weak relation implies strong relation}):
\begin{equation}\label{diagram: diagram 2 for lemma weak relation implies strong relation}
\begin{tikzcd}
 \y_1.\p_1 \arrow[rr,"v_1'",red ]  \arrow[rd,"v_1'\setminus \vb_2",magenta, dashed ] & & \x_1.\q_1 \arrow[rd, "\w_1",dashed] \arrow[green!80!blue, rr, "w_1'"] & &  \y_1'.\p_1'\\
&  \x_1'.\q_1' \arrow[ru, "v_2",cyan] \arrow[dashed, rd , "\w_1"] & &\phantom{\x_1.\q_1} \arrow[rd, "\vb_1", dashed] \arrow[ru, "w_1'\setminus \w_1",blue]& \\
&  &  {\x_3'.\q_3'} \arrow[ru, "v_2", yellow!70!red]  \arrow[rd, "v_1", yellow!70!red] & &  \x_3.\q_3\\
&  \x_2'.\q_2' \arrow[ru, "\w_2", dashed]  \arrow[rd, "v_1", cyan] & &  \phantom{\x_1.\q_1}  \arrow[ru, "\vb_2", dashed] \arrow[rd, "w_2'\setminus \w_2",blue]& \\
 \y_2.\p_2 \arrow[rr,"v_2'",red ] \arrow[ru,"v_2'\setminus \vb_1",magenta , dashed]  &  &\x_2.\q_2  \arrow[ru, "\w_2", dashed] \arrow[rr,"w_2'",green!80!blue ]& & \y_2'.\p_2' 
\end{tikzcd}.
\end{equation}

Indeed,
let $i,j$ be distinct elements in $\{1,2\}$.
By the assumption that 
 $\x_i.\q_i$  is contained in  a strongly diverse infinite walk,
we get the solid  arrow  ${\color{red}{\xlongrightarrow{v_i'}}}$,
where  $v_i'$ is a word such that
 $|v_i'|\geq \vb_{j}$.
Similarly, we get the solid  arrow  
${\color{green!80!blue}{\xlongrightarrow{w_i'}}}$,
where $w_i'$ is a word such that $|w_i'|\geq \w_i$.
By the  removal lemma (Lemma~\ref{lemma: removal lemma}) and the assumption that 
$|w_i'|\geq \w_i$,
we get the solid  arrow ${\color{blue}\xlongrightarrow{w_i'\setminus \w_i}}$ in Diagram~\eqref{diagram: diagram 2 for lemma weak relation implies strong relation}.
By the abelian property,   the assumption
that 
$|v_i'|\geq \vb_j$, and Lemma~\ref{lemma: the monoid $N^A$ acts invertibly on locally recurrent configurations}\eqref{item: monoid acts invertible on locally recurrent states 2}, we get the dashed  arrow $\color{magenta} \xdashrightarrow{v_i' \setminus \vb_j}$ in Diagram~\eqref{diagram: diagram 2 for lemma weak relation implies strong relation}.
Write $v_j:=v_i' \setminus (|v_i'|-\vb_j)$.
Note  that $|v_j|=\vb_j$ because $|v_i'|\geq \vb_j$, and in particular $\supp(|v_j|)=A_{<}$.
By the removal lemma, we get the solid (cyan) arrow 
${\color{cyan} \xlongrightarrow{v_j}}$ in Diagram~\eqref{diagram: diagram 2 for lemma weak relation implies strong relation}.
By the removal lemma and the fact that $\supp(|v_j|)=A_{<}$ and $\supp(\w_i)=A_{\geq}$  are disjoint sets,
we get the solid (yellow) arrow ${\color{yellow!70!red} \xlongrightarrow{v_j}}$ in Diagram~\eqref{diagram: diagram 2 for lemma weak relation implies strong relation}.

The conclusion of the proposition now follows from Diagram~\eqref{diagram: diagram 2 for lemma weak relation implies strong relation} and the transitivity  of the legal relation~(Diagram~\eqref{diagram: diagram transitivity}).
\end{proof}

We now build towards the proof of Proposition~\ref{proposition: recurrent component combinatorial description}.
We start by checking \eqref{item: recurrent combinatorial 1} implies \eqref{item: recurrent combinatorial 2} for subcritical and supercritical case.

\begin{lemma}\label{lemma: recurrent component subcritical supercritical}
Let $\Net$ be a finite, locally irreducible, and strongly connected subcritical (resp. supercritical) network.
Then any strongly diverse infinite walk in $\Net$ 
is a diverse backward (resp. forward) infinite path.
\end{lemma}
\begin{proof}
We present only the proof of the subcritical case, as the proof of the supercritical case is analogous.

Since $\Net$ is subcritical, a strongly diverse infinite walk
of $\Net$ is of the form
\[\cdots \xlongrightarrow{v} \x_{-2}.\q\xlongrightarrow{v} \x_{-1}.\q   \xlongrightarrow{v} \x.\q, \]
where $v$ is a word such that $\supp(|v|)=A$.
Note that the infinite execution above is a diverse backward infinite walk.
Hence it suffices to show that  this infinite walk is a path.

Suppose to the contrary that this infinite walk is not a path.
Then there exists a configuration $\x'.\q'$ and a nonempty word $w$ such that the  execution 
$\x'.\q' \xlongrightarrow{w} \x'.\q'$
is legal and is a subsequence of the  infinite walk above.
By Lemma~\ref{l. N locally irreducible}, we then have:
\[ P|w|=\x'+|w|-\x'=|w|.  \]
 The Perron-Frobenius theorem (Lemma~\ref{lemma: Perron-Frobenius theorem}\eqref{item: Perron-Frobenius spectral from inequality})
then implies that   $\lambda(P)=1$, contradicting the assumption that $\Net$ is subcritical.
This proves the claim.
\end{proof}

We will  use the following version of Dickson's lemma to check \eqref{item: recurrent combinatorial 1} implies \eqref{item: recurrent combinatorial 2} for critical case.
A sequence of vectors 
$\x_1,\x_2,\ldots $  in $\Z^A$  has a \emph{lower bound} if there exists $\x \in \Z^A$ such that $\x_i\geq \x$ for all $i\geq 1$.

\begin{lemma}[{\cite[Dickson's~lemma]{Dic13}}]\label{lemma: Dickson's lemma}
Let $\x_1,\x_2,\ldots $ be a sequence of vectors in $\Z^A$ that has a lower bound.
Then there exist integers $j,k\geq 1$ such that $\x_j \leq \x_{j+k}$. \qed 
\end{lemma}

Denote by $\nol $ the vector in $\Z^A$ with all entries being equal to $0$.

\begin{lemma}\label{lemma: recurrent component critical}
Let $\Net$ be a finite, locally irreducible, and strongly connected critical  network.
Then any  strongly diverse infinite walk in $\Net$ is a diverse cycle.
\end{lemma}
\begin{proof}
Since $\Net$ is critical, a strongly diverse infinite walk
of $\Net$ is of the form
\[ \x.\q \xlongrightarrow{w} \x_{1}.\q \xlongrightarrow{w} \x_{2}.\q \xlongrightarrow{w} \cdots \]
where $w$ is a word such that $\supp(|w|)=A$.
Hence it suffices to show $\x_1=\x$.

By Lemma~\ref{o. to legal properties}\eqref{o. nonnegativity},
the sequence $\x_0,\x_1,\ldots$ is lower bounded 
by the vector $\x \in \Z^A$ given by $\x(a):= \min \{\x_0(a),0\}$ ($a \in A$).
By Dickson's lemma,
there exist integers $j,k \geq 1$ such that $\x_j\leq \x_{j+k}$.

 Since  $\x_{j}.\q_j \xlongrightarrow{w^k} \x_{j+k}.\q_{j+k}$ and $\q_{j+k}=\q_j$, we have by Lemma~\ref{l. N locally irreducible} that
\[  (P-I)|w|=\frac{\x_{j+k}-\x_{j} }{k}\geq \nol. \]
Since $\Net$ is strongly connected and critical, it follows from the  Perron-Frobenius theorem (Lemma~\ref{lemma: Perron-Frobenius theorem}\eqref{item: Perron-Frobenius spectral from inequality}) that $(P-I)|w|=\nol$.
This implies that
 \[ \x_1-\x= (P-I)|w|=\nol, \]
as desired.
\end{proof}

We now present the proof of Proposition~\ref{proposition: recurrent component combinatorial description}.

\begin{proof}[Proof of Proposition~\ref{proposition: recurrent component combinatorial description}]

\eqref{item: recurrent combinatorial 1} implies \eqref{item: recurrent combinatorial 2}: This follows from Lemma~\ref{lemma: recurrent classes has a strongly diverse infinite walk}, Lemma~\ref{lemma: recurrent component subcritical supercritical}, and   Lemma~\ref{lemma: recurrent component critical}. 

\eqref{item: recurrent combinatorial 2} implies \eqref{item: recurrent combinatorial 3} is straightforward.

\eqref{item: recurrent combinatorial 3} implies \eqref{item: recurrent combinatorial 1}: 
We present only the proof of the subcritical case, as the proof of the other two cases are analogous.

By \eqref{item: recurrent combinatorial 3}, there exists a diverse infinite path in $\Net$ of the form
 \[\cdots \xlongrightarrow{v_3} \x_{-2}.\q\xlongrightarrow{v_2} \x_{-1}.\q   \xlongrightarrow{v_1} \x.\q, \]
where $v_1,v_2\ldots$ are words such that $\supp(|v_i|)=A$.
Note that $\x_i\neq \x_j$ for distinct $i$ and $j$ since the infinite walk above is a path.

Since $\x_{-(i+1)}.\q \xlongrightarrow{v_{i+1}} \x_{-i}.\q $    and $\supp(|v_{i+1}|)=A$  for any $i\geq 0$,
we have 
by Lemma~\ref{o. to legal properties}\eqref{o. nonnegativity}
that $\x_{-i}$ is a nonnegative vector for any $i \geq 0$. 
By Dickson's lemma,
there exist integers $j,k \geq 1$ such that $\x_{-j}\leq \x_{-(j+k)}$.

Write $v:=v_{k}v_{k-1}\ldots v_{j+1}$.
Now note that
\[  (I-P) |v|=\x_{-(j+k)}-\x_{-j}\geq \nol, \]
where the first equality is due to Lemma~\ref{l. N locally irreducible}.
Also note that $(I-P)|v|=\x_{-(j+k)}-\x_{-j}$ is not equal to $\nol$ since $\x_{-(j+k)}\neq \x_{-j}$ by assumption.
Since $\Net$ is strongly connected, it then follows from the Perron-Frobenius theorem (Lemma~\ref{lemma: Perron-Frobenius theorem}\eqref{item: Perron-Frobenius spectral from inequality}) that $\lambda(P)$ is strictly less than 1, as desired.
\end{proof}

 \section{Construction of the torsion group}
\label{subsection: construction of torsion group for all abelian networks}
In this section we define the torsion group for any  abelian network by building on results from \S\ref{subsection: recurrent components}.
The reader can use the networks from  Example~\ref{example: toppling networks}  to develop intuition  when reading this section.

\begin{definition}[Shift monoid]\label{definition: monoid action on recurrent classes}
Let $\Net$ be an abelian network.
The monoid $\N^A$ acts on $\Lrec(\Net)$ by 
\begin{align*}
\phi: \N^A &\to \End(\Lrec(\Net))\\
\phi(\n)(\overline{\x.\q}) &:=\overline{(\x+\n).\q}. 
\end{align*}
The \emph{shift monoid} is the monoid $\Mon(\Net):=\phi(\N^A)$.
\end{definition} 
It follows from Lemma~\ref{o. to legal properties}\eqref{o. to contagious}
that $\overline{(\x+\n).\q}$  does not depend on the choice of $\x.\q$, and is a recurrent component if $\overline{\x.\q}$ is recurrent.
Hence the monoid action in Definition~\ref{definition: monoid action on recurrent classes} is well-defined.

 Note that  $\Mon(\Net)$ is generated by the set $\{ \phi(|a|) \mid a \in A \}$, and hence is a finitely generated commutative monoid.
We denote by $\Grt(\Net)$ the Grothendieck group~(see \S\ref{subsection: commutative monoid that acts injectively}) of $\Mon$.
We remark  that   $\Mon(\Net)$, $\Grt(\Net)$,  and $\Lrec(\Net)$ can be infinite;
see  Example~\ref{example: torsion group for different abelian networks}\eqref{item: example torsion group of critical toppling network}.

\begin{definition}[Torsion group]\label{definition: torsion group}
Let $\Net$ be an abelian network.
The \emph{torsion group} of $\Net$ is
\[\Tor(\Net):=\tau(\Grt(\Net)), \]
the torsion subgroup of the Grothendieck group of $\Mon(\Net)$.
\end{definition}

\begin{definition}[Invertible recurrent component]\label{definition: invertible recurrent class}
Let $\Net$ be an abelian network.
A recurrent component $\overline{\x.\q}$ is \emph{invertible} if, for any $g \in \Tor(\Net)$ and any $\n,\n' \in \N^A$ such that $g=\overline{(\phi(\n),\phi(\n'))}$, there exists $\overline{\x'.\q'} \in \Lrec(\Net)$ such that 
\[\overline{(\x+\n).\q}=\overline{(\x'+\n').\q'}. \]
We denote by  $\Lrec(\Net)^\times$   the set of invertible recurrent components of $\Net$.
\end{definition}
Note that not all recurrent components are invertible; see Example~\ref{example: torsion group for different abelian networks}\eqref{item: example torsion group of critical toppling network}.


Assume throughout the rest of this section that   $\Net$  is a finite and locally irreducible abelian network, unless stated otherwise.

\begin{definition}[Action of $\Tor(\Net)$ on $\Lrec(\Net)^\times$]\label{definition: action of the torsion group on invertible recurrent classes}
Let $\Net$ be a finite and locally irreducible abelian network.
The group $\Tor(\Net)$ acts on $\Lrec(\Net)^\times$ by
\begin{align*}
\Tor(\Net) \times \Lrec(\Net)^\times &\to \Lrec(\Net)^\times\\
(g, \overline{\x.\q}) &\mapsto \overline{\x'.\q'},
\end{align*}
where $\overline{\x'.\q'}$ is as in Definition~\ref{definition: invertible recurrent class}.
\end{definition}
We will show later in  Lemma~\ref{lemma: the action of Mon is free, injective, and with kernel described}\eqref{item: action of the torsion group is well defined} that this group action is well-defined.
 Note that the action of $\Tor(\Net)$ is not defined for recurrent components that are not invertible.

We now state the main result of this section.
Recall the definition of the total kernel $K$ (Definition~\ref{definition: total kernel}) and the production matrix $P$ (Definition~\ref{definition: production matrix}).
Recall that the action of a  monoid $\Mon$  on a set $X$ is \emph{free} if,
for any $x \in X$ and $m,m' \in \Mon$, we have
$mx=m'x$ implies that $m=m'$. 
The action of $\Mon$ on $X$ is \emph{transitive} if $X$ is nonempty and for any $x,x' \in X$ there exists $m \in \Mon$ such that $x'=mx$.

\begin{theorem}\label{theorem: construction of torsion group for general abelian networks}
Let $\Net$ be a finite and locally irreducible abelian network.
Then
\begin{enumerate}
\item \label{item: theorem torsion group 1}   $\Lrec(\Net)^\times$ is nonempty. 
\item \label{item: theorem torsion group 2}  $\Tor(\Net)$ is a finite abelian group that acts freely on  $\Lrec(\Net)^\times$.
\item\label{item: theorem grothendieck group is equal to ZA/(I-P)K}  The 
map $\phi: \N^A \to \End(\Rec(\Net))$ induces  an isomorphism of abelian groups  
\[
 \Grt(\Net) \simeq \Z^A/(I-P)K. 
 \]
\end{enumerate}
\end{theorem}

We remark that Theorem~\ref{intro theorem: construction of torsion group for general abelian networks}, stated in the introduction, 
is a direct corollary of Theorem~\ref{theorem: construction of torsion group for general abelian networks}.

Note that the action of the torsion group on $\Lrec(\Net)^\times$ is in general not  transitive;
see Example~\ref{example: torsion group for different abelian networks}\eqref{item: example torsion group of critical toppling network}.
The torsion group is a generalization of  the  critical group for halting networks as defined in \cite{BL16c}.
We will discuss this in more details in \S\ref{subsection: torsion group of subcritical networks}.

\begin{example}\label{example: torsion group for different abelian networks}
Consider the toppling network $\Net_t$~(Example~\ref{example: toppling networks}) on the bidirected cycle $C_3$ with threshold
$t_{v_0}=t_{v_1}=t_{v_2}=:t$.
\begin{enumerate}[wide, labelwidth=!, labelindent=10pt]
\item \label{item: example torsion group of subcritical toppling network}
If $t=3$ (note that $\Net_3$ is subcritical), then
\begin{align*}
\Tor(\Net_3)=\Z^V \Bigg{/} \left \langle  \begin{bmatrix} 3\\-1\\-1 \end{bmatrix}, \begin{bmatrix} -1\\3\\-1 \end{bmatrix}, \begin{bmatrix} -1\\-1\\3 \end{bmatrix} \right \rangle_{\Z} =\Z_4 \oplus \Z_4.
\end{align*}
$\Net_3$ has sixteen recurrent components, namely all permutations of these five:

\begin{align*}
 \left\{\overline{\x.\q} \ \Bigg{|}\ \x=\begin{bmatrix}
0 \\ 0 \\ 0
\end{bmatrix}, \, \q\in \left\{ \begin{bmatrix}
0 \\ 1 \\ 2
\end{bmatrix}, 
\begin{bmatrix}
1 \\ 1 \\ 2
\end{bmatrix},
\begin{bmatrix}
0 \\ 2 \\ 2
\end{bmatrix},  \begin{bmatrix}
1 \\ 2 \\ 2
\end{bmatrix},  
\begin{bmatrix}
2 \\ 2 \\ 2
\end{bmatrix}  
 \right\} \right\}.
\end{align*}
All sixteen recurrent components of $\Net_3$ are invertible, and the action of $\Tor(\Net_3)$ on $\Lrec(\Net_3)^\times$ is free and transitive.

\item \label{item: example torsion group of critical toppling network}
If $t=2$ (note that $\Net_2$ is critical), then:
\begin{align*}
\Tor(\Net_2)=&\tau \left( \Z^V \Bigg{/} \left \langle  \begin{bmatrix} 2\\-1\\-1 \end{bmatrix}, \begin{bmatrix} -1\\2\\-1 \end{bmatrix}, \begin{bmatrix} -1\\-1\\2 \end{bmatrix} \right \rangle_{\Z} \right)\\
=&\tau(\Z_3 \oplus \Z)=\Z_3.
\end{align*}
The recurrent components of $\Net_2$ are given by 
\[\Lrec(\Net_2)=  \bigsqcup_{m\geq 3} \Lrec(\Net_2,m), \]
where 
\begin{align*}
  \Lrec(\Net_2,3)=\left\{  \overline{\x.\q} \ \Bigg{|} \ \q=\begin{bmatrix}
0 \\ 0 \\0 
\end{bmatrix}, \ \x\in \left \{ \begin{bmatrix}
0 \\ 1 \\2 
\end{bmatrix}, \begin{bmatrix}
0 \\ 2 \\1 
\end{bmatrix}\right \} \right \},
\end{align*}
and, for $m\geq 4$,
\begin{align*}
\Lrec(\Net_2,m)=
\left\{  \overline{\x.\q} \ \Bigg{|} \ \q=\begin{bmatrix}
0 \\ 0 \\0 
\end{bmatrix}, \, \x\in \left \{ \begin{bmatrix}
0 \\ 1 \\m-1 
\end{bmatrix}, \begin{bmatrix}
0 \\ 2 \\m-2 
\end{bmatrix}, \begin{bmatrix}
1 \\ 1 \\m-2 
\end{bmatrix}\right \} \right \}.
\end{align*}
The invertible recurrent components of $\Net_2$ are given by:
\[ \Lrec(\Net_2)^\times= \bigsqcup_{m\geq 4} \Lrec(\Net_2,m).\]
In particular, the two   recurrent components in $\Lrec(\Net_2,3)$ are not invertible, and hence the torsion group  does not act on them.

Note that the action of $\Tor(\Net_2)$ on $\Lrec(\Net_2)^\times$ is free but not transitive, as  each $\Lrec(\Net_2,m)$ for $m\geq 4$ is an orbit of this action.

\item \label{item: example torsion group of supercritical toppling network}
If $t=1$ (note that $\Net_1$ is supercritical), then
\begin{align*}
\Tor(\Net_1)=\Z^V \Bigg{/} \left \langle  \begin{bmatrix} 1\\-1\\-1 \end{bmatrix}, \begin{bmatrix} -1\\1\\-1 \end{bmatrix}, \begin{bmatrix} -1\\-1\\1 \end{bmatrix} \right \rangle_{\Z} =\Z_2 \oplus \Z_2.
\end{align*}
$\Net_1$ has four recurrent components:
\begin{align*}
\left\{  \overline{\x.\q} \ \Bigg{|} \ \q=\begin{bmatrix}
0 \\ 0 \\0 
\end{bmatrix}, \, \x\in \left \{ \begin{bmatrix}
1 \\ 0 \\ 0
\end{bmatrix}, \begin{bmatrix}
0 \\ 1 \\0 
\end{bmatrix}, \begin{bmatrix}
0 \\ 0 \\1 
\end{bmatrix}, \begin{bmatrix}
1 \\ 1 \\ 1
\end{bmatrix} \right \} \right \}.
\end{align*}
All four recurrent components of $\Net_1$ are invertible, and the action of $\Tor(\Net_1)$ on $\Lrec(\Net_1)^\times$ is free and transitive. \qedhere
\end{enumerate}
\end{example}

Our strategy of proving  Theorem~\ref{theorem: construction of torsion group for general abelian networks} is to apply  Proposition~\ref{proposition: abstract construction of torsion groups} to the setting of Theorem~\ref{theorem: construction of torsion group for general abelian networks}.
In order to do so, we need to check that the action of $\Mon(\Net)$ on $\Lrec(\Net)$ satisfies the conditions in Proposition~\ref{proposition: abstract construction of torsion groups}, and that requires the following technical lemma.

Recall the definition of injective action from Definition~\ref{definition: injective action}.
\begin{lemma}\label{lemma: the action of Mon is free, injective, and with kernel described}
Let $\Net$ be a finite and locally irreducible abelian network.
Then 
\begin{enumerate}
\item \label{item: the kernel of action of Mon described} For any $\n,\n' \in \N^A$, we have 
 $\phi(\n)=\phi(\n')$ if and only if $\n-\n' \in (I-P)K$;
\item \label{item: the action of Mon is free and injective} The action of  $\Mon(\Net)$ on $\Lrec(\Net)$ is free and injective; and
 \item \label{item: action of the torsion group is well defined}
 The action of $\Tor(\Net)$ on $\Lrec(\Net)^\times$ in Definition~\ref{definition: action of the torsion group on invertible recurrent classes} is well defined.
\end{enumerate}
\end{lemma}

\begin{proof}
Let $\x.\q$ be any configuration such that  $\overline{\x.\q}$ is recurrent.
For any $\n,\n' \in \N^A$,
\begin{align}\label{equation: equation in the lemma the action is free and kernel is described}
\begin{split}
&(\x+\n).\q  \rlarrow (\x+\n').\q\\
&\Longleftrightarrow \quad  (\x+\n).\q  \dashrlarrow (\x+\n').\q \qquad \text{(by Proposition~\ref{proposition: weak relation implies strong relation for recurrent class})}\\
&\Longleftrightarrow \quad  \n-\n' \in (I-P)K \qquad \text{(by Lemma~\ref{l. N locally irreducible})}.
\end{split}
\end{align}
Since the choice of $\x.\q$ is arbitrary, we then conclude that 
$\phi(\n)=\phi(\n')$ if and only if $\n-\n' \in (I-P)K$.
This proves part~\eqref{item: the kernel of action of Mon described}.

Let  $\x.\q$ and $\x'.\q'$ be any configurations such that  $\overline{\x.\q}$ and $\overline{\x'.\q'}$ are recurrent.
For any $\n \in \N^A$, 
\begin{align*}
& (\x+\n).\q \rlarrow  (\x'+\n).\q'  \\
  &\Longrightarrow \quad (\x+\n).\q \dashrlarrow  (\x'+\n).\q' \\
&\Longrightarrow \quad  \x.\q \dashrlarrow  \x'.\q'  \qquad \text{(by Lemma~\ref{o. to legal properties}\eqref{item: weak arrow if and only if})}\\
&\Longrightarrow \quad  \x.\q \rlarrow  \x'.\q' \qquad \text{(by Proposition~\ref{proposition: weak relation implies strong relation for recurrent class}).}
\end{align*}
Hence the action of $\Mon(\Net)$ on $\Lrec(\Net)$ is injective.
For any $\n,\n' \in \N^A$,
\begin{align*}
& (\x+\n).\q \rlarrow  (\x+\n').\q  \\
&\Longrightarrow \quad  \n-\n' \in (I-P)K \qquad \text{(by equation~\eqref{equation: equation in the lemma the action is free and kernel is described})}\\
&\Longrightarrow \quad  (\x'+\n).\q' \rlarrow  (\x'+\n').\q'   \qquad \text{(by equation~\eqref{equation: equation in the lemma the action is free and kernel is described})}.
\end{align*}
Since the choice of  $\x'.\q'$ is arbitrary, 
we then conclude that
$\phi(\n)(\x.\q)=\phi(\n')(\x.\q)$
implies that  $\phi(\n)=\phi(\n')$.
Hence the 
the action of $\Mon(\Net)$ on $\Lrec(\Net)$ is free.
This proves part~\eqref{item: the action of Mon is free and injective}.

Since $\Mon(\Net)$ acts on $\Lrec(\Net)$ injectively by 
 part~\eqref{item: the action of Mon is free and injective},
 it follows from Lemma~\ref{lemma: H-invertible elements let you define a group action}
 that the group action in Definition~\ref{definition: action of the torsion group on invertible recurrent classes} is well-defined.
 This proves part~\eqref{item: action of the torsion group is well defined}.
\end{proof}

We now present the proof of Theorem~\ref{theorem: construction of torsion group for general abelian networks}. 

\begin{proof}[Proof of Theorem~\ref{theorem: construction of torsion group for general abelian networks}]
Note that  action of $\Mon(\Net)$ on $\Lrec(\Net)$ is free and injective 
(by Lemma~\ref{lemma: the action of Mon is free, injective, and with kernel described}\eqref{item: the action of Mon is free and injective}), and that $\Lrec(\Net)$ is a nonempty set (by Corollary~\ref{corollary: the set of recurrent classes is nonempty}).
Part \eqref{item: theorem torsion group 1} and \eqref{item: theorem torsion group 2} now follow directly from Proposition~\ref{proposition: abstract construction of torsion groups}.

For part  \eqref{item: theorem grothendieck group is equal to ZA/(I-P)K}, note that $\Z^A$ is the Grothendieck group of $\N^A$ and $\Grt(\Net)$ is the Grothendieck group of $\Mon(\Net)$.
Also note that $\phi: \N^A \to \Mon(\Net)$ is a surjective monoid homomorphism.
By the universal  property of the Grothendieck group,
the map $\phi$ induces  a  surjective group homomorphism ${\phi}:\Z^A\to \Grt(\Net)$.
Also note that  
\[\ker(\phi)=\{ \z \in \Z^A \mid \phi(\z^+)=\phi(\z^-)\},\]
 where $\z^+$ and $\z^-$ are the positive part and the negative part of $\z$, respectively.  
The claim now follows from Lemma~\ref{lemma: the action of Mon is free, injective, and with kernel described}\eqref{item: the kernel of action of Mon described}.
\end{proof}


\section[Halting case]{Relations to the critical group in the halting case}\label{subsection: torsion group of subcritical networks}
 Consider  a finite, locally irreducible, and subcritical abelian network $\Subnet$.
 In this section we show that the torsion group   of $\Subnet$ is isomorphic to the critical group defined in~\cite{BL16c}.

We start by quoting a useful theorem from~\cite{BL16c}.
A configuration $\x.\q$ is \emph{stable} if $\x(a)\leq 0$ for all $a \in A$.
A configuration $\x.\q$ halts if there exists a stable configuration $\x'.\q'$ such that $\x.\q \longrightarrow \x'.\q'$.

\begin{theorem}[{\cite[Theorem~5.6]{BL16b}}]
\label{theorem: subcritical networks halts on all inputs}
Let $\Subnet$ be a finite, locally irreducible, and subcritical abelian network.
Then  every configuration $\x.\q$
in $\Subnet$ is a halting configuration. \qed 
\end{theorem}

\begin{lemma}\label{lemma: stabilization is well-defined for subcritical networks}
Let $\Subnet$ be a finite, locally irreducible, and subcritical abelian network. 
Then every component of the trajectory digraph 
 contains a unique stable configuration.
\end{lemma}
\begin{proof}
Let $\C$ be an arbitrary component of the trajectory digraph.
By Theorem~\ref{theorem: subcritical networks halts on all inputs},  there exists a stable configuration $\x.\q$ in   $\C$.

We now prove that $\x.\q$ is unique.
Let $\x'.\q'$ be another stable configuration in $\C$.
Then there exists $\y.\p$  such that $\x.\q \longrightarrow \y.\p$ and $\x'.\q' \longrightarrow \y.\p$.
Since $\x(a)\leq 0$ for all $a \in A$, it is necessary that $\x.\q=\y.\p$.
By symmetry $\x'.\q'=\y.\p$, and hence $\x.\q=\x'.\q'$. 
\end{proof}

We define
the  \emph{stabilization} $\st(\C)$ of a component $\C$ to be the unique stable configuration in $\C$.
Let $\Qcr$ be the set:
\[\Qcr:=\{\C \mid  \st(\C)=\nol.\q \text{ for some }  \q \in Q \}.\]
The set $\Qcr$ is in one-to-one correspondence with the state space $Q$ via $\overline{\nol.\q} \mapsto \q$, and in particular $\Qcr$ is finite.

The monoid $\N^A$ acts on $\Qcr$ by:
\begin{align*}
\Phi: \N^A &\to \End(\Qcr)\\
 \Phi(\n)(\overline{\nol.\q}) &:=\overline{\n.\q}.
\end{align*}
Note that $\st(\overline{\n.\q})=\nol.\q'$ for some $\q'\in Q$ since $\n\geq \nol$, and hence $\overline{\n.\q}$ is contained in $\Qcr$.

The \emph{global monoid} in the sense of~\cite{BL16c} is the monoid $\Msf(\Subnet):=\Phi(\N^A)$.
 Note that  $\Msf(\Subnet)$ is a finite commutative monoid as $\Qcr$ is finite.

Let $e \in \Msf(\Subnet)$ be the minimal idempotent of $\Msf(\Subnet)$~(see Definition~\ref{definition: minimal idempotent}).
The \emph{critical group} of $\Net$ in the sense of~\cite{BL16c}  is the group $e\Msf(\Subnet)$.

\begin{definition}[Recurrent state]\label{definition: recurrent state}
Let $\Subnet$ be a finite and locally irreducible subcritical network.
An element of $\Qcr$ is \emph{recurrent} in the sense of~\cite{BL16c}
if it is contained in $e\Qcr$.
A state $\q \in Q$ is \emph{recurrent} if its corresponding component in $\Qcr$ is a recurrent component.
\end{definition}

We now explain how these objects from \cite{BL16c}  fit into our work.
Recall that $\Lrec(\Subnet)$ is the set of recurrent components of $\Subnet$ (see Definition~\ref{definition: recurrent class}).

\begin{lemma}\label{lemma: the set of recurrent classes is a close subset of Q}
Let $\Subnet$ be a finite, locally irreducible, and subcritical abelian network.
Then      $\Lrec(\Subnet)$ is a closed subset of $\Qcr$ under the action of $\Msf(\Subnet)$.
\end{lemma}
\begin{proof}
We first show that
the set $\Lrec(\Subnet)$ is a subset of 
 $\Qcr$.
Let $\C$ be any recurrent component of $\Subnet$, and let
 $\x.\q:= \st(\C)$.
Since $\Subnet$ is subcritical and $\C$ is recurrent,  by Lemma~\ref{lemma: recurrent classes has a strongly diverse infinite walk} 
there exist a configuration $\x'.\q'$ and $w \in A^*$ such that 
 $\x'.\q' \xlongrightarrow{w} \x.\q$  and $|w|\geq \satu$.
 By Lemma~\ref{o. to legal properties}\eqref{o. nonnegativity} and the fact that $\x.\q$ is stable,
 we conclude that 
  that $\x=\nol$.
   This then implies that $\C$ is in $\Qcr$.

Let $\n$ be any nonnegative vector and let $\overline{\x.\q}$ be any recurrent component. 
It follows from Lemma~\ref{o. to legal properties}\eqref{o. to contagious} and the definition of recurrence that $\overline{(\x+\n).\q}$
 is a recurrent component.
This shows that $\Lrec(\Subnet)$ is closed under the action of $\Msf(\Subnet)$.
\end{proof}

Let $\eta:\Msf(\Subnet) \to \End(\Lrec(\Subnet))$  be  the monoid homomorphism induced by the action of $\Msf(\Subnet)$ on $\Lrec(\Subnet)$.
Note that the
shift monoid $\Mon(\Subnet)$ from Definition~\ref{definition: monoid action on recurrent classes} is the image of the global monoid $\Msf(\Subnet)$ under the map $\eta$.
We denote by $\epsilon$ the identity of element of $\Mon(\Subnet)$.

Recall  that
 the torsion group $\Tor(\Subnet)$   is 
 the  torsion subgroup of the Grothendieck group   of $\Mon(\Subnet)$,
and $\Tor(\Subnet)$   acts on the set of invertible recurrent components $\Lrec(\Subnet)^\times$ (see Definition~\ref{definition: invertible recurrent class}).

We now state a theorem which shows that, for a subcritical network,
the construction in \cite{BL16c} and our construction give rise to the same group.

\begin{theorem}\label{theorem: torsion group for BL and CL constructions are equal for subcritical networks}
Let $\Subnet$ be a finite, locally irreducible, and subcritical abelian network.
Then 
\begin{enumerate}
\item\label{item: subcritical torsion group BL=torsion group CL} $e\Msf(\Subnet) \simeq \Tor(\Subnet)$ by  the map $F:e\Msf(\Subnet) \to \Tor(\Subnet)$ defined by $em \mapsto\overline{(\eta(em),\epsilon)}$.
\item \label{item: subcritical recurrent BL=recurrent CL} $e\Qcr=\Lrec(\Subnet)=\Lrec(\Subnet)^\times$.
\item \label{item: subcritical isomorphism preserves action} The isomorphism $F:e\Msf(\Subnet) \to \Tor(\Subnet)$ preserves the action of 
$e\Msf(\Subnet)$ and $\Tor(\Subnet)$ on $e\Qcr=\Lrec(\Subnet)^\times$.
\end{enumerate}
\end{theorem}
 \begin{proof}
We first  check that the assumptions  
in Proposition~\ref{proposition: finite commutative monoid has a unique injective set} are satisfied.
 The action of $\Msf(\Subnet)$ on $\Qcr$ is faithful by definition.
  We now
 show that the action of $\Msf(\Subnet)$ on $\Qcr$ is irreducible.
 Let $\overline{\nol.\q}$ and  $\overline{\nol.\q'}$ be any two elements of  $\Qcr$.
 Since $\Subnet$ is locally irreducible, there exist $w,w' \in A^*$ such that  $t_w\q=t_{w'}\q'$.
Then there exist $\n,\n',\m \in \N^A$ such that 
$\n.\q \xlongrightarrow{w} \m.t_w(\q)$ 
 and 
 $\n'.\q' \xlongrightarrow{w'} \m.t_{w'}(\q')$.
 These two facts imply that
 $\Phi(\n)(\overline{\nol.\q})= \Phi(\n')(\overline{\nol.\q'})$,
 which proves irreducibility.
 Also note that the set $\Qcr$ is nonempty since $Q$ is  nonempty  by the definition of abelian networks.
 
Note that $\Lrec(\Subnet)$ is  nonempty (by Corollary~\ref{corollary: the set of recurrent classes is nonempty}), is a  closed subset of $\Qcr$ (by Lemma~\ref{lemma: the set of recurrent classes is a close subset of Q}), and  the action of  $\Msf(\Subnet)$  on it is injective   (by Lemma~\ref{lemma: the action of Mon is free, injective, and with kernel described}\eqref{item: the action of Mon is free and injective}).  
The theorem now follows from  Proposition~\ref{proposition: finite commutative monoid has a unique injective set}. 
 \end{proof}

\chapter{Critical Networks: Recurrence} \label{s. critical networks}
In this chapter we study critical networks in more detail, with a focus on their recurrent configurations and torsion group.
Examples of critical networks include sinkless rotor networks~(Example~\ref{e. rotor network}),  sinkless sandpile networks~(Example~\ref{e. sandpile network}), sinkless height-arrow networks~(Example~\ref{e. height-arrow}), arithmetical networks~(Example~\ref{e. arithmetical graphs}), and inverse networks~(Example~\ref{e. inverse}).

\section{Recurrent configurations and the burning test}\label{ss. recurrent configurations and burning test}
In this section we define the notion of recurrence for configurations of a critical network, and we outline a test to check for the recurrence of a configuration.

We assume throughout this section that  $\Net$ is a finite, locally irreducible, and strongly connected critical network unless stated otherwise.

Integral to our study of  critical networks is the notion of period vector, defined as follows.

Denote by $\Eig$ the (right) eigenspace of $\lambda(P)$ of the production matrix $P$ of $\Net$.
By  the Perron-Frobenius theorem (Lemma~\ref{lemma: Perron-Frobenius theorem}\eqref{item: Perron-Frobenius 5}), the vector space $\Eig$  is spanned by a positive integer vector.
Since the total kernel $K$ is a subgroup of $\Z^A$ of finite index~(Lemma~\ref{lemma: the total kernel is a subgroup of finite index}\eqref{item: the total kernel is a subgroup of finite index}),
the set $\Eig \cap \, K$ is equal to the $\Z$-span of a unique positive integer vector.

\begin{definition}[Period vector]
\label{definition: period vector}
Let $\Net$ be a finite, locally irreducible, and strongly connected critical network.
The \emph{period vector} $\rr$ of  $\Net$ is the unique positive vector that generates $\Eig \cap \, K$.
\end{definition}

The period vectors of some critical networks are shown in Table~\ref{table: table for critical networks with their exchange rate vector}.

\begin{remark}
We would like to warn the reader about the difference between    the period vector  in this paper and   in \cite{BL92, FL16}.
For the sandpile network on a strongly-connected digraph $G$,
the period vector  in \cite{BL92, FL16} is  
\[ \rr= \left( \frac{t(G,v)}{\gcd_{w \in V}(t(G,w))} \right)_{v \in V},\]
where $t(G,v)$ is the number of directed spanning trees of $G$ rooted toward $v$.
On the other hand, the period vector based on our definition is 
\[ \rr= \left( \frac{\outdeg(v)t(G,v)}{\gcd_{w \in V}(t(G,w))} \right)_{v \in V}.\]
This  is
because the former is the period vector 
for the Laplacian matrix $L_G$,
while  the latter is  the period vector for  the production matrix (which in this case is equal to $A_{G}D_G^{-1}$).
\hfill $\triangle$
\end{remark}

 {
    \renewcommand{\arraystretch}{1.4}
  \begin{table}[tb]
      \caption{A list of the period vectors and exchange rate vectors of some critical networks.
      Note that $t(G,v)$ is the number of directed spanning trees rooted toward $v$, and $t^*(G,v)$ is the number of directed spanning trees rooted away from $v$.
}

  \begin{threeparttable}
  {\small
  \begin{tabular}{|c|c|c|}
  \hline
  Critical network  & Period vector & Exchange rate vector\\
  $\Net$ on $G$  & $\rr$~(Definition~\ref{definition: period vector}) &   $\s$~(Definition~\ref{definition: exchange rate vector})\\
  \hline
  Height-arrow network & $ \left( \frac{\outdeg(v) t(G,v)}{\gcd_{w \in V}(t(G,w))} \right)_{v \in V}$ & $\satu$ \\
  \hline
  Row chip-firing network  & $(\indeg(v))_{v \in V}$ & $ \left( \frac{t^*(G,v)}{\gcd_{w \in V}(t^*(G,w))} \right)_{v \in V}$  \\
  \hline
  Arithmetical network $(\Dartm,\M, \bb)$ & ${\Dartm\bb}$ & depends on $\M$\\
  \hline 
McKay-Cartan network of $(\Gc,\gamma)$ & $  (\dim \gamma \dim \chi)_{\chi \in \text{Irrep}(\Gc)}$  & $(\dim \chi)_{\chi \in \text{Irrep}(\Gc)}$  
  \\
  \hline
  Inverse network & depends on $\Net$ & $\satu$\\
  \hline
  \end{tabular}
  }
    \label{table: table for critical networks with their exchange rate vector}
\end{threeparttable}
  \end{table}
  
  }

Recall the definition of $\rlarrow$ from Definition~\ref{definition: weak and strong relation}.

\begin{definition}[Recurrent configuration]\label{definition: recurrent configuration}
Let $\Net$ be a finite, locally irreducible, and strongly connected critical network.
A configuration $\x.\q$  is 
 \emph{recurrent} if both of the following conditions hold:
\begin{enumerate}
\item There exists a nonempty legal execution for $\x.\q$; and
\item For all  configurations  $\x'.\q'$  satisfying  $\x.\q \rlarrow \x'.\q'$, we have  $\x'.\q' \longrightarrow \x.\q$. \qedhere
\end{enumerate}
\end{definition}
Later in Lemma~\ref{lemma: relation between recurrent configurations and recurrent classes} we relate recurrent configurations to recurrent components from \S\ref{subsection: construction of torsion group for all abelian networks}.

In the next lemma we give two other equivalent definitions for recurrent configurations.
Recall that, for any $w \in A^*$,
we denote by  $|w|$ the vector in $\N^A$ that counts the number of occurrences of each letter in $w$.  
Also recall  the definition of $w \setminus \n$ ($\n \in \N^A$) from Definition~\ref{definition: removal}.

\begin{lemma}\label{l. three definitions of recurrence}
Let $\Net$ be a finite, locally irreducible, strongly connected, and critical abelian network, and let
 $\x.\q$ be a configuration of $\Net$.
 The following are equivalent:
\begin{enumerate}
\item \label{i. recurrence definition strongest}  $\x.\q$ is recurrent.
\item \label{i. recurrence definition classical} There exists a nonempty word $v \in A^*$ such that $\x.\q \xlongrightarrow{v}  \x.\q$.
\item \label{i. recurrence definition period vector} There exists a legal execution $w$ for $\x.\q$ such that $|w|=\rr$ and $t_{w} \q=\q$.
\end{enumerate}
\end{lemma}
\begin{proof}
\eqref{i. recurrence definition strongest} implies \eqref{i. recurrence definition classical}: 
By the first condition of recurrence,
there is a nonempty word $w'$ and a configuration $\x'.\q'$ such that
$\x.\q \xlongrightarrow{w'} \x'.\q'$.
Since $\x.\q$ is recurrent,  there exists $w'' \in A^*$ such that $\x'.\q' \xlongrightarrow{w''} \x.\q$.
Then $w'w''$ is a nonempty word such that $\x.\q \xlongrightarrow{w'w''} \x.\q$, as desired.

\eqref{i. recurrence definition classical} implies  \eqref{i. recurrence definition period vector}: 
By Lemma~\ref{l. N locally irreducible}, the word $v$  in (\ref{i. recurrence definition classical}) satisfies
$|v| \in K$ and $(I-P)|v|=\NN_{v}(\q)=\x-\x=\nol$.
By the definition of period vector, it follows that $|v|=k\rr$ for some positive $k$.
In particular $|v|$ is a positive vector, and hence $\q$ is locally recurrent by 
Lemma \ref{l. local recurrence}\eqref{i. local recurrence sufficient and necessary condition}.

Write  $w:=v\setminus (k-1)\rr$.
The removal lemma  (Lemma~\ref{lemma: removal lemma}) 
implies that  $\pi_{(k-1)\rr}(\x.\q) \xlongrightarrow{w} \x.\q$.
Note that  $\pi_{(k-1)\rr}(\x.\q)=\x.\q$ (since $\rr \in K$ and $\q$ is locally recurrent), $|w|=\rr$, and 
 $t_{w}\q=t_\rr\q=\q$.
 This proves the claim.

\eqref{i. recurrence definition period vector} implies  \eqref{i. recurrence definition strongest}: It suffices to show that if there exist $w_1,w_2 \in A^*$ and $\x'.\q'$, $\x''.\q''$ such that $\x.\q \xlongrightarrow{w_1} \x''.\q''$ and $\x'.\q' \xlongrightarrow{w_2} \x''.\q''$, then $\x'.\q' \longrightarrow \x.\q$.

Let $k$ be a positive integer such that $k|w|=k\rr\geq |w_1|$. (Note that $k$ exists because $\rr\geq \satu$.)
By the removal lemma,
\begin{center}
\begin{tikzcd}
\x.\q \arrow[rr,"w^k"]  \arrow[rd,"w_1"]& &  \x.\q\\
\x'.\q' \arrow[r,"w_2"] & \x''.\q'' \arrow[ru,"w^k \setminus |w_1|"]
\end{tikzcd}.
\end{center}
This shows that $\x'.\q' \longrightarrow \x.\q$, as desired.
\end{proof}

We remark that \cite[Definition~3.2]{HLM08} and \cite[Definition~13]{HKT15} use  condition   \eqref{i. recurrence definition classical} in Lemma \ref{l. three definitions of recurrence}  as the 
  definition of recurrent configurations for sinkless rotor networks  and  for sinkless sandpile networks  on a strongly connected digraph, respectively.

In the next lemma, we list several basic properties of recurrent configurations.

\begin{lemma}\label{p. properties of recurrent configurations}
Let $\Net$ be a finite, locally irreducible, strongly connected, and critical abelian network, and let 
  $\x.\q$ be  a recurrent configuration of $\Net$.
Then:
\begin{enumerate}
\item \label{i. recurrence implies local recurrence} The state $\q$ is locally recurrent.
\item \label{p. properties of recurrent configuration, x is nonzero} The vector $\x$ is in $\N^A \setminus \{\nol\}$.
\item \label{p. properties of recurrent configuration, adding to recurrent makes recurrent} The configuration $(\x+\n).\q$ is recurrent for all  $\n \in \N^A$.
\item  If $\x.\q \xlongrightarrow{} \x'.\q'$, then $\x'.\q'$ is also a recurrent configuration. 
\end{enumerate}
\end{lemma}
\begin{proof}
\begin{enumerate}[wide, labelwidth=!, labelindent=10pt]
\item By Lemma \ref{l. three definitions of recurrence}(\ref{i. recurrence definition period vector}), there is a positive vector $\w$ such that $t_\w\q=\q$.
By Lemma \ref{l. local recurrence}(\ref{i. local recurrence sufficient and necessary condition}),
the state $\q$ 
 is locally recurrent.

\item By Lemma \ref{l. three definitions of recurrence}(\ref{i. recurrence definition period vector}), there exists $w \in A^*$  such that $|w| \geq \satu$ and $\pi_{w}(\x.\q)=\x.\q$.
By Lemma \ref{o. to legal properties}(\ref{o. nonnegativity}), 
the vector $\x$ is nonnegative.
Since $w$ is a nonempty legal execution on $\x.\q$,
the vector $\x$ is nonzero.

\item By Lemma \ref{l. three definitions of recurrence}(\ref{i. recurrence definition classical}), there is a nonempty word $w\in A^*$
such that $\x.\q \xlongrightarrow{w} \x.\q$.
By Lemma \ref{o. to legal properties}(\ref{o. to contagious})    $(\x+\n).\q \xlongrightarrow{w} (\x+\n).\q$, and  hence $(\x+\n).\q$ is  recurrent by Lemma \ref{l. three definitions of recurrence}(\ref{i. recurrence definition classical}).

\item Let $w_1 \in A^*$ be a word such that $\x.\q \xlongrightarrow{w_1} \x'.\q'$.
By the  definition of recurrence there exists  $w_2\in A^*$ such that  $\x'.\q' \xlongrightarrow{w_2} \x.\q$.
By Lemma \ref{l. three definitions of recurrence}(\ref{i. recurrence definition classical}) there is a nonempty word $w_3 \in A^*$ such that  $\x.\q \xlongrightarrow{w_3} \x.\q$.
Now note that $w_2w_3w_1$ is a nonempty word and  $\x'.\q'\xlongrightarrow{w_2w_3w_1}\x'.\q'$.
 Hence $\x'.\q'$ is recurrent by Lemma \ref{l. three definitions of recurrence}(\ref{i. recurrence definition classical}). \qedhere
\end{enumerate}
\end{proof}

Here we present a  consequence of Lemma \ref{l. three definitions of recurrence} and Lemma \ref{p. properties of recurrent configurations}   that will be used  in Chapter \ref{s. abelian mobile agents}.
For  any $a \in A$  we say that a word $w$ is ${a}$\emph{-tight} if $|w|\leq \rr$  and $|w|(a)=\rr(a)$.

\begin{lemma}\label{c. tightness condition for recurrent configurations}
Let $\Net$ be a finite, locally irreducible, strongly connected, and critical abelian network.
A configuration $\x.\q$ is recurrent if and only if these two conditions are satisfied:
\begin{enumerate}
\item \label{i. tightness 1} The state $\q$ is locally recurrent; and 
\item \label{i. tightness 2} For each $a \in A$ there exists an $a$-tight legal execution for $\x.\q$.
\end{enumerate}
\end{lemma}

\begin{proof}
Proof of only if direction: Condition \eqref{i. tightness 1} follows from Lemma \ref{p. properties of recurrent configurations}(\ref{i. recurrence implies local recurrence}). 
For condition \eqref{i. tightness 2},  Lemma \ref{l. three definitions of recurrence}(\ref{i. recurrence definition period vector}) implies that there exists a legal execution $w$ for $\x.\q$ such that $|w|=\rr$.
Note that this  $w$ is an $a$-tight word for all $a \in A$, and condition \eqref{i. tightness 2}   follows.

Proof of  if direction: 
For each $a \in A$ let $w_a$ be an $a$-tight legal execution for $\x.\q$ given by condition (\ref{i. tightness 2}).
By applying the exchange lemma~(Lemma \ref{l. weak exchange lemma}) consecutively, there exists a legal execution $w$ for $\x.\q$ such that $|w|=\max_{a \in A} \{|w_a| \}$.
The tightness condition for all $a \in A$ then implies that $|w|=\rr$.
Since $\q$ is locally recurrent by condition \eqref{i. tightness 1}, we then have $t_{w}\q=t_\rr\q=\q$. 
 By Lemma \ref{l. three definitions of recurrence}(\ref{i. recurrence definition period vector}), we  conclude that   $\x.\q$ is recurrent.
\end{proof}

We now outline a recurrence test for configurations of critical networks, answering a question posed in~\cite{BL16c}.
This recurrence test is called the \emph{burning test}, named after a similar test for  sandpile  networks~\cite{Dhar90, Speer93, AB11}.

Given a configuration $\x.\q$ and a legal execution $w$ for $\x.\q$ , we say that $w$ is $\rr$-\emph{maximal} if 
\begin{enumerate}
\item $|w| \leq \rr$; and 
\item For all $a \in A$ either $|w|(a)= \rr(a)$ or $wa$ is not a legal execution for  $\x.\q$.
\end{enumerate}

\begin{theorem}[Critical burning test]
\label{t. recurrence test}
Let $\Net$ be a finite, locally irreducible, strongly connected, and critical abelian network.
Let  $\x.\q$ be a configuration of $\Net$, and  let $w\in A^*$ 
  be any $\rr$-maximal legal execution for $\x.\q$.
Then $\x.\q$ is recurrent if and only if  the word $w$ satisfies $|w|=\rr$ and $t_w\q=\q$.
\end{theorem}
\begin{proof}
 Proof of if direction: This  follows  directly from Lemma \ref{l. three definitions of recurrence}(\ref{i. recurrence definition period vector}).

Proof of only if direction:  
We first show that $|w|=\rr$.
By Lemma \ref{l. three definitions of recurrence}(\ref{i. recurrence definition period vector}) there 
is a legal execution $w'$ for $\x.\q$  such that $|w'|=\rr$.
By  the removal lemma (Lemma~\ref{lemma: removal lemma}),  the 
word $w'\setminus |w|$ is a legal execution for $\pi_{w}(\x.\q)$.
By the $\rr$-maximality of $w$,  we then have $w'\setminus |w|$ is the empty word, and hence $|w|=|w'|=\rr$.

By Lemma \ref{p. properties of recurrent configurations}\eqref{i. recurrence implies local recurrence} the state $\q$ is locally recurrent; hence  $t_w\q=t_{\rr}\q=\q$.
The proof is now complete.
\end{proof}

\begin{figure}[tb]
\centering
   \includegraphics[width=0.8\textwidth]{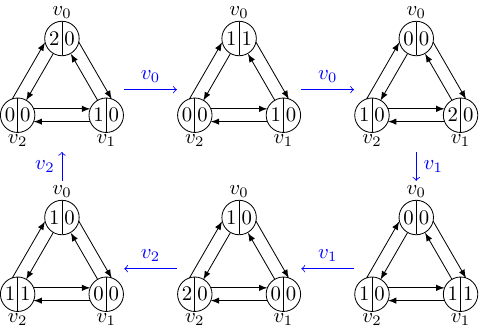}
   \caption{An instance of the burning test for the sinkless sandpile network on the bidirected cyle $C_3$.
      In the figure, the left part of  $v\in V$ records $\x(v)$, while the right part records $\q(v)$.
   The inputs are 
     $\x:=(2,1,0)^\top$, $\q:=(0,0,0)^\top$, and  $\rr=(2,2,2)^\top$. 
     The configuration $\x.\q$ is recurrent by the burning test.
    }  
   \label{figure:sandpile network burning test}
\end{figure}

Using  Theorem \ref{t. recurrence test}, we derive a  recurrence test for critical networks by constructing an $\rr$-maximal legal execution $w$ for $\x.\q$.
The test in its precise form is given in Algorithm \ref{a. burning test}.
See Figure~\ref{figure:sandpile network burning test} for an example of the burning test in action.

\begin{algorithm}\label{a. burning test}
 \caption{The burning test to check for recurrence of a configuration in a critical abelian network.}
\SetKwInOut{Input}{Input}\SetKwInOut{Output}{Output}
 \Input{ A critical network $\Net$, a configuration $\x.\q$. }
 \Output{TRUE if $\x.\q$ is recurrent, FALSE if $\x.\q$ is not recurrent.}
 $\q':=\q$\;
 $\x':=\x$\;
 $w:= \varnothing $ \;
 \While{$|w|(a) < \rr(a)$  and  $\x'(a)\geq 1$ for some $a \in A$}{
     $\x':=\x'+\NN(a,\q')-|a|$\;
   $\q':=t_a\q'$\;
   $w:=wa$.
  }
    \eIf{$|w|==\rr$ and $\q==\q'$}{
output TRUE.   }{
output FALSE.  }
\end{algorithm}

  {
The running time of the burning test is equal to  the sum of the entries of the period vector $\rr$, which can take exponential time with respect to $|A|$ 
(One example is the sandpile network on a bidirected path with edge multiplicities 2 to the left and 3 to the right;
see \cite[Figure~1]{FL16}).
 
 
In \S \ref{ss. cycle test}, we present a more efficient  recurrence test called the {``cycle test''}  for a subclass of critical networks called   agent networks.
}

 \section[Thief networks]{Thief networks of a critical network}\label{ss. abelian network with thief}
  In this section we relate the burning test  for critical networks~(Algorithm~\ref{a. burning test}) to the preexisting burning test for subcritical networks.
  
\begin{figure}[tb]
\centering
   \includegraphics[width=\textwidth]{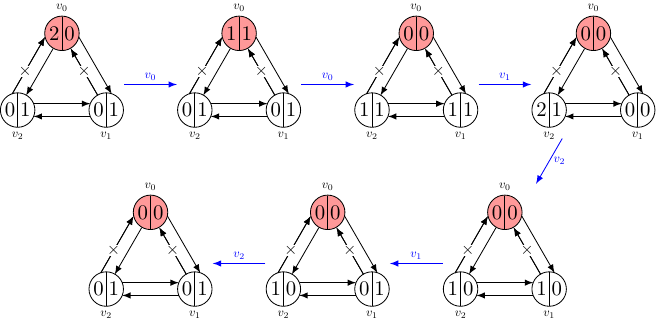}
   \caption{A subcritical burning test for the sandpile network  with   sink at $S=\{v_0\}$ (colored in red).   
   In the figure, the left part of  $v \in V$ records $\x(v)$, while the right part records $\q(v)$.
The  inputs for the test are $\q:=(0,1,1)^\top$ and $\kk:=(2,2,2)^\top$.
   (Note that  $(I-P)\kk=(2,0,0)^\top$ here.)
The state $\q$ is recurrent by the burning test.
    }  
   \label{figure: sandpile network burning test with sink}
\end{figure}

\begin{theorem}[Subcritical burning test~{\cite[Theorem~5.5]{BL16c}}]\label{theorem: burning test subcritical}
Let $\Subnet$ be a finite, locally irreducible, and subcritical abelian network with total kernel $K$ and production matrix $P$.
Let $\kk \in K$ be such that $\kk\geq \satu$ and $P\kk \leq \kk$.
Then $\q \in Q$ is recurrent if and only if $(I-P)\kk.\q\xlongrightarrow{} \nol.\q$. \qed
\end{theorem}  
See Figure~\ref{figure: sandpile network burning test with sink}
for an example of this burning test for sandpile networks with sinks.

The relation between these two burning tests  can be explained by using the notion of thief networks.   
  
  \begin{remark}
In this section we  often discuss two abelian networks at the same time.
When there is more than one network in the discussion, we will indicate in the notation which network we are referring to, e.g. $t_a^{\Net}$, $\NN_a^{\Net}$, $\pi^{\Net}_a$, $\Net$-recurrent, $\xlongrightarrow[\Net]{}$, etc.
\end{remark}

For   $R \subseteq A$ and  $\x \in \Z^A$, let  $\x_R$ denote the vector in $\Z^A$ for which  $\x_R(a):=\x(a)$ if $a \in R$ and $\x_R(a):=0$ if $a \notin R$.

  Let $\Net$ be an abelian network, and let $R \subseteq A$.
  The \emph{thief network based on} $\Net$ \emph{with messages restricted to} $R$ (\emph{thief network} $\Net_R$ for short)  
  is the abelian network (with the same underyling digraph as $\Net$) defined by:
  \begin{itemize}
  \item The alphabet  $A^{\Net_R}$, the state space $Q^{\Net_R}$ and the transition functions $(t_{a}^{\Net_R})_{a \in A}$ of $\Net_R$ are identical with those of $\Net$.
  \item For any $a \in A$ and $\q \in Q$, the message-passing vector  $\NN_{a}^{\Net_R}(\q)$ is equal to $(\NN_{a}^\Net(\q))_R$.
  \end{itemize}
  
  One  can think of  $\Net_R$ as a network of computers where 
  the wires    used for  transmitting letters from $A \setminus R$  are stolen by a wire thief.
   Hence all the  letters from $A \setminus R$ will not appear in the messages exchanged between computers in the network.

        Note that  $t_{a}^{\Net_R}$ and $\NN_{a}^{\Net_R}$ are defined even for $a \in A \setminus R$, so 
   $\Net_R$ retains the ability to process inputs with letters from $A\setminus R$.
  One can think of this to  mean that  the keyboards for the computers in the network  are  working fine and are not tampered  by  the wire thief.

The reader can use height-arrow networks with sinks~(Example~\ref{e. sandpile network with sinks})  as a running example  when reading this section.
Note that  
a height-arrow network with sinks at $S$~(Example~\ref{e. sandpile network with sinks}) is the thief network of the corresponding sinkless height-arrow network~(Example~\ref{e. height-arrow}) restricted to   $V\setminus S$.

We now relate the total kernel and the production matrix of $\Net_R$  to those of $\Net$.

Let  $M$ be a  matrix with rows  indexed by $A$.
For  $R \subseteq A$,
we denote by $M_R$ the matrix obtained by replacing the rows of $M$ indexed by 
$A \setminus R$ with the zero vector.

\begin{lemma}\label{l. thief network network properties}
Let $\Net$ be a finite and locally irreducible abelian network with total kernel $K$ and production matrix $P$, and  let $R \subseteq A$.
\begin{enumerate}
\item \label{i. total kernel and production matrix of thief network} The network $\Net_R$ is finite and locally irreducible,
the total kernel of $\Net_R$ is equal to $K$, and the production matrix of $\Net_R$ is equal to $P_R$.
\item \label{i. thief network is a subcritical network}  If $\Net$ is a  strongly connected critical network and $R \subsetneq A$, then $\Net_R$ is a subcritical network.
\end{enumerate}
\end{lemma}
\begin{proof}
\begin{enumerate}[wide, labelwidth=!, labelindent=10pt]
\item Since the transition functions  of $\Net_R$ are the same as those of $\Net$, the network $\Net_R$ is  finite and locally irreducible.
By  the same reason, the total kernel of $\Net_R$ is equal to $K$.

Since  $\NN_{a}^{\Net_R}(\q)=(\NN_{a}^{\Net}(\q))_R$ for all $a \in A$ and $\q \in Q$, it follows directly from the definition that
the production matrix of $\Net_R$ is equal to $P_R$.
\item Note that  $P$ is strongly connected (since $\Net$ is strongly connected),  $P_R \leq P$ (by definition), and $P_R\neq P$ (since $R \subsetneq A$).
The claim now
 follows directly from the  Perron-Frobenius theorem (Lemma~\ref{lemma: Perron-Frobenius theorem}\eqref{item: Perron-Frobenius lambda is strictly increasing}).
\qedhere 
\end{enumerate}
\end{proof}

We remark that the network
   $\Net_R$ is not strongly connected whenever $R \subsetneq A$, as some of the rows of $P_R$ are zero vectors.

Recall the definition of recurrent configurations for a critical network (Definition~\ref{definition: recurrent configuration}) and the definition of  recurrent states for a subcritical network~(Definition~\ref{definition: recurrent state}).
We now state the main results of this subsection, which are two propositions that relate the recurrent configurations of a critical network to the recurrent states of its thief networks.

Recall that the support  of $\x\in \Z^A$ is    $\supp(\x)=\{a \in A: \x(a)\neq 0 \}$.
\begin{proposition}\label{proposition: the state of a recurrent configuration is a recurrent state of its thief networks}
Let $\Net$ be a finite, locally irreducible,  strongly connected, and  critical abelian network.
Let  $\x \in \N^A \setminus \{\nol\}$ and  let $R:=A \setminus \supp(\x)$.
If $\x.\q$ is an $\Net$-recurrent configuration,
then $\q$ is an $\Net_R$-recurrent state.
\end{proposition}

We remark that the converse of 
Proposition~\ref{proposition: the state of a recurrent configuration is a recurrent state of its thief networks}
is false; see  Example~\ref{e. recurrence between sinkless and sink network is not two way}.
With that being said, we will present a 
 special family of critical networks for which the converse holds in Lemma~\ref{lemma: recurrent configurations of agent network is essentially a recurrent state of subcritical networks}.

\begin{example}\label{e. recurrence between sinkless and sink network is not two way}
Let $\Net$ be the  sinkless sandpile network~(Example~\ref{e. sandpile network}) on the bidirected cycle
$C_3$, and let  $R:=V \setminus \{v_0\}$.

Let $\x\in \Z^V$ and $\q \in (\Z_2)^V$ be given by:
\[\x:=(1,0,0)^\top 
\quad \text{ and } \quad 
\q:=(0,1,1)^\top.  \]

The state $\q$ is $\Net_R$-recurrent because it passes the  burning test in Theorem~\ref{theorem: burning test subcritical}, as shown in Figure~\ref{figure: sandpile network burning test with sink}.
On the other hand, note that $\x.\q \xlongrightarrow[\Net]{v_0} \nol.\q'$, where $\q':=(1,1,1)^\top$.
This shows that $\x.\q$ is an $\Net$-halting configuration, and hence $\x.\q$  is not  $\Net$-recurrent.
\end{example}   

Recall that $\rr$ denotes the period vector of a critical network $\Net$~(Definition~\ref{definition: period vector}).

     \begin{proposition}\label{proposition: recurrence for thief implies a very specific recurrence in the original network}
     Let $\Net$ be a finite, locally irreducible,  strongly connected, and  critical abelian network, and let $R \subsetneq A$.
     Then $\q \in Q$ is
    an $\Net_R$-recurrent state if and only if
     $(I-P_R)\rr.\q$ is an $\Net$-recurrent configuration.     
     \end{proposition}
In particular, checking for the recurrence of $\q \in Q$ in $\Net_R$  can be done by applying the critical burning test for $\Net$~(Algorithm~\ref{a. burning test}) on $(I-P_R)\rr.\q$,
and it can be shown that
 this test is equivalent to  the subcritical burning test for $\Net_R$~(Theorem~\ref{theorem: burning test subcritical}) with $\kk=\rr$.
The critical burning test for $\Net$ on $(I-P)\rr.\q$ can be derived from the subcritical burning test for $\Net_R$ in a similar manner. 

We now build towards the proof of  these two propositions, and we start with a technical lemma.
\begin{lemma}\label{p.  connection between nonthief and thief network}
Let $\Net$ be an abelian network and let $R \subseteq A$. 
\begin{enumerate}
\item \label{p. thief to nonthief} If $w\in A^*$ is an  $\Net_R$-legal execution for $\x.\q$, then $w$ is an $\Net$-legal execution for $\x.\q$.

\item \label{p. nonthief to thief} 
If  $w \in A^*$ is an $\Net$-legal execution for $\x.\q$,
then $w$ is an $\Net_R$-legal execution for
$(\x_R+\w_{A\setminus R}).\q$, where $\w:=|w|$.
\end{enumerate}
\end{lemma}

\begin{proof}

 Part~\eqref{p. thief to nonthief} follows from the inequality  $\NN_{a}^{\Net_R}(\q)\leq \NN_{a}^{\Net}(\q)$ for all $a \in A$ and $\q \in Q$.

 We now prove part~\eqref{p. nonthief to thief}.
Let $w=a_1\cdots a_\ell$.
For any  $i \in \{0,1,\ldots, \ell\}$ we write 
\[ w_i:=a_1\ldots a_i, \quad \x_i.\q_i:=\pi_{a_1\cdots a_i}^\Net(\x.\q), \quad \x_i'.\q_i':=\pi_{a_1\cdots a_i}^{\Net_R}((\x_R+\w_{A \setminus R}).\q).\]
It suffices to show that $\x'_{i-1}(a_i) \geq 1$ for all $i \in \{1,\ldots, \ell\}$.

Fix  $i \in \{1,\ldots, \ell\}$.
Then
\begin{align*}
&\x'_{i-1}(a_i)= \x_R(a_i)+\w_{A \setminus R}(a_i) +\NN_{w_{i-1}}^{\Net_R}(\q)(a_i)-|w_{i-1}|(a_i)\\
&=
\begin{cases}
|w|(a_i)-|w_{i-1}|(a_i)  & \quad \text{if } a_i \in A \setminus R;\\
\x(a_i)+\NN_{w_{i-1}}^{\Net}(\q)(a_i)-|w_{i-1}|(a_i)=\x_{i-1}(a_i)   & \quad \text{if } a_i \in  R.
\end{cases}
\end{align*}
Note that $|w|(a_i)-|w_{i-1}|(a_i)\geq 1$ because the $i$-th letter of $w$ is $a_i$.
Also note that $\x_{i-1}(a_i)\geq 1$ because $w$ is legal for $\x.\q$.
Hence we conclude that $\x'_{i-1}(a_i)\geq 1$, as desired.
\qedhere
\end{proof}

\begin{proof}[Proof of Proposition~\ref{proposition: the state of a recurrent configuration is a recurrent state of its thief networks}]
Note that by Lemma~\ref{l. thief network network properties}\eqref{i. thief network is a subcritical network} the network $\Net_R$ is  subcritical    since $R\subsetneq A$.
Also note that the period vector $\rr$ of $\Net$ satisfies $\rr\in K$, $\rr \geq \satu$, and $P_R\rr=\rr_{R}\leq \rr$.
Hence by Theorem~\ref{theorem: burning test subcritical}
  it suffices to show that $(I-P_R)\rr.\q \xlongrightarrow[{\Net_R}]{} \nol.\q$.

Since $\x.\q$ is $\Net$-recurrent, by Theorem \ref{t. recurrence test} there exists  $w \in A^*$ 
such that $\x.\q \xlongrightarrow[\Net]{w}\x.\q$ and $|w|=\rr$.
Since $\x_R=\nol$ by assumption,
the word $w$ is an $\Net_R$-legal execution for 
$\rr_{A \setminus R}.\q$ by Lemma~\ref{p.  connection between nonthief and thief network}\eqref{p. nonthief to thief}.
Now note that
\begin{align*}
 \pi_w^{\Net_R}(\rr_{A \setminus R}.\q)&=(\rr_{A \setminus R}+\NN_{w}^{\Net_R}(\q) -|w|).t_{w}\q \\
 &=(\rr_{A \setminus R}+(\NN_{w}^{\Net}(\q))_R-\rr).\q\\
 &=\nol.\q  \quad \text{(because }\NN_{w}^{\Net}(\q)=\rr).
\end{align*}
Also note that $\rr_{A \setminus R}=(I-P_R)\rr$.
Hence, we conclude that $(I-P_R)\rr.\q \xlongrightarrow[{\Net_R}]{w} \nol.\q$, as desired.  
 \end{proof}

     \begin{proof}[Proof of Proposition~\ref{proposition: recurrence for thief implies a very specific recurrence in the original network}]
The if direction follows from Proposition~\ref{proposition: the state of a recurrent configuration is a recurrent state of its thief networks} and the fact that  $\supp((I-P_R)\rr)=A \setminus R$.     
     
 We now prove the only if direction.
     Since $\q$ is $\Net_R$-recurrrent, by Theorem~\ref{theorem: burning test subcritical}
    there exists  $w \in A^*$  such that $(I-P_R)\rr.\q \xlongrightarrow[{\Agn_R}]{w} \nol.\q$.   
By Lemma \ref{l. N locally irreducible},
this  implies that $\NN_{w}^{ \Agn_R}(\q)=P_R|w|$.
Then 
\begin{equation}\label{equation: thef recurrence is essentially network recurrence}
(I-P_R)\rr= |w|-\NN_{w}^{ \Agn_R}(\q)=(I-P_R)|w|. 
\end{equation}
Since $P_R$ has spectral radius strictly less than 1~(by Lemma~\ref{l. thief network network properties}\eqref{i. thief network is a subcritical network}), the matrix $I-P_R$ is invertible.
It then follows from equation~\eqref{equation: thef recurrence is essentially network recurrence} that  $|w|=\rr$.

   By  Lemma \ref{p.  connection between nonthief and thief network}\eqref{p. thief to nonthief}, the word  $w$ is an $\Agn$-legal execution  for  $(I-P_R)\rr.\q$.
   Since  $|w|=\rr$ and $t_\rr^{\Agn}\q=t_w^{\Agn_R}\q=\q$,
    by Theorem \ref{t. recurrence test} we conclude that   $(I-P_R)\rr.\q$ is  an $\Agn$-recurrent configuration, as desired.
     \end{proof}

\section[Capacity and level]{The capacity and the level   of a configuration}\label{ss. level}
 In this section we define the capacity of a network and the level of a  configuration of a critical network.
 Those two notions will be used  later in \S\ref{subsection: torsion group for critical networks} to give a combinatorial description for  the invertible recurrent components  of a critical network.
   
     Let  $\Net$ be a finite, locally irreducible, and strongly connected abelian network.
     By  the  Perron-Frobenius theorem (Lemma~\ref{lemma: Perron-Frobenius theorem}\eqref{item: Perron-Frobenius 4})
     the  $\lambda(P^\top)$-eigenspace of $P^\top$ is spanned by a positive real  vector.
     
\begin{definition}[Exchange rate vector]\label{definition: exchange rate vector}
Let $\Net$ be a finite, locally irreducible, and strongly connected abelian network.
An \emph{exchange rate vector} $\s$ is a positive real vector that spans the $\lambda(P^\top)$-eigenspace of $P^\top$.
\end{definition}     
     The vector $\s$ measures the comparative value between any two letters in $\Net$, in a manner to be made precise soon.
     
Throughout this paper we fix an exchange rate vector $\s$.
In the case when $\lambda(P)=\lambda(P^\top)$ is rational, then  we    choose $\s$ to be an exchange rate vector 
 that is a positive integer vector and such that $\gcd_{a \in A} \s(a)=1$.
This choice of $\s$ exists and is unique by 
   the   Perron-Frobenius theorem (Lemma~\ref{lemma: Perron-Frobenius theorem}\eqref{item: Perron-Frobenius 5}). 
     The exchange rate vectors of some critical networks are
     shown in Table~\ref{table: table for critical networks with their exchange rate vector}.

Recall that a configuration $\x.\q$ \emph{halts} if $\x.\q \xlongrightarrow{} \x'.\q'$ for some $\x' \leq \nol$ and some $\q' \in Q$.

\begin{definition}[Capacity]\label{definition: capacity} 
Let $\Net$ be a finite, locally irreducible, and strongly connected abelian network.
The \emph{capacity} of a configuration $\x.\q$ and the \emph{capacity} of a state $\q$ are  
\[ \cpt(\x.\q):=\sup_{\z \in \Z^A} \{\s^\top \z : (\z+\x).\q \text{ halts} \}; \qquad \cpt(\q):=\cpt(\nol.\q), \]
respectively.
The \emph{capacity} of  $\Net$ is 
\[\cpt(\Net):= \max_{\q \in Q} \{\cpt(\q)\}. \qedhere\]
\end{definition}
In words, the capacity of a configuration is  the maximum number of letters (weighted according to the exchange rate vector) that can be absorbed by the configuration without causing the process to run forever.


The following is an  example that illustrates the notion of capacity.

\begin{example}
\label{e. height-arrow networks level}
First consider the   sinkless rotor network~(Example \ref{e. rotor network}).
In this network, processing a chip will  result in moving the chip to another vertex of the digraph.
So if there are a positive number of chips in the  network, then the process will run forever, as at any time stage there will always be some chips that can be moved around. 
Hence the capacity of a  sinkless rotor network is equal to zero.

On the other end of the scale, we have  sinkless sandpile networks~(Example~\ref{e. sandpile network}).
In this network, 
processing a chip means either moving the chip into the locker $\PP_v$ (if $\PP_v$ is not full),
or sending all stored chips in $\PP_v$ together with the processed chip to other vertices (if $\PP_v$ is already full).
Note that each locker $\PP_v$ can store at most $\outdeg(v)-1$ chips.
Therefore, if the total number of chips is strictly greater than 
$|E|-|V|=\sum_{v \in V} (\outdeg(v)-1)$, then at any time stage of the process there is always an unstored chip that can be processed.
Hence  the sandpile network has capacity at most $|E|-|V|$.
On the other hand, the configuration $\x.\q$ with $\x:=(\outdeg(v)-1)_{v \in V}$ and $\q:=\nol$ is a halting configuration, which implies  that the sandpile network has capacity  at least $\satu^\top \x=|E|-|V|$.
Hence we conclude that the capacity of a sinkless sandpile network is equal to $|E|-|V|$.

By an analogous argument, the capacity of a height-arrow network is equal to
$\sum_{v \in V} (\tau_v-1)$, which lies between
 the capacity of  rotor network and  sandpile network on the same digraph.
 \end{example}

The capacity of a subcritical network is infinite, as every  configuration
  halts in a subcritical network~(Theorem~\ref{theorem: subcritical networks halts on all inputs}). 
We now show that conversely, the capacity of a critical or supercritical network is always finite.

Recall that a configuration $\x.\q$ is stable if $\x\leq \nol$.
\begin{lemma}\label{p. saturation index and level is finite}
     Let  $\Net$ be a finite, locally irreducible, and strongly connected abelian network.
If $\Net$ is a critical or supercritical network, then $\cpt(\Net)< \infty$.
\end{lemma} 
\begin{proof}
Suppose to the contrary that the claim is false.
Then there exist configurations $\z_1.\q_1,\z_2.\q_2, \ldots$
and stable configurations $\z_1'.\q_1',\z_2'.\q_2', \ldots$
  such that  $\z_i.\q_i\xlongrightarrow{w_i} \z_i'.\q_i'$ for all $i\geq 1$ and  $\s^\top \z_i \to \infty$ as $i \to \infty$.

By the pigeonhole principle, there exists an infinite subset $J$ of $\Z_{\geq 1}$
such that $\q_j=\q_i$ and $\q'_j=\q'_i$ for all $i,j \in J$.
Fix an $j\in J$ and write $\lambda:=\lambda(P)$.
 Then for any $i \in J$,
\begin{align}
 \z_i- \z_j &= (\z'_i-\NN_{w_i}(\q_i) +|w_i|)-(\z'_j-\NN_{w_j}(\q_i) +|w_j|) \notag \\
&=(\z'_i+(I-P)|w_i|) -(\z'_j+(I-P)|w_j|) \ \ \text{(by Lemma \ref{l. N locally irreducible})}  \notag
\end{align}
Multiplying   $\s^\top$ to both sides of the equation above, we get
\begin{equation}\label{equation: capacity bound}
 \s^\top (\z_i- \z_j )=(\s^\top\z'_i+(1-\lambda)\s^\top|w_i|))- (\s^\top\z'_j+(1-\lambda)\s^\top|w_j|)
\end{equation}

Now note that $\s^\top \z'_i\leq 0$ since $\z'_i \leq \nol$, and $(1-\lambda)\leq 0$ by assumption.
Plugging this into  equation~\eqref{equation: capacity bound},
we get
\[\s^\top \z_i \leq \s^\top \z_j -\s^\top(\z'_j+(1-\lambda)|w_j|).  \] 
 This gives an upper bound for 
 $\s^\top \z_i$ that is independent of $i$,
 which contradicts the assumption that $s^\top \z_i \to \infty$ as $i \to \infty$.
\end{proof}

 \begin{definition}[Level]\label{definition: level}
 Let $\Net$ be a finite, locally irreducible, and strongly connected critical network.
 The \emph{level} of a state $\q$ and the \emph{level} of a configuration $\x.\q$  are 
 \[\lvl(\q):=\cpt(\Net)-\cpt(\q); \qquad \lvl(\x.\q):=\cpt(\Net)-\cpt(\x.\q),\]
 respectively.
 \end{definition}

Note that by the definition of capacity, we have
\[ \lvl(\x.\q)= \cpt(\Net)-\cpt(\q)+ \s^\top \x= \lvl(\q)+\s^\top \x.\]

For height-arrow networks,
 the level of a configuration $\x.\q$  is
equal to  $\sum_{v \in V} \x(v) +\q(v)$, 
the total number of chips (counting both stored and unstored chips) in the configuration.

  Here we list basic properties of the capacity (equivalently, level) of a configuration in  a critical network.
  
\begin{lemma}\label{p. level and capacity  properties}
     Let  $\Net$ be a finite, locally irreducible,  strongly connected, and critical abelian network.
\begin{enumerate}
\item \label{i. level can not grow}   If $\x.\q$ and $\x'.\q'$ are configurations such that  $\x.\q \dashrightarrow \x'.\q'$, then $\cpt(\x'.\q') \leq \cpt(\x.\q)$.
\item \label{i. level is equal} 
  If $\x.\q$ and $\x'.\q'$ are configurations such that  $\x.\q \dashrightarrow \x'.\q'$  and $\q \in \Loc(\Net)$, then $\cpt(\x.\q) = \cpt(\x'.\q')$.
\item \label{item: capacity of a state is between 0 and cpt(Net)}For any $\q \in Q$, we have $0 \leq \cpt (\q) \leq \cpt(\Net)$.
\item There exists $\q \in Q$ such that $\cpt(\q)=\cpt(\Net)$.
\item \label{i. capacity minimum is achieved by locally recurrent states} There exists $\q \in \Loc(\Net)$ such that $\cpt(\q)=0$.
\end{enumerate}
\end{lemma}  
\begin{proof}
\begin{enumerate}[wide, labelwidth=!, labelindent=10pt]
\item Let $\z \in \Z^A$ be any vector such that $(\z+\x').\q'$ halts.
Then there exists a stable configuration $\y.\p$ 
such that $(\z+\x').\q' \dashrightarrow \y.\p$.
By the transitivity of $\dashrightarrow$, 
we then have $(\z+\x).\q \dashrightarrow \y.\p$.
By the least action principle~(Corollary~\ref{c. least action principle}),
we  conclude that $(\z+\x).\q$ halts.
Hence 
\[\{\z \in \Z^A: (\z+\x').\q' \text{ halts}\} \subseteq \{\z \in \Z^A: (\z+\x).\q \text{ halts}\}, \]
 which implies that
$\cpt(\x'.\q') \leq \cpt(\x.\q)$.

\item By part \eqref{i. level can not grow}, it suffices to show that $\cpt(\x.\q) \leq \cpt(\x'.\q')$.
Let $w \in A^*$ be such that $\x.\q \xdashrightarrow{w} \x'.\q'$, and let $k$ be such that $k\rr \geq |w|$. (Note that $k$ exists because the period vector $\rr$ is positive.)
Then
\[\pi_{k\rr-|w|}(\x'.\q') = \pi_{k\rr-|w|}(\pi_{w}(\x.\q))=\pi_{k\rr}(\x.\q)= \x.\q,\]
where the last equality is because $\q$ is locally recurrent.
Hence we have $\x'.\q' \dashrightarrow \x.\q$, which then implies that
  $\cpt(\x.\q)\leq \cpt(\x'.\q')$ by  part \eqref{i. level can not grow}, as desired.

\item For any $\q \in Q$ the configuration $\nol.\q$ halts by definition, and hence 
 $\cpt(\q)\geq 0$.
The other inequality follows directly  from the definition of $\cpt(\Net)$.

\item This follows directly from the definition of $\cpt(\Net)$.

\item Let $\q$ be a locally recurrent state with minimum capacity among all locally recurrent states.
Let 
 $\z \in \Z^A$ be any vector such that $\z.\q$ halts.
By definition, there exists a stable configuration $\y.\p$ such that  
 $\z.\q \longrightarrow \y.\p$ and $\y\leq \nol$.

By Lemma~\ref{lemma: the monoid $N^A$ acts invertibly on locally recurrent configurations}\eqref{item: monoid acts invertible on locally recurrent states 1}, the state $\p$ is locally recurrent,
and hence $\cpt(\q)\leq \cpt(\p)$ by the minimality assumption. 
  On the other hand, by  part (\ref{i. level is equal})
 we have $-\s^\top \z+ \cpt(\q)=-\s^\top \y+\cpt(\p)$.
These two facts then imply  $\s^\top \z \leq \nol$.
 
   Since the choice of $\z$ is arbitrary,   we conclude that $\cpt(\q) \leq 0$. 
  By part~\eqref{item: capacity of a state is between 0 and cpt(Net)} it then follows that $\cpt(\q)=0$. \qedhere

\end{enumerate}
\end{proof}

   Lemma~\ref{p. level and capacity  properties}\eqref{i. level is equal}      implies that, in a critical network,  
   the level of a configuration  does not change over time,
   provided that the initial state of the configuration is locally recurrent. 
This distinguishes  critical networks from subcritical  and supercritical networks, 
where an analogous notion of level can decrease for the former, and increase for the latter.

\section[Stoppable levels]{Stoppable levels: When does the torsion group act transitively?}\label{subsection: torsion group for critical networks}
In this section we study the torsion group of a critical network in more detail.
  
  We start with the relationship between   recurrent components~(Definition~\ref{definition: recurrent class}) and  recurrent configurations~(Definition~\ref{definition: recurrent configuration}) of a critical network.

  \begin{lemma}\label{lemma: relation between recurrent configurations and recurrent classes}
       Let  $\Net$ be a finite, locally irreducible,  strongly connected, and critical abelian network.
A component $\C$ of the  trajectory digraph is a recurrent component if and only if $\C$ contains a recurrent configuration.
    \end{lemma}
\begin{proof}
Proof of if direction: Let $\x.\q$ be  a recurrent configuration  in $\C$. 
Let $\rr$ be the period vector of $\Net$ (Definition~\ref{definition: period vector}).
By Lemma~\ref{l. three definitions of recurrence}\eqref{i. recurrence definition period vector}, there exists $w \in A^*$ such that  $|w|=\rr$ and 
\[\cdots \xlongrightarrow{w}  \x.\q \xlongrightarrow{w} \x.\q \xlongrightarrow{w} \cdots.\]
This is a diverse infinite walk (Definition~\ref{definition: recurrent class})  in $\C$ (because $|w|=\rr\geq \satu$), and hence  $\C$ is a recurrent component.

Proof of only if direction:
By Proposition~\ref{proposition: recurrent component combinatorial description},
the recurrent component $\C$ contains a diverse cycle.
In particular, this implies that 
there exists a configuration $\x.\q$ in $\C$ and a nonempty word $w$ such that $\x.\q \xlongrightarrow{w}\x.\q$.
Now note that $\x.\q$ is a recurrent configuration by  Lemma~\ref{l. three definitions of recurrence}\eqref{i. recurrence definition classical}.
This proves the claim.
\end{proof}  
  
Note that a recurrent component may contain a non-recurrent configuration, as shown in the following example.

\begin{example}
 Consider the sinkless sandpile network $\Net$~(Example~\ref{e. sandpile network}) on the bidirected cycle $C_3$.    
 Let $\x \in \Z^V$ and $\q \in (\Z_2)^V$ be given by:
 \[\x:=(2,1,0)^\top
 \quad \text{ and }\quad
 \q:=(0,0,0)^\top.
 \]
 Note that $\x.\q$ is a recurrent configuration as it passes the burning test, as shown in  Figure~\ref{figure:sandpile network burning test}.

 Let $\x' \in \Z^V$ and $\q' \in (\Z_2)^V$ be given by:
 \[\x':=(1,2,-1)^\top
 \quad \text{ and }\quad
 \q':=(0,1,0)^\top.
 \]
The configuration $\x'.\q'$ is in the same component as $\x.\q$ since $\x'.\q' \xlongrightarrow{v_1} \x.\q$.
However, $\x'.\q'$ is not recurrent by Lemma~\ref{p. properties of recurrent configurations}\eqref{p. properties of recurrent configuration, x is nonzero} since $\x'$ has a negative entry.
\end{example}

  The \emph{level} of a recurrent component $\C$ is
\[ \lvl(\C):=\lvl(\x.\q),  \]
where $\x.\q$ is any recurrent configuration  in $\C$.
The value of $\lvl(\C)$ does not depend on the choice of $\x.\q$ as a consequence of Lemma~\ref{p. level and capacity  properties}\eqref{i. level is equal} and Lemma~\ref{p. properties of recurrent configurations}\eqref{i. recurrence implies local recurrence}.
For any  $m \in \N$
we denote by $\Lrec(\Net,m)$  the set of recurrent components of $\Net$ with level $m$.

\begin{definition}[Stoppable level]\label{definition: stoppable level}
Let $\Net$ be a finite, locally irreducible, and strongly connected critical network.
The set of  \emph{stoppable levels} of $\Net$ is
 \[\Stop(\Net):= \{ m \in \N \mid m=\lvl(\x.\q)  \text{ for some }\x \leq \nol \text{ and } \q \in \Loc(\Net)  \}. \qedhere\]
\end{definition}

\begin{example}
Let $\Net$ be  the row chip-firing network~(Example~\ref{e. arithmetical graphs}) from 
Figure~\ref{figure:row chip firing network example}.
The underlying digraph   $G$  has  two vertices $v_1$ and $v_2$,  with three edges directed from $v_1$ to $v_2$, and two edges directed from $v_2$ to $v_1$.

The production matrix and the exchange rate vector of this network are given by
\[ 
P=\begin{bmatrix}
0 & \frac{2}{3}\\
\frac{3}{2} & 0
\end{bmatrix},
\qquad 
\s=\begin{bmatrix}
3 \\
 2 
\end{bmatrix},
\]
respectively.
The state space is $Q=\Z_{2} \times \Z_3$, and the levels of the states are given by:
\begin{align*}
& \lvl\left(\begin{bmatrix}
0 \\ 0
\end{bmatrix}\right)=0, \quad 
\lvl\left(\begin{bmatrix}
0 \\ 1
\end{bmatrix}\right)=2, \quad
\lvl\left(\begin{bmatrix}
0 \\ 2
\end{bmatrix}\right)=4,\\
& \lvl\left(\begin{bmatrix}
1 \\ 0
\end{bmatrix}\right)=3, \quad 
\lvl\left(\begin{bmatrix}
1 \\ 1
\end{bmatrix}\right)=5, \quad
\lvl\left(\begin{bmatrix}
1 \\ 2
\end{bmatrix}\right)=7.
\end{align*}
The capacity of this network is then equal to $7$, and the set of stoppable levels  is  given by:
\begin{align*}
\Stop(\Net)=&  \{0,1,2,3,4,5,7\}. 
\end{align*}
(Note that 1 is a stoppable level because the configuration $\begin{bmatrix}
0 \\ -1
\end{bmatrix}. \begin{bmatrix}
1 \\ 0
\end{bmatrix}$ has level 1.)
\end{example}
  
\begin{lemma}\label{lemma: unstoppable levels}
       Let  $\Net$ be a finite, locally irreducible,  strongly connected, and critical abelian network.
       Then 
             \[ \Stop(\Net) \subseteq \{ 0,1,\ldots, \cpt(\Net) \}, \]
with equality if the exchange rate vector $\s$ has a coordinate equal to $1$.
\end{lemma}

  \begin{proof}
Let $\x.\q$ be any configuration such that $\x \leq \nol$ and $\q \in \Loc(\Net)$.
Then
\begin{align*}
\lvl(\x.\q)= \s^\top\x+ \lvl(\q) \leq \lvl(\q)\leq \cpt(\Net), 
\end{align*}  
where the last inequality is due to  Lemma~\ref{p. level and capacity  properties}\eqref{item: capacity of a state is between 0 and cpt(Net)}.
Since the choice of $\x.\q$ is arbitrary, the inequality above implies that any level greater than $\cpt(\Net)$ is unstoppable, proving the first part of the lemma.

For the second part of the lemma, note that:
  \begin{align*} 
     \Stop(\Net)
    =&\N \cap \{ \s^\top \x+ \lvl(\q) \mid \x \leq \nol \text{ and } \q \in \Loc(\Net)  \}\\
    \supseteq& \N \cap \{ \s^\top \x+ \cpt(\Net) \mid \x \leq \nol   \}  \qquad \text{(by Lemma~\ref{p. level and capacity  properties}\eqref{i. capacity minimum is achieved by locally recurrent states})}.\\
    =& \N \cap (\cpt(\Net)+\{\s^\top \x \mid \x \leq \nol\}) \\
        =& \N \cap (\cpt(\Net)+\{0,-1,-2,\ldots\})  \\
        =&\{0,\ldots, \cpt(\Net)\},
  \end{align*}
  where the second to last equality uses the hypothesis that $\s$ has a coordinate equal to $1$.
  \end{proof}

  \begin{remark}
  The condition that $\s$ has a coordinate equal to 1 is not necessary for $\Stop(\Net)$ to be equal to $\{ 0,1,\ldots, \cpt(\Net) \}$; as can be seen from the following example.
  \end{remark}
 
\begin{example}\label{example: stoppable level with no gap}
Let $G$ 
be the digraph  with  vertex set $\{v_1,v_2\}$, and with  three edges directed from $v_1$ to $v_2$, and two edges directed from $v_2$ to $v_1$.
Consider the network $\Net$ on $G$ with
the alphabet, state space, and state transition of the processor $\PP_{v}$ given by 
\begin{align*}
& A_{v}:=\{v\}, \qquad  Q_v:=\{0,1,\ldots, \indeg(v)-1\}, \qquad  T_v(i):=i+1 \quad (\text{mod }\indeg(v)).
\end{align*}
For each $v \in V$, fix a total order  $e_0^v,\ldots, e_{\outdeg(v)-1}^v$ on the outgoing edges of $v$.
The message-passing function of $\Net$ is given by:
\begin{align*}
 T_{e_j^{v_1}}(i,v_1):=&\begin{cases} v_2 & \text{ if } i= j=0; \text{ or if } i=1 \text{ and } j\in \{1,2\}; \\
\epsilon & \text{ otherwise.} \end{cases};\\
 T_{e_j^{v_2}}(i,v_2):=&\begin{cases} v_1 & \text{ if } i\in \{1,2\} \text{ and } j=i-1; \\
\epsilon & \text{ otherwise.} \end{cases}.
\end{align*}
See Figure~\ref{figure: stoppable level example} for an illustration of this process.

\begin{figure}[tb]
\centering
   \includegraphics[width=1\textwidth]{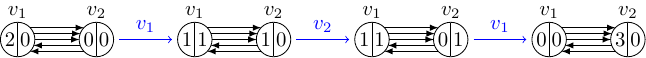}
   \caption{A three-step legal execution in the abelian network in Example~\ref{example: stoppable level with no gap}.
  In the figure,  the left part of a vertex records the number of letters waiting to be processed, and the right part records the state of the processor.}  
   \label{figure: stoppable level example}
\end{figure}

The production matrix and the exchange rate vector of $\Net$ are given by
\[ 
P=\begin{bmatrix}
0 & \frac{2}{3}\\
\frac{3}{2} & 0
\end{bmatrix},
\qquad 
\s=\begin{bmatrix}
3 \\
 2 
\end{bmatrix},
\]
respectively.
The levels of the states of $\Net$ are given by:
\begin{align*}
& \lvl\left(\begin{bmatrix}
0 \\ 0
\end{bmatrix}\right)=0, \quad 
\lvl\left(\begin{bmatrix}
0 \\ 1
\end{bmatrix}\right)=2, \quad
\lvl\left(\begin{bmatrix}
0 \\ 2
\end{bmatrix}\right)=1,\\
& \lvl\left(\begin{bmatrix}
1 \\ 0
\end{bmatrix}\right)=1, \quad 
\lvl\left(\begin{bmatrix}
1 \\ 1
\end{bmatrix}\right)=3, \quad
\lvl\left(\begin{bmatrix}
1 \\ 2
\end{bmatrix}\right)=2.
\end{align*}
The capacity of $\Net$ is then equal to $3$, and the set of stoppable levels  is  given by:
\begin{align*}
\Stop(\Net)=&  \{0,1,2,3\}. \qedhere 
\end{align*}
\end{example}

We now state the main result of this subsection, which is a refinement of Theorem~\ref{theorem: construction of torsion group for general abelian networks} for critical networks.

Recall that the torsion group $\Tor(\Net)$~(Definition~\ref{definition: torsion group}) acts on the set of invertible recurrent components $\Lrec(\Net)^\times$~(Definition~\ref{definition: invertible recurrent class}) using the action described in Definition~\ref{definition: action of the torsion group on invertible recurrent classes}.
Recall the definition of free and transitive  actions from \S\ref{subsection: construction of torsion group for all abelian networks}.
Let 
   $\Z^A_0:=\{\z \in \Z^A \mid \s^\top\z=0  \}$, and let
   $\phi: \N^A \to \End(\Lrec(\Net))$ be the monoid 
homomorphism from Definition~\ref{definition: monoid action on recurrent classes}.   

\begin{theorem}\label{theorem: torsion group for critical networks}
       Let  $\Net$ be a finite, locally irreducible,  strongly connected, and critical abelian network.
Then
\begin{enumerate}
\item \label{item: the explicit form of torsion group for critical networks} The map $\phi: \N^A \to \End(\Lrec(\Net))$  induces an isomorphism of abelian groups
\[ \Tor(\Net) \simeq \Z^A_0/(I-P)K. \]
\item \label{item: invertible rec classes split into rec classes with unstoppable levels}
$\displaystyle \Lrec(\Net)^\times=\bigsqcup_{m \in \N \setminus \Stop(\Net)} \Lrec(\Net,m)$.
\item \label{item: the action of torsion group of critical networks is free and transitive} For any $m \in \N \setminus \Stop(\Net)$, the action of the torsion group
\[\Tor(\Net) \times \Lrec(\Net,m) \to \Lrec(\Net,m) \]
is free and transitive.
\end{enumerate}
\end{theorem}
We remark that Theorem~\ref{intro theorem: torsion group for critical networks}, stated in the introduction, is a direct corollary of Theorem~\ref{theorem: torsion group for critical networks}\eqref{item: the action of torsion group of critical networks is free and transitive}.

As an application of  Theorem~\ref{theorem: torsion group for critical networks},  
we compute $(|\Lrec(\Net,m)|)_{m\geq \cpt(\Net)}$ for any  height-arrow network $\Net$.
This generalizes  \cite[Theorem 1]{Pham15},
which computes $|\Lrec(\Net,\cpt(\Net))|$  for a  sinkless rotor network $\Net$.

\begin{example}
Let $\Net$ be a locally irreducible sinkless height-arrow network~(Example~\ref{e. height-arrow}) on a strongly connected digraph $G$.
By Theorem~\ref{theorem: torsion group for critical networks}\eqref{item: the explicit form of torsion group for critical networks},
the torsion group of $\Net$ is isomorphic to
\[ \Tor(\Net)\simeq {\Z^V_0}/((\D_G-\A_G)\Z^{V}), \]
where $\D_G$ is the outdegree matrix of $G$, $\A_G$ is the adjacency matrix of $G$, and  $\Z^V_0=\{\z \in \Z^V \mid \satu^\top \z=0 \}$.
 By \cite[Theorem~2.10]{FL16},
the cardinality of $\Tor(\Net)$  is then equal to the \emph{Pham index}, 
\[\Pham(G):= \gcd_{v \in V}\{ t(G,v)\},\]
where $t(G,v)$ is the number of spanning trees of $G$ oriented toward $v$.
By Theorem~\ref{theorem: torsion group for critical networks}\eqref{item: the action of torsion group of critical networks is free and transitive}, this is also the number of recurrent components   of level $m$,
 where  $m$  is any integer greater than $\cpt(\Net)$.
\end{example}


  We now build toward  the proof of Theorem~\ref{theorem: torsion group for critical networks}, and we start with a technical lemma.

Recall the definition of the relation $\dashrlarrow$ and $\rlarrow$ (Definition~\ref{definition: weak and strong relation}).
  Also recall that   $\overline{\x.\q}$ denotes the component of the trajectory digraph (Definition~\ref{definition: component}) that contains the configuration $\x.\q$.

\begin{lemma}\label{lemma: technical lemma for invertible recurrent classes of critical networks}
       Let  $\Net$ be a finite, locally irreducible,  strongly connected, and critical abelian network.
For any $\x,\x' \in \Z^A$ and $\q,\q' \in \Loc(\Net)$,
\begin{enumerate}
\item\label{item: same level implies dashrightleftarrow} If $\lvl(\x.\q)=\lvl(\x'.\q')$, then there exist $\n,\n' \in \N^A$ such that $(\x+\n).\q \dashrightarrow (\x'+\n').\q'$ and $\s^\top\n=\s^\top \n'$. 

\item\label{item: dashrlarrow implis rightarrow if x.q is recurrent} If $\x.\q \dashrlarrow \x'.\q'$ and $\x.\q$ is a recurrent configuration,
then $\x'.\q' \longrightarrow \x.\q$.
\item \label{item: unstoppable implies nonhalting} If $\lvl(\x.\q) \in \N \setminus \Stop(\Net)$, then $\x.\q$ does not halt.
\item\label{item: nonhalting implies the class is recurrent} 
The component  $\overline{\x.\q}$ is a recurrent component if and only if $\x.\q$ does not halt.
\end{enumerate}
\end{lemma}  
  \begin{proof}
  \begin{enumerate}[wide, labelwidth=!, labelindent=10pt]
  \item By the local irreducibility of $\Net$,  there exist $w \in A^*$ and $\x'' \in \Z^A$ such that $\x.\q \xdashrightarrow{w} \x''.\q'$.
  By Lemma~\ref{p. level and capacity  properties}\eqref{i. level is equal}, we then have $\lvl(\x''.\q')=\lvl(\x.\q)=\lvl(\x'.\q')$.
  In particular, 
  we have $\s^\top(\x'-\x'')=0$.
  Let $\n$ and $\n'$ be the positive and the negative part of $\x'-\x''$, respectively.
  It  follows that $(\x+\n).\q\xdashrightarrow{w}(\x'+\n').\q'$ and 
  $\s^\top\n=\s^\top \n'$.
  
  \item Because $\x.\q \dashrlarrow \x'.\q'$,  there exist $w_1,w_2 \in A^*$ and a configuration $\y.\p$  such that 
  $\x.\q \xdashrightarrow{w_1} \y.\p$ and $\x'.\q' \xdashrightarrow{w_2} \y.\p$.
  Also note that by Lemma~\ref{l. three definitions of recurrence}\eqref{i. recurrence definition period vector} there is $w \in A^*$ such that $\x.\q \xlongrightarrow{w} \x.\q$ and $|w|=\rr$.
 
  Let $k$ be a positive number such that $k|w| \geq |w_2|$, and let $l$ be a positive number such that $l|w|\geq k|w|+|w_1|-|w_2|$. (Note that $k$ and $l$ exist because $\rr\geq \satu$.)
  Write $w':=w^l \setminus (k|w| +|w_1|-|w_2|)$.
We have
\begin{center}
\begin{tikzcd}
\x.\q \arrow[rr,"w^l"]  \arrow[rd,dashed, "w_1"]& &  \x.\q\\
\x'.\q' \arrow[r,dashed, "w_2"] & \y.\p \arrow[r,dashed, "w^k \setminus |w_2|"] & \pi_{w^k}(\x'.\q') \arrow[u,blue, "w'",swap]
\end{tikzcd},
\end{center}
  where the solid arrow ${\color{blue}\xlongrightarrow{w'}}$ is due to   the removal lemma (Lemma~\ref{lemma: removal lemma}).
Now note that since $\q'$ is locally recurrent, we have by Lemma~\ref{l. N locally irreducible}  that $\pi_{w^k}(\x'.\q')=\pi_{k\rr}(\x'.\q')=\x'.\q'$.
Hence we conclude that $\x'.\q'\xlongrightarrow{w'} \x.\q$,  as desired.

\item Let $\y.\p$ be any configuration such that $\x.\q \longrightarrow \y.\p$.
Since $\q$ is locally recurrent,
the state $\p$ is also locally recurrent  by   Lemma~\ref{lemma: the monoid $N^A$ acts invertibly on locally recurrent configurations}\eqref{item: monoid acts invertible on locally recurrent states 1}.
By Lemma~\ref{p. level and capacity  properties}\eqref{i. level is equal} we then have $\lvl(\y.\p)=\lvl(\x.\q)$.
Since $\lvl(\x.\q) \in \N \setminus \Stop(\Net)$, it then follows that $\y.\p$ is not a stable configuration.
Since the choice of $\y.\p$ is arbitrary, we then conclude that $\x.\q$ does not halt.

\item Proof of only if direction: Suppose to the contrary that $\x.\q$ halts.
Without loss of generality, we can assume that $\x.\q$ is a stable configuration (by replacing $\x.\q$ with its stabilization if necessary).

By Lemma~\ref{lemma: relation between recurrent configurations and recurrent classes}, the component $\overline{\x.\q}$ contains a recurrent configuration $\y.\p$.
Since $\x.\q \rlarrow \y.\p$ and $\y.\p$ is recurrent,
we  have  
$\x.\q \xlongrightarrow{}\y.\p$.
Since $\x.\q$ is stable, we then have $\x.\q=\y.\p$.
Hence $\x.\q$ is both stable and recurrent, which contradicts the definition of recurrence. 

Proof of if direction: Because $\x.\q$ does not halt, the component $\overline{\x.\q}$ contains a legal execution of the form:
\[ \y_0.\p  \xlongrightarrow{w_1} \y_1.\p \xlongrightarrow{w_2} \y_2.\p \xlongrightarrow{w_3} \cdots,   \]
for some $\p \in Q$, $\y_i \in \Z^A$, and nonempty words $w_{i+1} \in A^*$ ($i\geq 0$).
Note that for all $i \geq 0$ we have
\[\s^\top \y_i=\s^\top \y_0, \quad \text{ and }   \quad  \y_i(a)\geq \min(\y_0(a),0) \quad  \forall a \in A,  \]
by Lemma~\ref{l. N locally irreducible} and  Lemma \ref{o. to legal properties}\eqref{o. nonnegativity}, respectively.
This  implies that the set $\{\y_i \mid i \geq 0  \}$ is finite.
By the pigeonhole principle, there exist $j \in \N$ and $k \geq 1$  such that $\y_{j}=\y_{j+k}$.

Write $w:=w_{j+1} \cdots w_{k}$ and $\y:=\y_j=\y_{j+k}$.
It follows that $w$ is a nonempty word and $\y.\p\xlongrightarrow{w} \y.\p$.
By Lemma~\ref{l. three definitions of recurrence}\eqref{i. recurrence definition classical} the configuration $\y.\p$ is recurrent,
and then by Lemma~\ref{lemma: relation between recurrent configurations and recurrent classes} the component $\overline{\x.\q}=\overline{\y.\p}$ is a recurrent component.
\qedhere
  \end{enumerate}
  \end{proof}

We now  prove  Theorem~\ref{theorem: torsion group for critical networks}.
  
\begin{proof}[Proof of Theorem~\ref{theorem: torsion group for critical networks}]
\begin{enumerate}[wide, labelwidth=!, labelindent=10pt]
\item By Theorem~\ref{theorem: construction of torsion group for general abelian networks}\eqref{item: theorem grothendieck group is equal to ZA/(I-P)K},
it suffices to show that $\Z^A_0/(I-P)K$ is the torsion subgroup of $\Z^A/(I-P)K$.

By definition of $\Z^A_0$, the group $(I-P)K$ is a subgroup of $\Z^A_0$.
Since $K$ is a subgroup of $\Z^A$ of finite index~(Lemma~\ref{lemma: the total kernel is a subgroup of finite index}\eqref{item: the total kernel is a subgroup of finite index})  and $P$ is strongly connected,
it follows from the  Perron-Frobenius theorem (Lemma~\ref{lemma: Perron-Frobenius theorem}\eqref{item: Perron-Frobenius 4}) that 
the $\R$-span of $(I-P)K$ has dimension $|A|-1$.
Since the $\R$-span of $\Z^A_0$ also has dimension $|A|-1$,
we conclude that 
the quotient group $\Z^A_0/(I-P)K$ is finite.

Since $\gcd_{a \in A} \s(a)=1$,  there exists $\s' \in \Z^A$ such that $\s^\top \s'=1$.
Then
\[\frac{\Z^A}{(I-P)K}=\frac{\Z^A_0}{(I-P)K} \, \oplus \Z\s' \simeq \frac{\Z^A_0}{(I-P)K} \, \oplus \Z, \]
and it  follows that 
$\tau(\Grt(\Net))=\Z^A_0/(I-P)K$, as desired.

\item Proof of the $\supseteq$  direction:
Let $\C$ be any recurrent component with level in $\N \setminus \Stop(\Net)$.
By part \eqref{item: the explicit form of torsion group for critical networks} and Definition~\ref{definition: invertible recurrent class},
it suffices to show that, for any 
$\n,\n' \in \N^A$ such that $\n-\n'\in \Z^A_0$,
there exists a recurrent component $\C'$ such that 
$\phi(\n)(\C)=\phi(\n')(\C')$.

By Lemma~\ref{lemma: relation between recurrent configurations and recurrent classes}, the recurrent component $\C$ contains a recurrent configuration $\x.\q$. 
In particular, $\q$ is locally recurrent by Lemma~\ref{p. properties of recurrent configurations}\eqref{i. recurrence implies local recurrence}.
Write $\x':=\x+\n-\n'$.
Since $\n-\n' \in \Z^A_0$, it follows that $\lvl(\x'.\q)=\lvl(\x.\q)$.
In particular, we have  $\lvl(\x'.\q) \in \N \setminus \Stop(\Net)$.

By Lemma~\ref{lemma: technical lemma for invertible recurrent classes of critical networks}\eqref{item: unstoppable implies nonhalting},
we then have $\x'.\q$ is a nonhalting configuration.
By Lemma~\ref{lemma: technical lemma for invertible recurrent classes of critical networks}\eqref{item: nonhalting implies the class is recurrent},
we then have $\overline{\x'.\q}$ is a recurrent component.
The claim now follows by taking $\C':=\overline{\x'.\q}$.

Proof of the $\subseteq$  direction: Let $\x.\q$ be a recurrent configuration such that 
$\overline{\x.\q} \in \Lrec(\Net)^\times$.
It follows from Lemma~\ref{p. properties of recurrent configurations}\eqref{p. properties of recurrent configuration, x is nonzero} and Lemma~\ref{p. level and capacity  properties}\eqref{item: capacity of a state is between 0 and cpt(Net)} that $\lvl(\x.\q)\geq 0$.

 Suppose to the contrary that
$\lvl(\x.\q)$ is in  $ \Stop(\Net)$.
Then there exist $\x'\leq \nol$ and $\q'\in \Loc(\Net)$ such that $\lvl(\x.\q)=\lvl(\x'.\q')$.
By Lemma~\ref{lemma: technical lemma for invertible recurrent classes of critical networks}\eqref{item: same level implies dashrightleftarrow}, there exist $\n,\n'\in \N^A$ such that $(\x+\n).\q \dashrightarrow (\x'+\n').\q'$ and $\n-\n' \in \Z^A_0$.

Since $\overline{\x.\q}$ is an invertible recurrent component and $\n-\n' \in \Z^A_0$, by
part \eqref{item: the explicit form of torsion group for critical networks} and Definition~\ref{definition: invertible recurrent class}
there exists a recurrent configuration
 $\y.\p$  such that
$\phi(\n)(\overline{\x.\q})=\phi(\n')(\overline{\y.\p})$.
Then
\begin{align*}
 & \phi(\n)(\overline{\x.\q})=\phi(\n')(\overline{\y.\p}) \quad \text{and} \quad  (\x+\n).\q \dashrightarrow (\x'+\n').\q'\\
& \Longrightarrow \quad  (\y+\n').\p \dashrlarrow (\x'+\n').\q'\\
& \Longrightarrow \quad  \y.\p \dashrlarrow \x'.\q' \qquad  \text{(by Lemma~\ref{o. to legal properties}\eqref{item: weak arrow if and only if})}\\
& \Longrightarrow \quad  \x'.\q' \longrightarrow \y.\p \qquad \text{(by Lemma~\ref{lemma: technical lemma for invertible recurrent classes of critical networks}\eqref{item: dashrlarrow implis rightarrow if x.q is recurrent})}\\
& \Longrightarrow \quad  \x'.\q'=\y.\p \qquad \text{(since } \x'\leq \nol). 
\end{align*}
In particular we have  $\x'.\q'$ is a recurrent configuration.
However, this contradicts 
 the assumption that $\x'.\q'$ is stable, and the proof is complete.

\item It follows from part \eqref{item: the explicit form of torsion group for critical networks} that the action of  $\Tor(\Net)$ preserves the level of  invertible recurrent component it acts on.
By part \eqref{item: invertible rec classes split into rec classes with unstoppable levels}, 
it then follows that
the group $\Tor(\Net)$ acts on $\Lrec(\Net,m)$ for all $m \in\N \setminus  \Stop(\Net)$.
The freeness of the action follows from Theorem~\ref{theorem: construction of torsion group for general abelian networks}.

We now prove the transitivity of the action.
Let $m \in \N \setminus \Stop(\Net)$. 
We first show that $\Lrec(\Net,m)$ is nonempty.
Let $\q \in \Loc(\Net)$, and let $\x \in \Z^A$ such that $\s^\top \x=m-\lvl(\q)$ (Note that $\x$ exists because $\gcd_{a \in A} \s(a)=1$).
It follows that $\x.\q$ is a configuration with level $m \in \N \setminus \Stop(\Net)$.
By  Lemma~\ref{lemma: technical lemma for invertible recurrent classes of critical networks}\eqref{item: unstoppable implies nonhalting},  $\x.\q$ is a nonhalting configuration.
By Lemma~\ref{lemma: technical lemma for invertible recurrent classes of critical networks}\eqref{item: nonhalting implies the class is recurrent}, the component $\overline{\x.\q}$ is a recurrent component.
Hence  $\Lrec(\Net,m)$ is nonempty.

Let $\overline{\x'.\q'}$ be any recurrent component with level $m$.
 By Lemma~\ref{lemma: relation between recurrent configurations and recurrent classes}   we can assume that $\x'.\q'$ is a recurrent configuration. 
In particular, $\q'$ is locally recurrent by Lemma~\ref{p. properties of recurrent configurations}\eqref{i. recurrence implies local recurrence}.
By Lemma~\ref{lemma: technical lemma for invertible recurrent classes of critical networks}\eqref{item: same level implies dashrightleftarrow} there exist $\n,\n' \in \N^A$ such that $(\x+\n).\q \dashrlarrow (\x'+\n').\q'$ and $\n-\n' \in \Z^A_0$.
By Lemma~\ref{p. properties of recurrent configurations}\eqref{p. properties of recurrent configuration, adding to recurrent makes recurrent}
both $\overline{(\x+\n).\q}$ and  $\overline{(\x'+\n').\q'}$ are recurrent components.
By Proposition~\ref{proposition: weak relation implies strong relation for recurrent class},  we then conclude that $\overline{(\x+\n).\q}=\overline{(\x'+\n').\q'}$. 
Now note that
\[\phi(\n)(\overline{\x.\q})=  \overline{(\x+\n).\q}=\overline{(\x'+\n').\q'}=\phi(\n')(\overline{\x'.\q'}).\]
Since the choice of $\overline{\x'.\q'}$ is arbitrary, we conclude that the action is transitive, as desired. \qedhere  
\end{enumerate}
\end{proof}

\chapter{Critical Networks: Dynamics}\label{section: critical networks dynamics}

In this chapter we study the dynamics of critical networks in more detail, with a focus on the activity and the legal executions of a configuration.

\section{Activity as a component invariant}\label{subsection: activity}
In this section we show that the activity  of a configuration (as defined below) is a component invariant for a large family of update rules that includes the parallel update.

\begin{definition}[Update rule]\label{definition: update rule}
Let $\Net$ be an abelian network.
An \emph{update rule}  of $\Net$ is an assignment of a  word $u(\x.\q) \in A^*$
to    each configuration $\x.\q$  such that
  $u(\x.\q)$  is a legal execution for $\x.\q$.
\end{definition} 
Described in words, an update rule tells the network how to process any given input  configuration.

We refer to the word $u(\x.\q)$ assigned to $\x.\q$ as  the \emph{update word} for $\x.\q$.
     The \emph{update function} $U: \Z^A\times Q \to \Z^A\times Q$ is the function that maps a configuration $\x.\q$ to its updated configuration  $\pi_{u(\x.\q)}(\x.\q)$.  
In order to simplify the notation,
we  use $u$ instead of $u(\x.\q)$ to denote the  update word for $\x.\q$.
 For any $i\geq 1$, we use $u_{i}$ to denote the  update word for $U^{i-1}(\x.\q)$.
     The words $u'$ and $(u_i')_{i \geq 1}$ for the configuration $\x'.\q'$ are defined similarly.

Recall that, for any $w \in A^*$,
we denote by  $|w|$ the vector in $\N^A$ that counts the number of occurrences of each letter in $w$.

\begin{definition}[Activity vector]\label{definition: activity vector}
Let $\Net$ be a finite, locally irreducible, strongly connected, and critical abelian network.
The \emph{activity vector} of a configuration $\x.\q$ w.r.t. a given update rule $u$ is
\[ \act_u(\x.\q)=\lim_{n \to \infty} \frac{1}{n}\sum_{i=1}^{n} |u_{i}|.  \qedhere \]
\end{definition}  
Described in words, the activity vector records the average number of times a letter is  processed when  $\x.\q$ is the input configuration.

Note that the limit in Definition~\ref{definition: activity vector} exists and is finite.
This is because the sequence $\x.\q, U(\x.\q), U^2(\x.\q), \ldots$ is eventually periodic (as $\{U^i(\x.\q)\}_{i \geq 0}$ is finite by criticality).

We are mainly interested in update rules that satisfy 
 these two properties:
\begin{enumerate}[{
label=\textnormal{({H\arabic*})},
ref=\textnormal{H\arabic*}}]
\item \label{item: hypothesis nonhalting} For any configuration $\x.\q$ such that $\x\in \N^A \setminus \{ \nol\}$, the  update word $u$ for $\x.\q$ is a nonempty word.
\item \label{item: hypothesis monotonicity} {For any $a\in A$ and  any configurations $\x.\q$ and $\x'.\q'$ such that $\x.\q \xlongrightarrow{a} \x'.\q'$, the  update words $u$ for $\x.\q$ and $u'$ for  $\x'.\q'$ satisfy $|u|\leq |a|+ |u'|$.}
\end{enumerate}

The following are several   examples of update rules on the sinkless sandpile network~(Example~\ref{e. sandpile network}) that satisfy \eqref{item: hypothesis nonhalting} and \eqref{item: hypothesis monotonicity}.
\begin{example}
[Parallel update~\cite{BLS91, BG92}]\label{example:  parallel update}
 The \emph{parallel update} on the sinkless sandpile network is the rule where every unstable vertex (i.e. $v \in V$ such that $\x(v)+\q(v)\geq \outdeg(v)$) of the input configuration is fired  once (i.e. sends one chip along every outgoing edge).
 Described formally,
the  update word $u$ for $\x.\q$ is a word that satisfies
 \[ |u|(v)=\min \{\x(v),\outdeg(v) \} \qquad (v \in V).  \]
See Figure~\ref{figure:parallel update} for an illustration of this update rule.

\begin{figure}[tb]
\centering
   \includegraphics[width=1\textwidth]{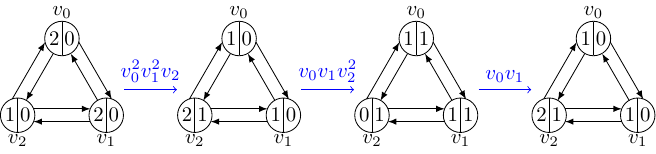}
   \caption{A three-step parallel update in the sinkless sandpile network on the bidirected cycle $C_3$.
  In the figure,  the left part of a vertex records the number of letters waiting to be processed, and the right part records the state of the processor.
Note that these configurations has activity $(1,1,1)^\top$,
as the last two steps of this update form a periodic two-step update   where every letter is fired twice.    
  }  
   \label{figure:parallel update}
\end{figure}

The parallel update  satisfies \eqref{item: hypothesis nonhalting} by definition, and satisfies \eqref{item: hypothesis monotonicity} by the following computation.
Let $\db\in \Z^V$ be given by $\db(v):=\outdeg(v)$ ($v \in V$).
Then for any $v \in V$ and any configuration $\x.\q$ and $\x'.\q'$ such that $\x.\q \xlongrightarrow{v} \x'.\q'$,
\begin{align*}
|v|+|u'|=& |v|+ \min \{ \x', \db\}=|v|+ \min \{ \x+P|v|-|v|, \db\}\\
\geq&  |v|+ \min \{ \x-|v|, \db\} 
\geq  \min \{ \x, \db\} \\
=&|u|.
\end{align*}

We remark that a variant of the parallel update rule where a vertex is being fired until it is stable (i.e.,  $|u|(v)=\x(v)$ for all $v \in V$)
also satisfies \eqref{item: hypothesis nonhalting} and \eqref{item: hypothesis monotonicity}.
\end{example}

\begin{example}[Sequential update]\label{example: sequential update}
Fix a total order  $v_0,\ldots, v_{n-1}$ on the vertices of $G$.
The \emph{sequential update} on the sinkless sandpile network is the rule where the vertices $v_0,\ldots, v_{n-1}$ are checked in this order,
and each of them is fired once during the checking process if it is found to be unstable.
Described formally,
the  update word $u=v_0^{k_0}v_1^{k_1}\ldots v_{n-1}^{k_{n-1}}$ for $\x.\q$ satisfies:
\begin{align*}
k_i:=& \min \{ \x_{i-1}(v), \outdeg(v)\} \qquad (i \in \{0,\ldots,n-1\}),
\end{align*}
where $\x_i.\q_i$ is the configuration $\pi_{k_0|v_0|+\ldots +k_{i}|v_i|}(\x.\q)$.
See Figure~\ref{figure:sequential update} for an illustration of this update rule.

\begin{figure}[tb]
\centering
   \includegraphics[width=1\textwidth]{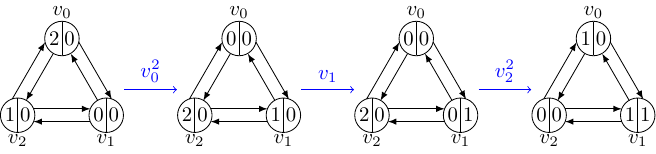}
   \caption{A breakdown of one-step  sequential update in the sinkless sandpile network on the bidirected cycle $C_3$.
  Note that vertex $v_2$ is fired (i.e., sending chips to its neighbor) even though it is initially stable (i.e., has less chips than its outgoing edge).  
  }  
   \label{figure:sequential update}
\end{figure}

The sequential update satisfies \eqref{item: hypothesis nonhalting} by definition, and satisfies~\eqref{item: hypothesis monotonicity} by a computation similar to Example~\ref{example:  parallel update}.

Unlike  the parallel update, here  a vertex  can potentially be fired even if the vertex is  stable in the input configuration.
This is because the vertex might acquire additional chips from other vertices that are checked before it; see Figure~\ref{figure:sequential update}.

We remark that a mix of the parallel update and the sequential update on a partition $V_0 \cup \ldots \cup V_{k-1}$ of $V$ (i.e.,  check $V_0,\ldots V_{k-1}$ in that order, and then apply the parallel update on  $V_i$ when it is being checked) also satisfies \eqref{item: hypothesis nonhalting} and \eqref{item: hypothesis monotonicity}. 
\end{example}

\begin{example}[Savings update]
Fix a nonempty subset  $S \subseteq V$.
The \emph{savings update} works as follow:
\begin{itemize}
\item If there exists an  unstable  vertex in $V \setminus S$,
 then apply the parallel update on $V \setminus S$.
 \item Otherwise, apply the parallel update  on $S$.
\end{itemize} 
Described in words, the vertices in $S$ are acting  as  saving accounts that are used only when all other accounts are running out of funds.
See Figure~\ref{figure:savings update} for an illustration of this update rule.

\begin{figure}[tb]
\centering
   \includegraphics[width=1\textwidth]{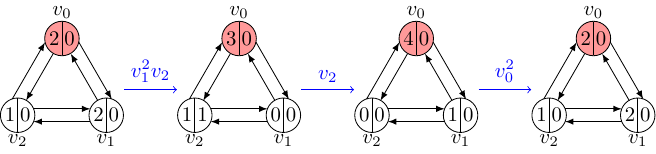}
   \caption{A three-step  savings update in the sinkless sandpile network on the bidirected cycle $C_3$, with $v_0$ as the distinguished vertex.
    Note $v_0$ is not fired in the first step even though it is unstable.
Also note that these configurations 
  has activity $(\frac{2}{3},\frac{2}{3},\frac{2}{3})^\top$,
  as every letter is fired twice in this (periodic) three-step  update.   
    }  
   \label{figure:savings update}
\end{figure}

Unlike the parallel and sequential updates, here it is possible for a vertex in $S$ to not fire even if it is unstable (i.e., when there exists another unstable vertex in $V \setminus S$), as can be seen from Figure~\ref{figure:savings update}. 

The savings update rule satisfies \eqref{item: hypothesis nonhalting} by definition,
 and satisfies \eqref{item: hypothesis monotonicity} when $S=\{v\}$
by the following argument:
Let $v \in V$ and let $\x.\q$ and $\x'.\q'$ be  configurations such that $\x.\q \xlongrightarrow{v}\x'.\q'$.
There are three possible scenarios:
\begin{itemize}
\item All vertices are stable in $\x.\q$.
In this scenario, no vertices are fired during the update of $\x.\q$ and $\x'.\q'$, and  
\eqref{item: hypothesis monotonicity} is vacuously true.

\item $V \setminus \{v\}$ is unstable in  $\x.\q$. 
In this scenario, \eqref{item: hypothesis monotonicity} can be verified by  the same computation  in Example~\ref{example:  parallel update}.

\item $V \setminus \{v\}$ is stable, and $v$ is unstable in  $\x.\q$.
In this scenario, the vertex $v$ is fired during the update of $\x.\q$.
Now note that, by the savings update rule, 
 either  $v$ is fired during the update of $\x'.\q'$, or $v$ is alredy fired during  the transition from $\x.\q$ to $\x'.\q'$.
 In either case, the inequality in  \eqref{item: hypothesis monotonicity} holds.
\end{itemize}

We would like to warn the reader that \eqref{item: hypothesis monotonicity} is not satisfied when $|S|\geq 2$; see Figure~\ref{figure: savings update counterexample}. 
\begin{figure}[tb]
\centering
   \includegraphics[width=0.5\textwidth]{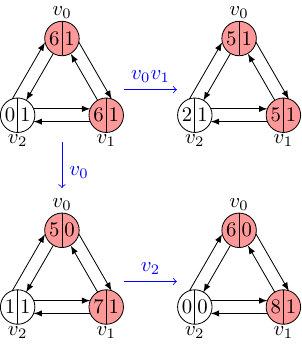}
   \caption{The horizontal arrows are savings updates in the sinkless sandpile network on the bidirected cycle $C_3$, with $S=\{v_0,v_1\}$.
The update word $u$ for the top-left configuration     is $v_0v_1$,
and the update word $u'$ for the bottom-left configuration is $v_2$.
The bottom-left configuration can be reached from the top-left configuration by executing the letter $v_0$.
 Note that $|u|=(1,1,0)^\top$ and $|v_0|+|u'|=(1,0,1)^\top$, so the inequality in \eqref{item: hypothesis monotonicity} is not satisfied.}  
   \label{figure: savings update counterexample}
\end{figure}
\end{example}
We remark that changing the update rule will usually result in changing the activity vector; see Example~\ref{figure:parallel update} and Example~\ref{figure:savings update}.

We now present the main result of this section.
Recall the definition of the relation $\rlarrow$  from Definition~\ref{definition: weak and strong relation}.

\begin{figure}[tb]
\centering
   \includegraphics[width=0.7\textwidth]{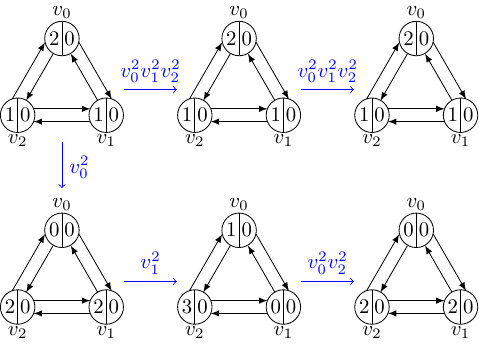}
   \caption{The horizontal arrows are  update rules  in the sinkless sandpile network on the bidirected cycle $C_3$.
The update word $u$ for the top-left configuration   is $v_0^2v_1^2v_2^2$,
the update word $u'$ for  the bottom-left configuration  is $v_1^2$, and the update word for the bottom-middle configuration is $v_0^2v_2^2$.
The bottom-left configuration can be reached from the top-left configuration by executing the letter $v_0^2$, and yet the former has activity  $(1,1,1)^\top$ while the latter has  activity  $(2,2,2)^\top$.
 Note that $|u|=(2,2,2)^\top$ and $|v_0^2|+|u'|=(2,2,0)^\top$, so  \eqref{item: hypothesis monotonicity} is not satisfied.}  
   \label{figure: activity invariance counterexample}
\end{figure}

\begin{proposition}\label{proposition: activity vector}
Let $\Net$ be a finite, locally irreducible, strongly connected, and critical abelian network.
 If the given update rule $u$ on $\Net$ satisfies \eqref{item: hypothesis nonhalting} and  \eqref{item: hypothesis monotonicity},
then $\x.\q \rlarrow \x.\q'$ implies $\act_u(\x.\q) = \act_u(\x'.\q')$.
\end{proposition}
Note that the conclusion of Proposition~\ref{proposition: activity vector} can fail when  the hypotheses are not satisfied;
see Figure~\ref{figure: activity invariance counterexample}.

We now build towards  the proof of Proposition~\ref{proposition: activity vector}.
We start with the following  lemma that extends the conclusion in \eqref{item: hypothesis monotonicity} from letters to words.

\begin{lemma}\label{lemma: stronger monotonicity hypothesis}
Let $\Net$ be an  abelian network.
If the given  update rule on $\Net$  satisfies   \eqref{item: hypothesis monotonicity},
then for any $w\in A^*$ and any  $\x.\q$ and $\x'.\q'$ such that $\x.\q \xlongrightarrow{w}\x'.\q'$,
we have 
\[ |u|\leq |w|+|u'|. \] 
\end{lemma}
\begin{proof}
Write $w=a_1\ldots a_\ell$.
 Let $\x_j.\q_j:=\pi_{a_1\ldots a_j}(\x.\q)$ $(j\in \{0,\ldots, \ell\})$,
 and let $w_{j+1}$ be the  update word for $\x_j.\q_j$.
 Then by \eqref{item: hypothesis monotonicity},
 \begin{align*}
 |u|&=|w_1|\leq |a_1|+|w_2|\leq |a_1|+|a_2|+|w_3| \leq \ldots  \\
 &\leq |a_1|+\ldots +|a_\ell|+ |w_{\ell+1}|= |w|+|u'|.
 \end{align*}
This proves the lemma.
\end{proof}

We will use the following technical lemma in the proof of Proposition~\ref{proposition: activity vector}.
Recall  the definition of $w \setminus \n$ ($w \in A^*, \n \in \N^A$) from Definition~\ref{definition: removal}.

\begin{lemma}\label{lemma: activity commutative diagram}
Let $\Net$ be an abelian network, and with 
 a given  update rule  that satisfies   \eqref{item: hypothesis monotonicity}.
Let $w \in A^*$ and let $\x.\q$ and $\x'.\q'$
 be configurations such that $\x.\q \xlongrightarrow{w}\x'.\q'$.
 Then  we have the following commutative diagram:
 \begin{center}
 \begin{tikzcd}
\x.\q \arrow[r,"u_1"]  \arrow[d, "w_0"]& U(\x.\q) \arrow[r,"u_2"]  \arrow[d, "w_1"] &  U^2(\x.\q) \arrow[d, "w_2"] \arrow[r,"u_3"] & \cdots \\
\x'.\q' \arrow[r, "u_1'"] & U(\x'.\q') \arrow[r, "u_2'"] & U^2(\x'.\q')  \arrow[r,"u'_3"] & \cdots
\end{tikzcd},
\end{center}
where $w_i$ is given by:
 \[w_i:=\begin{cases} w & \text{if } i=0;\\
w_{i-1}u_{i}'\setminus |u_{i}| & \text{if } i\geq 1.\end{cases}  \]
\end{lemma}
\begin{proof}
It suffices to show that $U^i(\x.\q)\xlongrightarrow{w_i} U^i(\x'.\q')$ for all $i\geq 0$.
We will prove this claim by induction on $i$.
The base case $i=0$ holds since $\x.\q \xlongrightarrow{w}\x'.\q'$ by assumption.
Now assume that $U^{i}(\x.\q)\xlongrightarrow{w_{i}} U^{i}(\x'.\q')$.
By Lemma~\ref{lemma: stronger monotonicity hypothesis}, we have  
$|u_{i+1}|\leq |w_{i}|+|u_{i+1}'|$.
By the removal lemma (Lemma~\ref{lemma: removal lemma}), 
we then have 
$U^{i+1}(\x.\q)\xlongrightarrow{w_{i+1}} U^{i+1}(\x'.\q')$, as desired.
\end{proof}

We now present the proof of Proposition~\ref{proposition: activity vector}.
\begin{proof}[Proof of Proposition~\ref{proposition: activity vector}]
Let $\x.\q$ and $\x'.\q'$ be any two configurations in the same component of the trajectory digraph of $\Net$.
Note that the  infinite sequence 
\begin{equation}\label{equation: periodic}
\x'.\q' \xlongrightarrow{u_1'} U(\x'.\q') \xlongrightarrow{u_2'} U^2(\x'.\q') \xlongrightarrow{u_3'} \ldots
\end{equation}
is eventually periodic since the set $\{U^i(\x'.\q') \mid i\geq 0 \}$ is finite (as $\Net$ is a critical network).
Also note that $\x'.\q'$ and $U^i(\x'.\q')$ have the same activity vector  by Definition~\ref{definition: activity vector}.
Hence (by replacing $\x'.\q'$ with $U^i(\x'.\q')$ for sufficiently large $i$ if necessary) we can without loss of generality assume that the sequence in equation~\eqref{equation: periodic} is periodic.

Note that 
by \eqref{item: hypothesis nonhalting},
we have either   $\x'\leq \nol$   or  the update word $u_0'$ for $\x'.\q'$ is nonempty.
In the former scenario, we have $\x.\q \longrightarrow \x'.\q'$ by Definition~\ref{definition: weak and strong relation} (since the empty word is the only legal execution for $\x'.\q'$).
In the latter scenario, we have $\x'.\q'$ is a recurrent configuration by
Lemma~\ref{l. three definitions of recurrence}\eqref{i. recurrence definition classical} (as a consequence of equation~\eqref{equation: periodic} being a periodic sequence).
The recurrence of $\x'.\q'$ then implies that $\x.\q \longrightarrow \x'.\q'$ by Definition~\ref{definition: recurrent configuration}.
In both  scenarios, we have $\x.\q \longrightarrow \x'.\q'$.

We now apply Lemma~\ref{lemma: activity commutative diagram} to
$\x.\q \longrightarrow \x'.\q'$,
and let  
 $w_0,w_1,w_2,\ldots \in A^*$ be words  from Lemma~\ref{lemma: activity commutative diagram}.
Note that, for any $i\geq 1$, we have $|u_{i}|\leq |w_{i-1}|+|u_{i}'|$ by Lemma~\ref{lemma: stronger monotonicity hypothesis}.
This  implies that, for any $i\geq 1$ 
\[|w_{i}|=|w_{i-1}u_i' \setminus |u_i||=|w_{i-1}|+|u_{i}'|- |u_{i}|.\] 
Hence, for any $n\geq 0$,
\begin{align*}
 \sum_{i=1}^{n} |u_i|
 =&\sum_{i=1}^{n} \left(|w_{i-1}| +|u_{i}'|- |w_{i}| \right)     \qquad \text{(by Lemma~\ref{lemma: activity commutative diagram})}\\
=& |w_0|-|w_n|+  \sum_{i=1}^{n} |u_i'| \\
\leq& |w_0|+  \sum_{i=1}^{n} |u_i'|.
\end{align*}
Since the equation above holds for all $n\geq 0$,
it then follows from Definition~\ref{definition: activity vector}  that $\act_u(\x.\q) \leq \act_u(\x'.\q')$.
By symmetry we then conclude that $\act_u(\x.\q)=\act_u(\x'.\q')$, as desired.
\end{proof}

\section[Near uniqueness of legal executions]{Near uniqueness of legal executions}
  \label{subsection: confluence}
 
In this section we estimate the proportion of any letter in a legal execution, up to an additive constant. 
 
We assume throughout this section that  $\Net$ is a finite, locally irreducible, and strongly connected critical network.
 
Let $p(\cdot,\cdot)$ be the $A \times A$ matrix given by 
  \[ p(a,b):=\frac{\s(b)}{\s(a)}P(b,a),  \]
 where $P$ is the production matrix   (Definition~\ref{definition: production matrix}) and $\s$ is the 
   exchange rate vector  of $\Net$ (i.e. the unique positive integer vector for which $\s P=\s$ and $\gcd_{a \in A} \s(a)=1$).
 Since $P$ is a nonnegative matrix, and $\s P=\s$ by the assumption that $\Net$ is critical,
 it follows that  $p(\cdot, \cdot)$
 is a probability transition matrix for a Markov chain on $A$.

  For letters $a,b,z \in A$ ,
  let $\Green_{z}(b, a)$ be the expected number of visits to $a$ strictly before hitting $z$, when the Markov chain starts at $b$.
  Let $\vb_{a,z} \in \R^A_{\geq 0}$ be the vector 
  \[ \vb_{a,z}(\cdot):=\frac{\s(\cdot)}{\s(a)} \Green_{z}(\cdot, a).  \]
    
In the special case that $\Net$ is a sandpile or rotor network on an undirected graph, the above quantities have familiar interpretations in terms of random walk and electrical networks (see, for example, \cite[chapter 2]{LP16}): 
$\s = \satu$ and $p$ is the transition matrix for simple random walk, 
    $\mathfrak{G}_{z}$ is the \emph{Green function} for the random walk absorbed at $z$,
    $\vb_{a,z}$ is the \emph{voltage function} for the unit current flow from $a$ to $z$,
    and the quantity $\frac{\vb_{a,z}(a)}{\deg(a)}$ is the \emph{effective resistance} 
   $R_{\text{eff}}(a,z)$ between $a$ and $z$.

Recall that  $\NN_{w}(\q) \in \N^A$ is the vector  that records  numbers of letters generated  by  executing $w$  at state $\q$.
  For any $\q,\q' \in \Loc(\Net)$, let $\diff_{a,z}(\q,\q')$ be given by
  \[ \diff_{a,z}(\q,\q'):=  \vb_{a,z}^\top (P|w|-\NN_{w}(\q)), \]
  where $w$ is  any (not necessarily legal) execution that 
sends $\q$ to $\q'$.
Note that $w$ exists because $\Net$ is locally irreducible and finite,
and also note that $P|w|-\NN_{w}(\q)$ does not depend on the choice of $w$ by Lemma~\ref{l. N locally irreducible}.

We now present the main result of this section.
Recall that $\rr$ is the period vector of $\Net$ (Definition~\ref{definition: period vector}), and $\satu$ is the vector $(1,\ldots,1)^\top$.
For any  $\n \in \N^A$, we denote by $||\n||$ the sum $\sum_{a \in A} \n(a)$.

  \begin{theorem}\label{theorem: confluence bound}
Let $\Net$ be a finite, locally irreducible, strongly connected, and critical network, and let $\q,\q' \in \Loc(\Net)$.
Then for any legal execution $w$ that sends $\x.\q$ to $\x'.\q'$,
\[  - \frac{||\mathbf{c}||}{||\rr||}\rr(a) -  \rr(a)  < |w|(a)-\frac{\ell}{||\rr||}\rr(a)< \rr(a) +\mathbf{c}(a)  \qquad \forall a \in A.
 \]
 where $\ell$ is the length of the execution $w$, and $\mathbf{c} \in \R^A$ is the vector given by 
 \[ \mathbf{c}(a):=  \max_{z \in A} \left( {\vb_{a,z}^\top (\x-\x') + \diff_{a,z}(\q',\q)} \right).\]
  \end{theorem}

Note that the vector $\mathbf{c}$ can be upper bounded by a positive vector that depends only on $\x.\q$ (as $\x'$ is lower bounded by the negative part of $\x$  by Lemma~\ref{o. to legal properties}\eqref{o. nonnegativity}, and there are only finitely many choices for $\q'$).
In particular, Theorem~\ref{theorem: confluence bound} implies that all legal executions of a configuration  of a given length  are equal up to permutation and  an additive constant that does not depend on the executions.

We now build towards the proof of Theorem~\ref{theorem: confluence bound}.
We will start with the following  lemma
relating $|w|(a)$ and $|w|(z)$.

\begin{lemma}\label{lemma: resistance bound}
Let $\Net$ be a finite,locally irreducible, strongly connected, and critical network, and let $\q,\q' \in \Loc(\Net)$.
Then for   any $a,z \in A$ and any legal execution $w$ sending $\x.\q$ to $\x'.\q'$,
we have:
\[ |w|(a)=  {\vb_{a,z}^\top (\x-\x') + \diff_{a,z}(\q', \q)} +\frac{\rr(a)}{\rr(z)} |w|(z).\]
\end{lemma}

\begin{proof}
Note that, if $a=z$, then the lemma follows immediately from the fact that $\vb_{a,a}$ is the zero vector.
Therefore, it suffices to prove the lemma for when $a$
 is not equal to $z$.
 
By a direct computation, we have
\begin{equation}
  (I-P^\top)\vb_{a,z}(b)=\begin{cases} 1 & \text{ if } b=a;\\
  -\frac{\rr(a)}{\rr(z)} &\text{ if } b=z;\\
0 & \text{ if } b \in A \setminus \{a,z\}.  \end{cases}
\end{equation}
In particular, this implies that 
\begin{align}\label{equation: voltage 3}
\vb_{a,z}^\top (I-P)|w|=|w|(a)-\frac{\rr(a)}{\rr(z)}|w|(z).
\end{align}

Let $w'$ be a word such that $t_{w'}(\q')=\q$. 
Note that we have $\pi_{ww'}(\x.\q)=\pi_{w'}(\x'.\q')=(\x'+\NN_{w'}(\q')-|w'|).\q$.
By Lemma~\ref{l. N locally irreducible}, we then have
\begin{align}\label{equation: voltage 1}
(I-P)(|w|+|w'|)=&\x-(\x'+\NN_{w'}(\q')-|w'|) \notag,
\end{align}
which is equivalent to 
\[
(I-P)|w|=(\x-\x')+ (P|w'|-\NN_{w'}(\q')).\]
Together with equation~\eqref{equation: voltage 3}, this implies  that:
\begin{align*}
 |w|(a)-\frac{\rr(a)}{\rr(z)}|w|(z) &=  \vb_{a,b}^\top(\x-\x')+ \diff_{a,b}(\q',\q).
\end{align*}
This proves the lemma.
\end{proof}

\begin{remark}
Lemma~\ref{lemma: resistance bound} implies the following inequality
from \cite[Proposition~4.8]{HLM08}: If $\Net$ is the sandpile network on an undirected graph and $\x.\q$ is a configuration such that $\x\geq \nol$ and $\q=(0,\ldots,0)^\top$, then any legal execution $w$ for $\x.\q$ that does not contain the letter $z$ satisfies
\begin{equation}\label{equation: inequality resistance}
 \ell \leq 2 |E| \,  ||\x||  \max_{a \in A}    R_{\text{eff}}(a,z),
 \end{equation}
where $\ell$ is the length of the execution $w$.
Indeed, this is because for all $a \in A$:
 \begin{align*}
 \begin{split}
 |w|(a)=& {\vb_{a,z}^\top (\x-\x') + \diff_{a,z}(\q', \q)} \qquad \text{(by Lemma~\ref{lemma: resistance bound})}\\
 \leq&  \vb_{a,z}^\top (\x-\x')  \qquad \text{(since $\diff_{a,z}(\q', \q)\leq 0$ if $\q=(0,\ldots,0)$)}\\
 \leq& \vb_{a,z}^\top \x \qquad \text{(since $\x'\geq \nol$ if $w$ is legal)} \\ \leq &{\vb_{a,z}(a) ||\x||}  \qquad \text{(since $\vb_{a,z}(b)\leq \vb_{a,z}(a)$ for all $b \in A$)}\\
 =&   \deg(a)  R_{\text{eff}}(a,z)  ||\x||.
 \end{split}
 \end{align*}
Equation~\eqref{equation: inequality resistance} now follows by 
summing the inequality $|w|(a)\leq \deg(a)  R_{\text{eff}}(a,z)  ||\x||$ over all letters in $A$.
\end{remark}

We now present the proof of Theorem~\ref{theorem: confluence bound}.
\begin{proof}[Proof of Theorem~\ref{theorem: confluence bound}]
Let $k$ be the largest nonnegative integer such that $k\rr \leq |w|$. 
Write $w':=w \setminus k\rr$.
Note that $w'$ is a legal execution for $\x.\q$ by the removal lemma (Lemma~\ref{lemma: removal lemma}).
Also note that, 
by the maximality assumption, there exists $z \in A$ such that $|w'|(z)<\rr(z)$.
By Lemma~\ref{lemma: resistance bound}, we then have for all $a \in A$: 
\begin{equation*}
|w'|(a)< {\vb_{a,z}^\top (\x-\x') + \diff_{a,z}(\q',\q)} +\rr(a)\leq \mathbf{c}(a)+\rr(a).
\end{equation*}
This implies that, for all $a \in A$,
\begin{align}\label{equation: bound 1}
k \rr(a) \leq |w|(a)< (k+1)\rr(a) +\mathbf{c}(a).
\end{align}
Summing
 equation~\eqref{equation: bound 1} over all letters in $A$, 
we get:
\[ k||\rr| \leq \ell < (k+1)||\rr||+||\mathbf{c}||, \]
which implies that
\begin{align}\label{equation: bound 2}
  \frac{\ell}{||\rr||}-\frac{||\mathbf{c}||}{||\rr||} -1   < k\leq  \frac{\ell}{||\rr||}. 
\end{align}
The proposition now follows from equation~\eqref{equation: bound 1} and \eqref{equation: bound 2}.
\end{proof}

\chapter{Rotor and Agent Networks}\label{s. abelian mobile agents}
An \emph{abelian mobile agent network}~\cite[Example 3.7]{BL16a}, or \emph{agent network} for short, 
is an  abelian network in which every processor $\PP_v$ produces one letter of output for each letter of input. 
Formally, an agent network is an abelian network such that for all $a \in A$ and $\q \in Q$ we have  $\satu^\top\NN_a(\q)=1$   (Recall that $\NN_a(\q) \in \N^A$ is the vector   recording the number of letters of each type that are produced when  the network in state $\q$ processes the letter $a$).

Examples of agent networks include  sinkless rotor networks~(Example \ref{e. rotor network}) and  inverse networks~(Example \ref{e. inverse}), while non-examples include  sinkless sandpile networks~(Example \ref{e. sandpile network}) and  arithmetical networks~(Example~\ref{e. arithmetical graphs}).

Any agent network is a critical network.
Indeed, by the definition of agent networks,
for any $\q \in Q$ and any $w \in A^*$, 
\[
\satu^\top \NN_{w}(\q)= \sum_{a \in A} |w|(a)=\satu^\top|w|,
\]
where $|w|\in \N^A$ is the vector  that counts the number of occurrences of each letter in $w$.  
This implies that the production matrix $P$ satisfies
\begin{equation}\label{equation: satu is the left eigenvector of the production matrix of agent networks}
 \satu ^\top P =\satu. 
 \end{equation}
By  the  Perron-Frobenius theorem (Lemma~\ref{lemma: Perron-Frobenius theorem}\eqref{item: Perron-Frobenius 2}), the spectral radius $\lambda(P)$ is equal to 1.
Hence an agent network  is a critical network.

We  assume throughout this chapter  that the agent network  we are working with is finite, locally irreducible, and strongly connected, unless stated otherwise.

Special to  agent networks is the notion of rotor digraph.
\begin{definition}[Rotor digraph]\label{definition: rotor digraph}
Let $\Net$ be an agent network.
For  $\q \in \Loc(\Net)$,
the  \emph{rotor digraph} $\varrho_\q$ is the digraph 
\[V(\varrho_\q):=A, \qquad E(\varrho_\q):= \{ (a,a_\q) \ | \ a \in A\}, \]
where $a_\q$ is the letter produced when the network $\Net$ in state $t_a^{-1}(\q)$ processes the letter $a$.
\end{definition}

Rotor digraphs belong to a special family of digraphs called cycle-rooted forests, defined as follows.
A \emph{cycle-rooted tree} is the disjoint union of a directed tree rooted at a vertex $r$ and an edge with source vertex $r$.
 Note that a cycle-rooted tree contains a unique directed cycle, and for every vertex $v$ in the digraph there is a  directed path from $v$ to the cycle.
A \emph{cycle-rooted forest} is a  disjoint union of cycle-rooted trees.
 Equivalently, a cycle-rooted forest is a digraph 
 in which every vertex has outdegree equal to 1.

The following are two examples of rotor digraphs.  
 
\begin{example}\label{e. rotor for rotor networks}
Consider the sinkless rotor network~(Example \ref{e. rotor network}) on the bidirected cycle $C_4$.

Let $\q \in \Pi_{k \in \Z_4} \Out(v_k)$ be the state given by
\[\q(k):=(v_k,v_{k+1})  \qquad (k \in \Z_4).  \]
See Figure~\ref{figure:rotor_digraph_example} for an illustration.

\begin{figure}[tb]
\centering
   \includegraphics[width=0.7\textwidth]{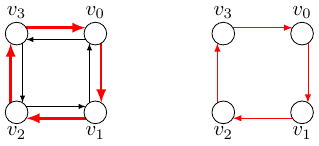}
   \caption{The figure on the left is the  state $\q:=((v_k,v_{k+1}))_{k \in \Z_4}$ (given by the (red) thick edges) of a sinkless rotor network, and the figure on the right is the  rotor digraph  of $\q$.
    }  
   \label{figure:rotor_digraph_example}
\end{figure}

On processing the letter $v_k$, the state $T_{v_k}^{-1}((v_k,v_{k+1}))=(v_k,v_{k-1})$  produces the letter $v_{k+1}$,
and therefore  the rotor digraph $\varrho_\q$ contains the edge $(v_k,v_{k+1})$.
This gives us the rotor digraph $\varrho_\q$   in Figure~\ref{figure:rotor_digraph_example}. 

By a similar reasoning,
for a sinkless rotor network on an arbitrary digraph $G$,
the rotor digraph $\varrho_\q$ of any state $\q$ is given by
\[V(\varrho_\q)=V(G), \qquad E(\varrho_q)=\{ \q(v) \mid v \in V(G)  \}.  \]
In particular,  if $G$ is a simple digraph, then the state $\q$ is  determined by its rotor digraph $\varrho_\q$.
This is not true for arbitrary agent networks, as shown in the next example. 
\end{example}

 \begin{example}\label{e. rotor for inverse networks}
Consider the inverse network~(Example~\ref{e. inverse}) on the bidirected cycle $C_3$
with  period $m_{v_k}=6$ for all $v_k \in V$ 
and with the message-passing function in Table~\ref{table: messsage passing function for inverse network to compute the rotor}.
{    \renewcommand{\arraystretch}{1.2}
\begin{table}[tb]
\caption{The message-passing function for the processor $\PP_{v_k}$ ($k \in \Z_3$).
The $(q,\alpha)$-th entry of the table represents the letter produced when a processor in state $q$ processes  the letter $\alpha$.
  }
{\small
\begin{tabular}{|c|c |c |c|c|c|c|}
\hline
 \backslashbox{$A_{v_k}$ }{$Q_{v_k}$}  & 0 & 1 &2 &3 &4 &5\\\hline
$a_{v_k}$  & $a_{v_{k+1}}$ & $a_{v_{k+1}}$  & $a_{v_{k+1}}$ & $a_{v_{k+1}}$ & $a_{v_{k+1}}$ & $b_{v_{k+1}}$    \\
\hline 
$b_{v_k}$  & $a_{v_{k+1}}$ & $b_{v_{k+1}}$ & $b_{v_{k+1}}$ & $b_{v_{k+1}}$ & $b_{v_{k+1}}$ & $b_{v_{k+1}}$   \\
\hline 
\end{tabular}
}
  \label{table: messsage passing function for inverse network to compute the rotor}
\end{table}
}

\begin{figure}[tb]
\centering
   \includegraphics[width=0.5\textwidth]{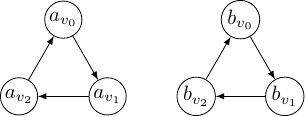}
   \caption{For the inverse network on the bidirected cycle $C_3$, 
   the rotor digraph of the state  $\q:=(1,1,1)$ is a disjoint union of two directed triangles.
Note that the state    $\q':=(2,2,2)$ has the same rotor digraph.
    }  
   \label{figure:rotor digraph inverse network}
\end{figure}

The states $\q:=(1,1,1)$ and $\q':=(2,2,2)$ have the same rotor digraph,
as shown in Figure~\ref{figure:rotor digraph inverse network}. 
However,  on processing the   input $b_{v_0}b_{v_0}$,
\begin{itemize}
\item The network at state $\q$ produces  $b_{v_1}a_{v_1}$ as output; while
\item The network at state $\q'$ produces  $b_{v_1}b_{v_1}$ as output.
\end{itemize}
Hence  a state is not determined by its rotor digraph in this inverse network.
 \end{example}
 
This chapter  is structured as follows.
In \S\ref{ss. cycle test} we derive an efficient recurrence test for agent networks.
In \S\ref{subsection: counting recurrent classes and configurations of agent networks} and \S\ref{subsection: determinantal formula recurrent configurations} we apply the methods developed in \S\ref{ss. abelian network with thief} to count
  the recurrent components and recurrent configurations of an agent network, respectively.

\section[Cycle test]{The cycle test for recurrence}\label{ss. cycle test}
In this section we present a recurrence test for agent networks that is more efficient than the burning test in \S\ref{ss. recurrent configurations and burning test}.

A \emph{directed walk} in the rotor digraph $\varrho_\q$ 
is a sequence $a_1,\ldots, a_{\ell+1} \in A^*$  such that $(a_i,a_{i+1}) \in E(\varrho_\q)$ for $i\in \{1,\ldots, \ell\}$.
A \emph{directed path} in $\varrho_\q$ is a directed walk in which all $a_i$'s are  distinct except possibly for $a_1$ and $a_{\ell+1}$.
A \emph{directed cycle} in $\varrho_\q$ is a directed path in which $a_1=a_{\ell+1}$.

Recall that the support  of $\x\in \Z^A$ is    $\supp(\x)=\{a \in A: \x(a)\neq 0 \}$.
 \begin{theorem}[Cycle test]\label{theorem: cycle test}
Let $\Agn$ be a finite, locally irreducible, and strongly connected   agent network.
A configuration $\x.\q$  is recurrent if and only if all these conditions are satisfied:
\begin{enumerate}
[{
label=\textnormal{(C{\arabic*})},
ref={C\arabic*}}]
\item \label{i. cycle test 1} The vector $\x$ is nonnegative;
\item \label{i. cycle test 2}           The state $\q$ is locally recurrent; and
\item \label{i. cycle test 3} Every directed cycle of the rotor digraph $\varrho_\q$ contains a vertex in $\supp(\x)$.
\end{enumerate}
\end{theorem}

We remark that Theorem~\ref{intro theorem: cycle test} in \S\ref{subsection: intro rotor-routing} is the special case of Theorem~\ref{theorem: cycle test} when $\Net$ is a sinkless rotor network (so that $\varrho_\q = \q$).
\begin{figure}[tb]
\centering
   \includegraphics[width=0.7\textwidth]{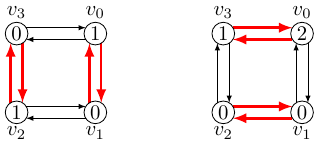}
   \caption{Two configurations in the  sinkless rotor network on the bidirected cycle $C_4$. The circled number by vertex $v_i$ indicates the number of chips $\x(v_i)$, and the (red) thick outgoing edge from $v_i$ records the rotor $\q(v_i)$.
   By the cycle test, the configuration on the left is recurrent while the configuration on the right is not recurrent.
    }  
   \label{figure:rotor_network_recurrent}
\end{figure}

{
Theorem~\ref{theorem: cycle test}  answers the question posed in~\cite{BL16c} for a characterization of recurrent configurations of agent networks.


The cycle test is often much more computationally efficient than the burning test~(Algorithm~\ref{a. burning test}).
In particular, for a sinkless rotor network on an $n$-vertex directed graph, conditions \eqref{i. cycle test 1}-\eqref{i. cycle test 3} can be checked in time linear in $n$.
}

 The following is  a corollary of Theorem \ref{theorem: cycle test} that we will use later in \S\ref{subsection: counting recurrent classes and configurations of agent networks}.
 
\begin{corollary}\label{c. one james bond}
 Let $\Agn$ be a finite, locally irreducible, and strongly connected   agent network.
Let  $\x$ and $\x'$  be nonnegative vectors such that $\supp(\x)=\supp(\x')$.
For any $\q \in Q$, the configuration $\x.\q$  is recurrent if and only if $\x'.\q$ is recurrent. \qed
\end{corollary}

 We now build toward the proof of Theorem~\ref{theorem: cycle test}, and we start with two technical lemmas.
Recall that, for any $w \in A^*$, we denote by $|w|$ the vector in $\N^A$ that counts the occurrences of each letter in $w$.

\begin{lemma}\label{lemma: paths in rotor digraphs}
 Let $\Agn$ be a finite and  locally irreducible agent network.
 Let $\q \in \Loc(\Net)$ and let 
  $a_1 \ldots  a_{\ell+1}$ be a directed path  in $\varrho_\q$.
 Write $w':=a_1\ldots a_\ell$ and $\q':=t_{a_1}^{-1}\cdots t_{a_\ell}^{-1}\q$,
then
\[|a_1|.\q' \xlongrightarrow{w'} |a_{\ell+1}|.\q. \]
\end{lemma}
\begin{proof}
We prove the claim by induction on $\ell$.
When $\ell=0$, 
the claim is true since $\w'$ is the empty word,
$a_1=a_{l+1}$, and $\q'=\q$.

We now prove the claim for when $\ell\geq 1$.
Write $w'':=a_2\ldots a_{\ell+1}$ and $\q'':=t_{a_2}^{-1}\cdots t_{a_\ell}^{-1}\q$.
By the induction hypothesis we have $|a_2|.\q'' \xlongrightarrow{w''} |a_{\ell+1}|.\q$.
Since $a_1$ is a legal execution for $|a_1|.\q'$
, 
it then suffices to show that $\pi_{a_1}(|a_1|.\q')=|a_2|.\q''$.

Now note that
\begin{align*}
 \NN_{w'}(\q')= \NN_{a_1}(\q')+\NN_{w''}(\q'')= \NN_{a_1}(\q')
 +|a_3| +\cdots+  |a_{\ell+1}|,
\end{align*}
where the last equality is due to $\pi_{w''}(|a_2|.\q'') = |a_{\ell+1}|.\q$.
Also note that
   \begin{align*}
   \begin{split}
   \NN_{w'}(\q')=&\NN_{a_{1}  \cdots a_\ell}(t_{a_1}^{-1}    \cdots  t_{a_\ell}^{-1} \q)\\
   =&\NN_{a_{2}  \cdots  a_\ell a_1}(t_{a_2}^{-1}    \cdots  t_{a_\ell}^{-1}  t_{a_1}^{-1}\q) \ \ \ \text{(by the abelian property (Lemma \ref{l. abelian enumerate}\eqref{l. abelian property}))}\\
=  & \NN_{a_{2} \cdots a_\ell}(t_{a_2}^{-1}   \cdots  t_{a_\ell}^{-1}  t_{a_1}^{-1}\q)    +\NN_{a_1}(t_{a_1}^{-1}\q)\\
\geq& \NN_{a_1}(t_{a_1}^{-1}\q)= |a_2|,
\end{split}
\end{align*}
where the last equality is because  $(a_1,a_2)$ is an edge in $\varrho_\q$.
These two equations then imply that
\begin{equation}\label{equation: equation 5 in paths in rotor digraphs}
\NN_{a_1}(\q')
 +|a_3| +\cdots+  |a_{\ell+1}| \geq |a_2|.  
\end{equation}

Now note that  $a_2 \notin \{a_3,\ldots,a_{\ell+1}\}$ since  $a_1\ldots a_{\ell+1}$ is a directed path in $\varrho_\q$.
It then follows from equation~\eqref{equation: equation 5 in paths in rotor digraphs} that $\NN_{a_1}(\q') \geq |a_2|$.  
Since $\Net$ is an agent network, we  conclude that $\NN_{a_1}(\q') = |a_2|$.
It then follows that $\pi_{a_1}(|a_1|.\q')=|a_2|.\q''$, and the proof is complete.
\end{proof}

Recall that $\rr$ denotes the period vector of $\Net$ (Definition~\ref{definition: period vector}).
Also recall  the definition of $w \setminus \n$ ($w \in A^*, \n \in \N^A$) from Definition~\ref{definition: removal}.

\begin{lemma}\label{p. solitary walk period}
 Let $\Agn$ be a finite, locally irreducible, and strongly connected   agent network.
Then  for any $\q \in \Loc(\Agn)$ and any $a \in A$
there
 exists  a legal   execution   $w$ for $|a|.\q$ such that $|w|(a)=\rr(a)+1$ and $|w|\leq \rr+|a|$.
\end{lemma}
 \begin{proof}
Fix  a letter $a\in A$. 
 Let $w'=a_1\cdots a_\ell$ be a word of maximum length such  that
  $w'$ is a legal  execution for   $|a|.\q$  and $|w'|\leq \rr$.

Write $a':=\NN_{a_\ell}(t_{a_1\cdots a_{\ell-1}}\q)$  and $w:=w'a'$.  
It follows that
$w$ is  a legal execution for $|a|.\q$.
Note that $|w|(a')=\rr(a')+1$, as otherwise 
 we would have $|w|\leq \rr$ and that  contradicts the maximality of $w$.
  Also note that  $|w|=|w'|+|a'|\leq \rr+|a'|$.

We  now show that $a'=a$.
Since $\Agn$ is an agent network and $w'$ is a legal execution for $|a|.\q$, we have $\NN_{a_i}(t_{a_1 \cdots a_{i-1}}\q)=|a_{i+1}|$ for any $i\in \{1,\ldots,\ell-1\}$.
Hence
\begin{align*}
\NN_{w'}(\q)=& \sum_{i=1}^\ell \NN_{a_i}(t_{a_1 \cdots a_{i-1}}\q)=
  \sum_{i=1}^{\ell-1} |a_{i+1}| +|a'|=|w|-|a_1|.
\end{align*}
Then
\begin{align}\label{equation: solitary walk}
|a_1|=|w|-\NN_{w'}(\q)\geq& |w|-\NN_{\rr}(\q)=|w|-\rr,
\end{align}
where the inequality is due to $|w'|\leq \rr$ and the monotonicity property (Lemma~\ref{l. abelian enumerate}\eqref{l. monotonicity}), and the last equality is due to $\q \in \Loc(\Net)$.
Since  $|w|(a')=\rr(a')+1$,
 equation~\eqref{equation: solitary walk} implies that $|a_1|(a') \geq 1$,
and hence we have $a_1=a'$.

Now note that  $a_1=a$ because $w=a_1\cdots a_\ell$ is a legal execution  for $|a|.\q$.
Hence  $a'=a_1=a$, and it then follows that $w$ satisfies the property in the lemma.
  \end{proof}

We now present the proof of Theorem \ref{theorem: cycle test}.
Recall that  a word $w \in A^*$ is called $a$-tight
if $|w| \leq \rr$ and $|w|(a)=\rr(a)$.

\begin{proof}[Proof of Theorem \ref{theorem: cycle test}]
 Proof of   if  direction: 
 Since $\q$ is locally recurrent by \eqref{i. cycle test 2},
 by Lemma \ref{c. tightness condition for recurrent configurations} it suffices to show that
 for each $a \in A$ there exists an $a$-tight legal execution $w$ for $\x.\q$.

Fix a letter $a \in A$.
Let $a_1, \ldots,  a_{\ell+1}$  be a directed path of minimum length in $\varrho_\q$ such that $a_1=a$
and $a_{\ell+1} \in \supp(\x)$. 
Note that such a directed path exists by   \eqref{i. cycle test 3}.
Write $w':=a_1\cdots a_{\ell}$ and  $\q':=t_{a_1}^{-1}  \cdots  t_{a_{\ell}}^{-1}\q$.
Note  that  $|a|.\q' \xlongrightarrow{w'} |a_{\ell+1}|.\q$ by  Lemma~\ref{lemma: paths in rotor digraphs}.
 Also note that  $|w'|(a)=1$  and $|w'|\leq \satu \leq \rr$ by the minimality assumption.

By Lemma \ref{p. solitary walk period}, there exists   an  legal execution $w''$ for $|a|.\q'$ such that $|w''|(a)= \rr(a)+1$ and $|w''|\leq \rr+|a|$.
Write $w:=w'' \setminus |w'|$.
By  the removal lemma (Lemma~\ref{lemma: removal lemma}),
$w$ is a legal execution for $|a_{\ell+1}|.\q$.
 Since  $\x\in \N^A$~(by \eqref{i. cycle test 1}) and $a_{\ell+1} \in \supp(\x)$,
by Lemma \ref{o. to legal properties}\eqref{o. to contagious} we conclude that $w$ is  a legal execution for $\x.\q$ .

We now show that $w$ is $a$-tight.
Note that
\begin{align}
      |w|=&\max (|w''|,|w'|)-|w'| \notag\\
      \leq& \max (|w''|,|w'|)-|a|  \quad \text{(since } |w'|(a)=1) \label{equation: cycle test equation 1}\\
      \leq&\rr+|a|-|a|  \quad \text{(since } |w''| \leq \rr+|a| \text{ and } |w'| \leq \rr)  \label{equation: cycle test equation 2}\\
       =& \rr. \notag
\end{align}
Also note that  we have equality for the $a$-th coordinate in equation~\eqref{equation: cycle test equation 1} (because $|w'|(a)=1$) and equation~\eqref{equation: cycle test equation 2} (because $|w''|(a)=\rr(a)+1$).
Hence we conclude that  $|w|\leq \rr$ and $|w|(a)=\rr(a)$,
i.e.,  the word  $w$ is $a$-tight.
This completes the proof.

 Proof of  only if direction:
 It suffices to show that \eqref{i. cycle test 3} holds, as \eqref{i. cycle test 1} and \eqref{i. cycle test 2}
 follow from Lemma \ref{p. properties of recurrent configurations}.
Let $a_1, \ldots, a_{\ell+1}$ be any directed cycle in $\varrho_\q$.
Note that $a_{\ell+1}=a_1$ by assumption.
 We need to show that  $\{a_1,\ldots, a_\ell\} \cap \, \supp(\x)$ is nonempty.

 By Theorem \ref{t. recurrence test}, there exists a legal execution $w$ for $\x.\q$ such that $|w|=\rr$ and $\x.\q \xlongrightarrow{w}\x.\q$.
Write  $\n:=\rr-\sum_{i=1}^\ell |a_i|$ and  $w':=w\setminus \n$.
 Note that $\n$ is a nonnegative vector (because $\rr\geq \satu$ and $a_1,\ldots, a_\ell$ are  distinct), and $w'$ is a permutation of the word $a_1\ldots a_\ell$.
   Write $\x'.\q':=\pi_{\n}(\x.\q)$.
  By the removal lemma,
  we have $\x'.\q' \xlongrightarrow{w'}\x.\q$.
  
 Since  $w'$ is  legal for $\x'.\q'$ and $w'$ is a permutation of $a_1\ldots a_l$, we have 
 $\supp(\x')\cap     \{a_1, \ldots, a_\ell\}$ is nonempty.
On the other hand, since $\pi_{w'}(\x'.\q')=\x.\q$,  we have  
 \begin{align*}
  \x=&\x'+\NN_{w'}(\q')-|w'| 
=\x'+|a_{\ell+1}|-|a_{1}| \qquad \text{(by Lemma~\ref{lemma: paths in rotor digraphs})}\\
=&\x'.
 \end{align*}
 In particular, we have $\supp(\x)=\supp(\x')$.
Hence we conclude that $\supp(\x) \cap \{a_1,\ldots, a_\ell\}$ is nonempty, as desired. 
\end{proof}

 \section{Counting recurrent components}

 \label{subsection: counting recurrent classes and configurations of agent networks}
 
In this section we turn to the problem of counting the number of  recurrent components of an agent network.

We start with the following lemma.
Recall the definition of capacity from Definition~\ref{definition: capacity}.
Also recall that a configuration $\x.\q$ is stable if $\x\leq \nol$,
and is halting if there exists  a stable configuration $\x'.\q'$
such that $\x.\q \longrightarrow \x'.\q'$.
\begin{lemma}\label{l. agent network is critical networks with capacity 0}
Let $\Net$ be a finite, locally irreducible, and strongly connected critical network.
\begin{enumerate}
\item If $\Net$ is an agent network, then $\cpt(\Net)=0$.
\item \label{item: agent network capacity 0} If  $\cpt(\Net)=0$ and all states of $\Net$ are locally recurrent, then $\Net$ is an agent network.
\end{enumerate}
\end{lemma}

\begin{proof}
\begin{enumerate}[wide, labelwidth=!, labelindent=10pt]
\item  
By  equation~\eqref{equation: satu is the left eigenvector of the production matrix of agent networks}  the exchange rate vector $\s$  (Definition~\ref{definition: exchange rate vector})  of an agent network is equal to $\satu$.
By the definition of capacity,
it suffices to show that
any configuration $\x.\q$  of $\Agn$ with $\satu^\top \x>0$ does not halt.

Let $w \in A^*$ be any word and  let $\x'.\q'$ be any configuration such that 
   $\x.\q \xlongrightarrow{w}\x'.\q'$.
   Then
   \[\satu^\top \x' =\satu^\top( \x+\NN_{w}(\q)-|w|)=\satu^\top \x +\satu^\top \NN_{w}(\q)-\satu^\top|w|=\satu^\top \x>0,\]
   where the third equality is due to $\Net$ being an agent network.
Hence $\x'.\q'$ is not a stable configuration.
Since the choice of $w$ and $\x'.\q'$ is arbitrary, this shows that $\x.\q$ does not halt, as desired.

\item Since $\cpt(\Net)=0$,  for any $a \in A$ and $\q \in Q$
 the configuration $|a|.\q$ does not halt.
 In particular the letter $a$ is not a complete execution for $|a|.\q$,
 and hence $\satu^\top \NN_a(\q)\geq 1$.
Therefore,  for all $w \in A^*$ and $\q \in Q$ we have  $\NN_w(\q)\geq \satu^\top |w|$, and the equality is achieved only if 
$\satu^\top\NN_{w'}(\q)=\satu^\top |w'|$ for all  $w' \in A$ satisfying $|w'|\leq |w|$. 

Let $\rr$ be the period vector of $\Net$.
Note that for any $\q \in Q$,
\begin{align*}
\satu^\top \rr=\satu^\top P\rr= \satu^\top \NN_{\rr}(\q)\geq \satu^\top \rr,
\end{align*}
where the second equality is due to the assumption that $\q \in \Loc(\Net)=Q$, and 
 the  inequality is due to the conclusion in the previous paragraph.
Since equality happens in the equation above and $\rr\geq \satu$,  we conclude that $\satu^\top\NN_{a}(\q)=1$ for all $a \in A$.
Hence $\Net$ is an agent network.  \qedhere
\end{enumerate}
\end{proof}

\begin{remark}
The condition in Lemma~\ref{l. agent network is critical networks with capacity 0}\eqref{item: agent network capacity 0} that every state in $\Net$ is locally recurrent is necessary.
Indeed, let $\Net$ be a network with  states $Q:=\{\q_1,\q_2\}$,  with alphabet  $A:=\{a\}$, 
and with  transition functions given by
\[  t_a(\q_1)=\q_2; \quad  \NN_a(\q_1)=2|a|;\qquad t_a(\q_2)=\q_2; \quad  \NN_a(\q_2)=|a|.  \]
This network has capacity zero, and yet is not an agent network since $ \satu^\top \NN_a(\q_1)=2$.
\end{remark}

Recall that for any  $m \in \N$, the set   $\Lrec(\Agn,m)$ denotes the set of recurrent components (Definition~\ref{definition: recurrent class}) with level $m$.
Also recall that $\Tor(\Agn)$ denotes the torsion group of $\Agn$ (Definition~\ref{definition: torsion group}).

\begin{proposition}
Let $\Net$ be a finite, locally irreducible, and strongly connected agent network.
Then
\label{proposition: number of recurrent classes of agent networks}
\begin{equation*}
|\Lrec(\Agn,m)|=\begin{cases}
0 & \text{ if } m= 0;\\
|\Tor(\Agn)| & \text{ if } m\geq 1. 
\end{cases}
\end{equation*}
\end{proposition}
\begin{proof}
By Lemma~\ref{p. properties of recurrent configurations}\eqref{p. properties of recurrent configuration, x is nonzero} the level of a  recurrent configuration is  strictly positive, and by Lemma~\ref{lemma: relation between recurrent configurations and recurrent classes} the same is true for recurrent components.
This proves the case when $m=0$.

We now prove the case when $m\geq 1$.
Since $\cpt(\Agn)=0$ by Lemma~\ref{l. agent network is critical networks with capacity 0} and $\s=\satu$ by equation \eqref{equation: satu is the left eigenvector of the production matrix of agent networks},
we have $\Stop(\Agn)=\{0\}$ by  Lemma~\ref{lemma: unstoppable levels}.
 Theorem~\ref{theorem: torsion group for critical networks}\eqref{item: the action of torsion group of critical networks is free and transitive}
then implies that $|\Lrec(\Agn,m)|=|\Tor(\Agn)|$ for all $m \geq 1$, as desired.
\end{proof}

\begin{remark}
As a comparison to Proposition~\ref{proposition: number of recurrent classes of agent networks},
the quantity $|\Lrec(\Net,m)|$
for the sinkless sandpile network (which is a non-agent network) on an undirected graph $G$
 is   the number of spanning trees of $G$ with {external activity}  at most  $m-|E|$~\cite[Theorem~1.3]{Chan18}. 
The assumption that $G$ is an undirected graph  can be relaxed to that $G$ is an Eulerian digraph; see \cite{Chan18}.
\end{remark}

 \section[Determinantal formula]{Determinantal generating functions for recurrent configurations} 
\label{subsection: determinantal formula recurrent configurations}

We now turn to the problem of counting the recurrent configurations of an agent network. We will derive two versions of a multivariate generating function identity.

The first identity counts recurrent configurations according to the number of chips at each vertex. 
For any  $\n \in \N^A$ and $m \in \N$, we write
\begin{align*}
\Rec(\Agn,\n):=& \{ \x.\q \mid   \x.\q \text{ is } \Agn\text{-recurrent and } \x=\n\}.
\end{align*}
Let $(z_a)_{a \in A}$ be indeterminates indexed by $A$.
We denote by $I(z)$ the $A \times A$ diagonal  matrix 
with $I(z)(a,a):=\frac{1}{1-z_a}$ ($a \in A$).

\begin{theorem}[Determinantal formula for agent networks]
\label{t. enumerate determinant}
Let $\Agn$ be a finite, locally irreducible, and strongly connected agent network.
Then, in the ring of formal power series with $(z_a)_{a \in A}$ as indeterminates, we have the following identity:  
\[|\Z^A/K| \,  \det \left(   I(z) -P \right)= \sum_{\n \in \N^A } |\Rec(\Agn,\n)| z^\n.\]
\end{theorem}

The second identity is a refinement of Theorem~\ref{t. enumerate determinant} for the special case of sinkless rotor networks,
which involves
 edge variables that keep track of the rotor configuration.

For a digraph $G$, which may have multiple edges, 
let $(y_e)_{e \in E}$ and $(z_v)_{v \in V}$ be indeterminates indexed by edges of $G$ and by vertices of $G$, respectively.
We denote by $A_G(y)$ the weighted adjacency matrix indexed by $V$ given by $A_G(y)(u,v):=\sum_{e} y_e$, where the sum is taken over all edges with source vertex $v$ and target vertex $u$.
We denote by $D_G(y,z)$ the diagonal matrix indexed by $V$ with $D_G(y,z)(v,v):=\frac{1}{1-z_v}\sum_{e \in \Out(v)} y_e$.
We denote by $\Z[y][[z]]$ the ring of formal power series in the $(z_v)_{v \in V}$ variables whose coefficients are polynomials in the $(y_e)_{e \in E}$ variables.

\begin{theorem}[Master determinant for rotor networks]
\label{theorem: determinantal formula for sinkless rotor networks}
Let $\Agn$ be a  sinkless  rotor network on a strongly connected digraph $G$.
Then, in the ring $\Z[y][[z]]$ we have the following identity of formal power series:
\[   \det \left( D_G(y,z) -\A_G(y) \right)=\sum_{\x.\q  \in \Rec(\Net) } z^\x  \, y_{\q}, \]
where  $y_\q:=\prod_{v \in V} y_{\q(v)}$.  
\end{theorem} 

We remark that this identity is a refinement of the matrix-tree theorem: to count the number $t(G,r)$ of spanning trees oriented toward $r$, set $z_v = 0$ for all $v \neq r$ and compare coefficients of $z_r$. The term $z_r y_{\q}$ appears in the sum on the right side if and only if $\q$ is a unicycle with $r$ contained in its unique cycle. The number of such unicycles is $\outdeg(r) t(G,r)$.
Theorem~\ref{theorem: determinantal formula for sinkless rotor networks} can be compared to the determinants that enumerate cycle-rooted spanning forests \cite[Theorem~1]{Forman93} and their oriented counterparts \cite[Theorem~6]{Kenyon11}.

We remark that Theorem~\ref{intro theorem: determinantal formula for sinkless rotor networks} in \S\ref{subsection: intro torsion group of abelian networks} is a direct corollary of Theorem~\ref{theorem: determinantal formula for sinkless rotor networks}
by substituting $y_e=1$ for all $e \in E$ and $z_v=z$ for all $v \in V$.

We now build towards the proof of these two theorems.
We start with a  lemma that refines Proposition~\ref{proposition: the state of a recurrent configuration is a recurrent state of its thief networks} for agent networks.

Recall the definition of thief networks $\Net_R$ from \S\ref{ss. abelian network with thief}.
Also recall the definition of recurrence for configurations~(Definition~\ref{definition: recurrent configuration})
and states~(Definition~\ref{definition: recurrent state}). 
   \begin{lemma}\label{lemma: recurrent configurations of agent network is essentially a recurrent state of subcritical networks}
   Let $\Agn$ be a finite, locally irreducible, and strongly connected agent network. 
Let  $\x \in \N^A \setminus \{\nol\}$ and  let $R:=A \setminus \supp(\x)$.
Then    
  $\x.\q$ is an $\Agn$-recurrent configuration if and only if  $\q$ is an $\Agn_R$-recurrent state.
   \end{lemma}

   \begin{proof}
Let $\rr$ be the period vector of $\Net$.   
   Note that $\supp((I-P_R)\rr)=A \setminus R=\supp(\x)$.
   By Corollary  \ref{c. one james bond},
   the configuration $\x.\q$ is $\Net$-recurrent if and only if 
   $(I-P_R)\rr.\q$ is $\Net$-recurrent.
   The lemma now follows from Proposition~\ref{proposition: recurrence for thief implies a very specific recurrence in the original network}.
   \end{proof}

The following   corollary of  Lemma~\ref{lemma: recurrent configurations of agent network is essentially a recurrent state of subcritical networks}  generalizes the characterization of recurrent states for  rotor networks with sinks  in  \cite[Lemma~3.16]{HLM08}.

\begin{corollary}\label{corollary: characterization of recurrent states for thief networks of an agent network}
   Let $\Agn$ be a finite, locally irreducible, and strongly connected agent network, and let $R \subsetneq A$.
   Then  $\q \in \Loc(\Net)$ is an $\Net_R$-recurrent state if and only if  every directed cycle in the rotor digraph $\varrho_\q$ contains a vertex in $R$. 
\end{corollary}
\begin{proof}
The corollary follows by applying Theorem~\ref{theorem: cycle test} and Lemma~\ref{lemma: recurrent configurations of agent network is essentially a recurrent state of subcritical networks} to the configuration $\satu_R.\q$.
\end{proof}

We now quote a result from \cite{BL16c} that counts the number of recurrent states in a subcritical network.
\begin{lemma}[{\cite[Theorem 3.3]{BL16c}}]\label{lemma: lemma from BL16c that we cite}
Let $\Subnet$ be a finite, locally irreducible, and  subcritical  abelian network with  total kernel $K$ and production matrix $P$.
Then
the  number of recurrent states of $\Subnet$ is equal to $|\Z^A/K| \det(I-P)$. \qed
\end{lemma}

We now present the proof of Theorem~\ref{t. enumerate determinant}.
For an $A \times A$  matrix $M$    and  $R \subseteq A$,
we denote by $\det(M;R)$  the determinant of the matrix  obtained from deleting the rows and columns of $M$ indexed by $A \setminus R$.

\begin{proof}[Proof of Theorem~\ref{t. enumerate determinant}]
Since $\Rec(\Net,\nol)=\varnothing$  by Lemma~\ref{p. properties of recurrent configurations}\eqref{p. properties of recurrent configuration, x is nonzero}, we have
\begin{align*}
\sum_{\n \in \N^A} |\Rec(\Agn,\n)| z^\n &=\sum_{R\subsetneq A} \sum_{\substack{\n \in \N^A ; \\ \supp(\n)=A \setminus R}} |\Rec(\Agn,\n)| z^\n.
\end{align*}
Then 
\begin{align*}
&\sum_{\n \in \N^A} |\Rec(\Agn,\n)| z^\n =\sum_{R \subsetneq A}|\Rec(\Agn_R)  |  \, \prod_{a \in A \setminus R}\frac{z_a}{(1-z_a)}  \quad 
\text{ (by Lemma~\ref{lemma: recurrent configurations of agent network is essentially a recurrent state of subcritical networks})}\\
 = &\sum_{R \subsetneq A} |\Z^A/K| \,  \det(I-P_R)  \, \prod_{a \in A \setminus R}\frac{z_a}{1-z_a}   \quad \text{(by Lemma~\ref{lemma: lemma from BL16c that we cite})}\\
 = &|\Z^A/K| \sum_{R \subsetneq A}  \,  \det(I-P;R)   \det \left( I(z) - I;{A \setminus R} \right) \\
= & |\Z^A/K| \, \det \left( I-P+I(z) - I\right)=|\Z^A/K| \, \det \left(I(z) - P\right). \qedhere
\end{align*}
\end{proof}

We now build towards the proof of Theorem~\ref{theorem: determinantal formula for sinkless rotor networks}.
A key ingredient in the refinement  is the following extended version of the matrix tree theorem.

Let $S$ be a  subset of $V$.
 A subgraph $\F$ of $G$ is a \emph{directed forest rooted at} $S$ if 
every vertex in $S$ has outdegree 0, 
 every vertex in $V \setminus S$ has outdegree 1, and the underlying graph of $\F$ has no cycles.

\begin{lemma}[Extended matrix tree theorem~\cite{Chaiken82}]
\label{lemma: extended matrix tree theorem}
Let $G$ be a  digraph, and let $S$ be a  subset of $V$.
Then
\[ \det(D_G(y,\nol)-A_G(y);V\setminus S)=\sum_{\F} \prod_{e \in E(\F)} y_e, \]
where the sum is taken over all directed forests of $G$ rooted at $S$. \qed
\end{lemma}

\begin{remark}
The standard  matrix tree theorem (i.e., when $y_e=1$ for all $e \in E$) can be derived from Theorem~\ref{theorem: cycle test} and Theorem~\ref{theorem: determinantal formula for sinkless rotor networks} by applying the operator 
 $\frac{\partial^{|S|}}{(\partial z_v)_{v \in S}} \biggl |_{z=\nol}$ to the equation in Theorem~\ref{theorem: determinantal formula for sinkless rotor networks} for when $\Net$ is  a sinkless rotor network on $G$.
\end{remark}

We now present the proof of Theorem~\ref{theorem: determinantal formula for sinkless rotor networks}.

\begin{proof}[Proof of Theorem~\ref{theorem: determinantal formula for sinkless rotor networks}]
We have
\begin{align*}
\sum_{\x.\q  \in \Rec(\Net) } z^\x \, y_{\q}= \sum_{S  \subseteq A} \sum_{\substack{\x.\q  \in \Rec(\Net);\\ \supp(\x)=S}} z^\x \, y_{\q}.
\end{align*}
Note that $\Net$ is strongly connected since $G$ is strongly connected.
By Theorem~\ref{theorem: cycle test}, a configuration $\x.\q$ with $\supp(\x)=S$ is recurrent if and only if 
the digraph $\F$ given by 
\[V(\F)=V(G),  \qquad E(\F)=\{\q(v) \mid v \notin S  \},\]
 is a directed forest rooted at $S$.
 It then follows that
 \begin{align*}
&\sum_{\x.\q  \in \Rec(\Net) } z^\x  \, y_{\q}\\
=& \sum_{S  \subseteq A}  \det(D_G(y,\nol)-A_G(y);V \setminus S)  \prod_{v \in S} \sum_{e \in \Out(v)} \frac{ y_e z_v}{1-z_v} \quad \text{(by Lemma~\ref{lemma: extended matrix tree theorem})}\\
=&\sum_{S  \subseteq A}  \det(D_G(y,\nol)-A_G(y);V \setminus S) \det( D_G(y,z)-D_G(y);S)\\
=& \det(D_G(y,\nol)-A_G(y)+D_G(y,z)-D_G(y))\\
=&\det(D_G(y,z)-A_G(y)). \qedhere
\end{align*}
\end{proof}

%
%
%

\chapter{Concluding Remarks}
 We conclude with  a few directions for future research.

\section[Unified notion of recurrence]{A unified notion of recurrence and burning test}
We have seen the definition of recurrent states~(Definition \ref{definition: recurrent state}) and recurrent configurations~(Definition \ref{definition: recurrent configuration}) for subcritical and critical networks, respectively, which play a central role in the dynamics of abelian networks.
In both cases we have a burning test (Theorem~\ref{theorem: burning test subcritical} for subcritical, and Theorem~\ref{t. recurrence test} for critical networks) to check  recurrence.

A natural next step would be to extend the definition of recurrence to supercritical networks and beyond.

\begin{question}
Give a definition of recurrence for all (finite, locally irreducible, strongly connected) networks that specializes to Definition \ref{definition: recurrent state}   and  Definition \ref{definition: recurrent configuration} for subcritical and critical networks, respectively. 
\end{question}

This unified definition of recurrence should come with a burning test that specializes to Theorem~\ref{theorem: burning test subcritical} and 
Theorem~\ref{t. recurrence test} for subcritical and critical networks, respectively. 

\section[Forbidden subconfiguration test]{Forbidden subconfiguration test for recurrence}

For a configuration $\x.\q$ of a sandpile network on a simple Eulerian digraph $(V,E)$, a nonempty set $U \subset V$ is called a \emph{forbidden subconfiguration} \cite{Dhar90} if
	\[ \x(u) + \q(u) < \# \{v \in U \,:\, (v,u) \in E \} \]
for all $u \in U$. Likewise, let us define a \emph{forbidden subconfiguration} in a rotor network as a set $U$ such that either
	\begin{enumerate}
	\item $U = \{u\}$ and $\x(u) < 0$; or
	\item the rotors $\{\q(u) \,:\, u \in U\}$ form an oriented cycle, and $\x(u)=0$ for all $u \in U$.
	\end{enumerate}
By the critical burning test (Theorem~\ref{t. recurrence test}) in the sandpile case, and the cycle test (Theorem~\ref{theorem: cycle test}) in the rotor case, $\x.\q$ is a recurrent configuration if and only if it has no forbidden subconfigurations. It would be interesting to characterize the forbidden subconfigurations of other critical networks, such as the McKay-Cartan networks.

\begin{question}
Give a recurrence test for sinkless height-arrow networks on Eulerian digraphs
that specializes to the forbidden subconfiguration test  for sandpile and rotor networks.
\end{question}  

%
%
%

\section[Counting recurrent configurations]{Number of recurrent configurations in a recurrent component}
\begin{figure}[tb]
\centering
   \includegraphics[width=0.7\textwidth]{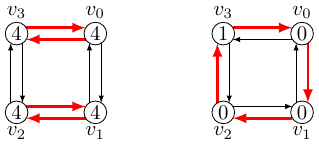}
   \caption{Two recurrent configurations in the sinkless rotor network on the bidirected cycle $C_4$, where  the circled number on $v\in V$ records $\x(v)$ and the (red) thick outgoing edge  of $v$ records the state $\q(v)$.
The configuration on the left has weight $2$ modulo $4$ (due to the $2$ counterclockwise red edges) and the configuration on the right has weight  $3$ modulo $4$ (due to the chip at $v_3$).}  
   \label{figure:rotor network weight}
\end{figure}

Consider the sinkless rotor network~(Example \ref{e. rotor network}) on the bidirected cycle $C_n$. 
The \emph{weight function} $\wt:E \to \Z_n$ for edges of $C_n$ is defined by 
\[\wt(e):=\begin{cases}1 & \text{ if } e=(v_k,v_{k-1}) \text{ for some } k \in \Z_n;\\
0 &\text{ otherwise.}
\end{cases}
\]
The \emph{weight function} $\wt:\Z^A\times Q \to \Z_n$ for configurations of $\Net$ is defined by
\[ \wt(\x.\q):= \sum_{k \in \Z_n} \x(v_k)k +\wt(\q(v_k)) \mod n. \]
See Figure~\ref{figure:rotor network weight} for  examples.

One can check that any execution in this network leaves the weight unchanged (i.e.,  $\wt(\x.\q)=\wt(\x'.\q')$ if $\x.\q \dashrightarrow\x'.\q'$).
In particular, weight depends only on the component a configuration is contained in.
One can also check that, for any positive $m$ and $i \in \Z_n$, there exists a unique recurrent component that has  total number of chips $m$  and weight $i$.
We denote this recurrent component by $\C_{n,m,i}$.



{    \renewcommand{\arraystretch}{1.2}
\begin{table}[tb]
\caption{ Counts of the number of recurrent configurations in some   recurrent components of the sinkless rotor network on the bidirected cycle $C_n$.
The $(i,n)$-th entry of the table corresponds to the recurrent component with weight   $i$ and total number of chips $n$.
 }
{\footnotesize
\begin{tabular}{|c|c |c |c|c|c|c|c|c|}
\hline
 \backslashbox{$n$ }{$i$}  & 0 & 1 &2 &3 &4 &5 &6&7  \\\hline
3  & 26 & 24 & 24  & & & & &  \\
\hline 
4  & 122 & 120 & 118 & 120   & & & &  \\
\hline 
5  & 642 & 640 & 640 & 640 & 640 &  & &  \\
\hline 
6  & 3630 & 3624 & 3624 & 3630 & 3624 &  3624&   & \\
\hline
7  & 21394 & 21392 & 21392 & 21392 & 21392 &  21392& 21392 &   \\
\hline
8  & 130090& 130080 & 130072 & 130080 & 130086 &  130080 & 130072 &130080   \\
\hline
\end{tabular}
}
 \label{tbl. cardinality of recurrent classes of cycles}
\end{table}
}

Let $r(\C_{n,m,i})$ denote the number of recurrent configurations in the recurrent component $\C_{n,m,i}$. 
Table~\ref{tbl. cardinality of recurrent classes of cycles} shows the values $r(\C_{n,n,i})$   for  small $n$.
An intriguing feature of this table is the near equality of entries in each row.  How fast does $\max_{i,j \in \Z_n} |r(\C_{n,n,i})-r(\C_{n,n,j})|$ grow?

In some cases the equality is exact: Data for small $m, n$ support the following conjecture.

\begin{conjecture}\label{conj. symmetry of recurrent classes}
For $n\geq 3, m\geq 1$ and $i,j \in \Z_n$,
we have $r(\C_{n,m,i})=r(\C_{n,m,j})$   whenever 
$\gcd(n,m,i)=\gcd(n,m,j)$.
\end{conjecture}
 
 The  case when $i-j$ is divisible by $\gcd(n,m)$ is a consequence of rotational symmetry, but the general case seems more mysterious.

\section*{Acknowledgement}
The authors would like to thank Darij Grinberg for careful proofreading of earlier drafts, and anonymous referees for helpful comments.

\backmatter
\bibliographystyle{amsalpha}
\bibliography{critical}
\printindex
\end{document}